\documentstyle[12pt,twoside,epsf]{report}  
\newcommand{\ncm}{\newcommand}
\renewcommand{\theequation}{\thesection.\arabic{equation}}
\newcommand{\sectiona}[1]{\setcounter{equation}{0}\section{#1}}
\newcommand{\aanhangsel}{\def\thesection{\thechapter.\Alph{section}}
\setcounter{section}{0}}
\ncm{\oH}{\bar{H}}
\ncm{\us}{\quad\mbox{using}\quad}
\ncm{\ra}{\rightarrow}
\ncm{\ot}{\otimes}
\ncm{\DH}{D(H)}
\ncm{\DW}{D^{\omega}(H)}
\ncm{\TH}{T(\oH)}
\ncm{\ssc}{\displaystyle}
\ncm{\oq}{\eta} 
\ncm{\im}{\imath}
\ncm{\ba}{\begin{array}}
\ncm{\ea}{\end{array}}
\ncm{\ul}{\underline}
\ncm{\ol}{\overline}

\def\hoek{\hbox{\vrule height 2.5ex depth 0pt \vrule width 2.5ex height .4pt
 depth 0pt}}
\def\haak#1#2{
\mathop{\hoek\llap{\vbox to 2.5ex{ \vfil
\hbox{$\scriptstyle#1$\hskip 2.8ex} \vfil}}}
\limits_{#2} }
\def\hook#1#2{\setbox0=\hbox{$\scriptstyle#1$}
\hskip\wd0\haak{\box0}{#2}}
\ncm{\str}{\rule{0cm}{3.5mm}}
\ncm{\om}{\omega}
\ncm{\ep}{\epsilon}
\newlength{\extraspace}
\newlength{\extraspaces}
\newcommand{\be}{\begin{equation}
\addtolength{\abovedisplayskip}{\extraspaces}
\addtolength{\belowdisplayskip}{\extraspaces}
\addtolength{\abovedisplayshortskip}{\extraspace}
\addtolength{\belowdisplayshortskip}{\extraspace}}
\newcommand{\ee}{\end{equation}}

\makeatletter

\def\eqnfourarray{\stepcounter{equation}%
  \def\@currentlabel{\p@equation\theequation}\global\@eqnswtrue \m@th
  \global\@eqcnt\z@ \tabskip\@centering
  \let\\\@eqncr \let\@@eqncr\@@eqnfourcr
  $$\everycr{}\halign to\displaywidth \bgroup
  \hskip\@centering $\displaystyle \tabskip\z@skip {##}$\@eqnsel &%
  \global\@eqcnt\@ne \hskip\tw@\arraycolsep \hfil ${##}$\hfil &%
  \global\@eqcnt\tw@ \hskip\tw@\arraycolsep $\displaystyle {##}$\hfil &%
  \global\@eqcnt\thr@@ \hskip\tw@\arraycolsep \hfil \hbox{##}%
    \tabskip\@centering &%
  \global\@eqcnt 4 \hbox\bgroup \hss ##\egroup
    \tabskip\z@skip \cr}

\def\endeqnfourarray{\@@eqncr \egroup
  \global\advance\c@equation\m@ne $$\global\@ignoretrue}

\def\@@eqnfourcr{\let\reserved@a\relax
  \ifcase\@eqcnt \def\reserved@a{& & & &}%
    \or \def\reserved@a{& & &}%
    \or \def\reserved@a{& &}%
    \or \def\reserved@a{&}%
    \else \let\reserved@a\@empty
      \@latex@error{Too many columns in eqnfourarray environment}\@ehc
  \fi
  \reserved@a \if@eqnsw\@eqnnum \stepcounter{equation}\fi
  \global\@eqnswtrue \global\@eqcnt \z@\cr}

\makeatother

\newcommand{\bea}{\begin{eqnarray}
\addtolength{\abovedisplayskip}{\extraspaces}
\addtolength{\belowdisplayskip}{\extraspaces}
\addtolength{\abovedisplayshortskip}{\extraspace}
\addtolength{\belowdisplayshortskip}{\extraspace}}
\newcommand{\eea}{\end{eqnarray}}
\newcommand{\beas}{\begin{eqnarray*}
\addtolength{\abovedisplayskip}{\extraspaces}
\addtolength{\belowdisplayskip}{\extraspaces}
\addtolength{\abovedisplayshortskip}{\extraspace}
\addtolength{\belowdisplayshortskip}{\extraspace}}
\newcommand{\eeas}{\end{eqnarray*}}
\ncm{\Z}{{\bf Z}}
\ncm{\al}{\alpha}
\ncm{\bt}{\beta}
\ncm{\gm}{\gamma}
\ncm{\dl}{\delta}
\ncm{\varep}{\varepsilon}
\ncm{\zt}{\zeta}
\ncm{\et}{\eta}
\ncm{\th}{\theta}
\ncm{\kp}{\kappa}
\ncm{\lm}{\lambda}
\ncm{\rh}{\rho}
\ncm{\hl}{\hline}
\ncm{\sg}{\sigma}
\ncm{\ta}{\tau}
\ncm{\ph}{\phi}
\ncm{\phv}{\varphi}
\ncm{\ch}{\chi}
\ncm{\ps}{\Phi}
\ncm{\nn}{\nonumber}
\title{ \vspace{-5cm}
{\flushleft {\normalsize November 1995
\hfill PAR--LPTHE 95-46, ITFA-95-20, hep-th/9511201}}\\[2.5cm]
DISCRETE GAUGE THEORIES\thanks{Lectures presented by F.A. Bais
at the CRM-CAP Summer School `Particles and Fields 94', Banff,
Alberta, Canada, August 16-24, 1994}\vspace{1.5cm}}
\author{Mark de Wild Propitius\thanks{e-mail: mdwp@lpthe.jussieu.fr}\\
\normalsize{ {\em
Laboratoire de Physique Th\'eorique et Haute Energies}}\thanks{Laboratoire 
associ\'e No. 280 au CNRS}\\
\normalsize{{\em Universit\'e Pierre et Marie Curie - PARIS VI}} \\
\normalsize{{\em Universit\'e Denis Diderot - PARIS VII}} \\
\normalsize{{\em 4 place Jussieu, Boite 126, Tour 16, 1$^{er}$ \'etage}}\\
\normalsize{{\em F-75252 Paris CEDEX 05, France}} \\ \\
F. Alexander Bais\thanks{e-mail: bais@phys.uva.nl}\\
\normalsize{{\em Instituut voor Theoretische Fysica}} \\
\normalsize{{\em Universiteit van Amsterdam}} \\
\normalsize{{\em Valckenierstraat 65, 1018XE Amsterdam, 
The Netherlands}}} 
\date{}
\begin{document}
\maketitle
\begin{abstract}
In these lecture notes, we present a self-contained treatment of planar gauge 
theories broken down to some finite residual gauge group $H$ via the Higgs 
mechanism. The main focus is on the discrete $H$ gauge theory describing the 
long distance physics of such a model. The spectrum features global $H$ 
charges, magnetic vortices and dyonic combinations. Due to the Aharonov-Bohm 
effect, these particles exhibit topological interactions. Among other things, 
we review the Hopf algebra related to this discrete $H$ gauge theory, which 
provides an unified description of the spin, braid and fusion properties of 
the particles in this model. Exotic phenomena such as flux metamorphosis, 
Alice fluxes, Cheshire charge, (non)abelian braid statistics, the generalized 
spin-statistics connection and nonabelian Aharonov-Bohm scattering are 
explained and illustrated by representative examples. 
\end{abstract}
\newpage

\tableofcontents

\chapter*{Broken symmetry revisited}
\addcontentsline{toc}{chapter}{Broken symmetry revisited}

Symmetry has become one of the major  
guiding principles in physics during the twentieth century.
Over the past ten decades, we have  gradually
progressed from external to internal, 
from global to local, from finite to infinite,
from ordinary to supersymmetry and quite recently arrived at the notion
of Hopf algebras or quantum groups.

In general, a physical system  consists of 
a finite or infinite number of degrees of freedom which may or may not 
interact. The dynamics is prescribed by a set of 
evolution equations which follow from varying the action 
with respect to the different degrees of freedom.
A symmetry then corresponds to  a group of 
transformations on the space time coordinates and/or the 
degrees of freedom that leave the action 
and therefore also the evolution equations invariant.
External symmetries have to do  with invariances (e.g.\ Lorentz invariance)
under transformations on the space time  coordinates.
Symmetries not related to transformations of space time 
coordinates are  called internal symmetries. We also discriminate 
between global symmetries and local symmetries. 
A global or rigid symmetry transformation is the same 
throughout space time and usually leads to a conserved quantity.
Turning a global symmetry into a local symmetry, i.e.\ 
allowing the symmetry transformations to vary continuously 
from one point in space time to another,
requires the introduction of additional {\em gauge} degrees 
of freedom mediating a force. 
It is this so-called gauge principle that has eventually led 
to the extremely successful standard model 
of the strong and electro-weak interactions between the 
elementary particles based on the local gauge group
$SU(3) \times SU(2) \times U(1)$.

The use of symmetry considerations has been extended significantly 
by the observation that a symmetry of the action is {\em not} 
automatically a symmetry of the groundstate of a physical system.
If the action is invariant under some symmetry group $G$
and the groundstate only under a subgroup $H$ of $G$, the 
symmetry group $G$ is said to be spontaneously broken down to $H$.
The symmetry is not completely lost though, for the broken generators 
of $G$ transform one groundstate into another.

The physics of a broken global symmetry is quite different from a broken 
local (gauge) symmetry.
The signature of a broken continuous {\em global} symmetry group $G$ 
in a physical system is  the occurrence 
of massless scalar degrees of freedom, the so-called Goldstone bosons. 
Specifically, each broken generator of $G$
gives rise to a massless Goldstone boson field. 
Well-known realizations of Goldstone bosons 
are the long range spin waves in a ferromagnet,
in which  the rotational symmetry is broken below the Curie temperature
through the appearance of spontaneous magnetization.  
An application in particle physics is the low energy physics of 
the strong interactions, where the spontaneous
breakdown of (approximate) chiral symmetry leads to 
(approximately) massless pseudoscalar particles such as the pions.

In the case of a broken {\em local} (gauge) symmetry, in contrast, 
the would be massless Goldstone bosons conspire with the massless 
gauge fields to form massive vector fields. 
This celebrated phenomenon is known as the Higgs mechanism. 
The canonical example in condensed matter physics 
is the ordinary superconductor. 
In the phase transition from the normal to the superconducting phase,
the $U(1)$ gauge symmetry is spontaneously broken 
to the finite cyclic group $\Z_2$ by a condensate of Cooper pairs.
This leads to a mass $M_A$ for the photon field in the 
superconducting medium as witnessed by the Meissner effect:
magnetic fields are expelled from a superconducting region and 
have a characteristic penetration depth which in proper units 
is just the inverse of the photon mass $M_A$.
Moreover, the Coulomb interactions among external electric 
charges in a superconductor are of finite range $\sim 1/M_A$.
The Higgs  mechanism also plays a key role in the
unified theory of weak and electromagnetic interactions, that is, the
Glashow-\-Weinberg-\-Salam model where the product gauge group
$SU(2)\times U(1)$ is broken to the $U(1)$ subgroup of
electromagnetism. In this context, the massive vector particles 
correspond to the $W$ and $Z$ bosons mediating the short range weak 
interactions. 
More speculative applications of the Higgs mechanism 
are those where the
standard model of the strong, weak and electromagnetic interactions is
embedded in a grand unified model with a large simple gauge group. The
most ambitious attempts invoke supersymmetry as well.

In addition to the aforementioned  characteristics in the spectrum of
fundamental excitations, there are in general 
other fingerprints of a broken symmetry in a physical system.
These are usually called {\em topological excitations} or 
just {\em defects} and correspond to collective degrees of freedom 
carrying `charges' or quantum numbers which are conserved for 
topological reasons, not related to a manifest symmetry of the action.  
(See, for example, the references~\cite{cola, mermin, presbook, raja} 
for reviews).
It is exactly the appearance of these
topological charges which renders the corresponding collective
excitations stable.  
Topological excitations may manifest themselves
as particle-like, string-like or planar-like objects (solitons), 
or have to be interpreted as quantum mechanical tunneling processes
(instantons).  Depending on the model in which they occur,
these excitations carry evocative names like kinks, domain walls, vortices,
cosmic strings, Alice strings, monopoles, skyrmions, texture,
sphalerons and so on.
Defects are crucial for a full understanding of
the physics of systems with a broken symmetry and lead to a host
of rather unexpected and exotic phenomena which are in general of a 
nonperturbative nature.

The prototypical example of a topological defect is the 
Abrikosov-Nielsen-Olesen flux tube 
in the type~II superconductor with broken $U(1)$ gauge 
symmetry~\cite{abri, niels}.
The topologically conserved quantum number characterizing these defects 
is the magnetic flux, which indeed can only take discrete values. 
A  beautiful but unfortunately not yet observed  example in particle physics 
is the 't Hooft-Polyakov monopole~\cite{thooftmon, polyamon} occurring 
in any  grand unified model in which  a simple gauge group $G$ is
broken to a subgroup $H$ containing the electromagnetic $U(1)$
factor. Here, it is the quantized magnetic charge carried by these monopoles
that is conserved for topological reasons. In fact, 
the discovery that these models support magnetic monopoles 
reconciled the
two well-known arguments for the quantization of electric
charge, namely Dirac's argument based on the existence of a
magnetic monopole~\cite{dirac} and 
the obvious fact that the $U(1)$ generator should
be compact as it belongs to a larger compact gauge group.

An example of a model with a broken global symmetry supporting 
topological excitations is  the effective sigma model describing the
low energy strong interactions for the mesons.  That is, the
phase with broken chiral symmetry mentioned before.  One may add a
topological term and a stabilizing term to the action and obtain a
theory that features topological particle-like objects called
skyrmions, which have exactly the properties of the baryons. 
See reference~\cite{skyrme} and also~\cite{fink, witsk}.
So, upon extending the effective model for the Goldstone bosons, 
we recover the
complete spectrum of the underlying strong interaction model
(quantum chromodynamics) and its low energy dynamics. Indeed,
this picture leads to an attractive phenomenological model for
baryons.

Another area of physics where defects  may play a fundamental
role is cosmology.
See for instance reference~\cite{brand} for a recent review. 
According to the standard cosmological hot big bang scenario, 
the universe cooled down through a sequence of 
local and/or global symmetry breaking phase transitions in a very early stage. 
The question of the actual formation of defects in these phase 
transitions is of prime importance. It has been argued, for instance,
that magnetic monopoles might have been produced copiously.
As they tend to dominate the mass in the universe, however, 
magnetic monopoles are notoriously hard to accommodate and if indeed formed,
they have to be `inflated away'. 
Phase transitions that see the production of (local or global) 
cosmic strings, on the 
other hand, are much more interesting.
In contrast with magnetic monopoles, the presence of
cosmic strings does not lead to cosmological disasters
and according to an attractive but still speculative theory cosmic
strings may even have 
acted as seeds for the formation 
of galaxies and other large scale structures in the present day universe.

Similar symmetry breaking 
phase transitions are extensively studied in condensed matter physics. 
We have already mentioned the transition 
from the normal to the superconducting phase in superconducting 
materials of type~II, which may give rise to
the formation of magnetic flux tubes. 
In the field of low temperature physics, there also exists a great body of 
both theoretical and experimental work 
on the transitions from the normal to the 
many superfluid phases of helium-3 in which  
line and point defects arise in a great variety, 
e.g.~\cite{volovik}. 
Furthermore, in uniaxial 
nematic liquid crystals,  point defects, 
line defects and texture arise
in the transition from the disordered to the ordered phase
in which the rotational global symmetry group $SO(3)$ is broken down 
to the semi-direct product group $U(1) \times_{\rm s.d.} \Z_2$.
Bi-axial nematic crystals, in turn, exhibit a phase 
transition in which the global rotational symmetry group is broken down  
to the product group $\Z_2 \times \Z_2$ yielding 
line defects labeled by the elements of the (nonabelian)
quaternion group $\bar{D}_2$, e.g.~\cite{mermin}. 
Nematic crystals are cheap materials and as compared to helium-3,
for instance, relatively easy to work with in the laboratory. 
The symmetry breaking phase transitions typically appear at temperatures
that can be reached by a standard  kitchen oven, 
whereas the size of the occurring defects 
is such that these can be seen  by means of a simple microscope. 
Hence, these materials form an easily accessible experimental playground 
for the investigation of defect producing phase transitions
and as such may partly mimic the physics of the early universe in the 
laboratory.
For some recent ingenious experimental 
studies on the formation and the dynamics of topological 
defects in nematic crystals making use of high speed film 
cameras, the interested reader is referred to~\cite{bowick, turok}.

From a theoretical point of view,
many aspects of topological defects have been studied and understood.  
At the classical level, one may roughly sketch
the following programme. One first uses simple topological
arguments, usually of the homotopy type, to see whether a given model
does exhibit topological charges. Subsequently, one may try to prove
the existence of the corresponding classical solutions by functional
analysis methods or just by explicit construction of particular
solutions. On the other hand, one may in many cases determine the
dimension of the solution or moduli space and its dependence on the
topological charge using index theory.  
Finally, one may attempt to determine the
general solution space more or less explicitly.  In this respect,
one has been successful in varying degree. 
In particular, the self-dual
instanton solutions of Yang-Mills theory on $S^4$ have been
obtained completely.

The physical properties of topological defects can be probed by
their interactions with the ordinary particles or excitations
in the model.  This amounts to investigating
(quantum) processes in the background of the defect. In particular, one
may calculate the one-loop corrections to the various quantities
characterizing the defect, which involves studying 
the fluctuation operator.  
Here, one naturally has to distinguish the modes with zero eigenvalue
from those with nonzero eigenvalues. The nonzero modes generically
give rise to the usual renormalization effects, such as mass and coupling
constant renormalization. The zero modes, which often arise as a
consequence of the global symmetries in the theory, lead to
collective coordinates.  Their quantization yields a semiclassical
description of the spectrum of the theory in a given topological
sector,  including the external quantum numbers of the soliton 
such as its energy and momentum and its internal quantum numbers such
as its electric charge, e.g.\ \cite{cola,presbook}.

In situations where the residual gauge group $H$ is nonabelian, 
the analysis outlined in the previous paragraph is rather subtle. 
For instance, 
the naive expectation that a soliton can
carry internal electric charges which form representations of the
complete unbroken group $H$ is wrong. As only
the subgroup of $H$ which commutes with the topological charge can be
globally implemented, these internal charges form representations 
of this so-called centralizer subgroup. (See~\cite{balglob, nelson, nelsonc} 
for the case of magnetic monopoles and~\cite{spm, balglob2} 
for the case of magnetic vortices). 
This makes the full spectrum of topological and
ordinary quantum numbers in such a  broken phase rather intricate.

Also, an important effect on the spectrum and the interactions 
of a theory with a broken gauge group 
is caused by the introduction of additional
topological terms in the action, such as a nonvanishing
$\theta$ angle in 3+1 dimensional space time 
and the Chern-Simons term in 2+1 dimensions.  
It has been shown by Witten that 
in case of a nonvanishing $\theta$ angle, for example,
magnetic monopoles carry electric charges
which are shifted by an amount proportional to $\theta/2\pi$ and their
magnetic charge~\cite{thetaw}.

Other results are even more surprising. A broken gauge theory only
containing bosonic fields
may support topological excitations (dyons),
which on the quantum level carry half-integral spin and are fermions,
thereby realizing the counterintuitive possibility to make
fermions out of bosons~\cite{thora, jare}. 
It has subsequently been argued by Wilczek~\cite{wilcchfl} 
that in 2+1 dimensional space time
one can even have topological excitations, 
namely flux/charge composites, which behave as anyons~\cite{leinaas}, 
i.e.\ particles with fractional spin and quantum statistics 
interpolating between bosons and fermions. 
The possibility of anyons in two
spatial dimensions is not merely of academic interest, as many systems 
in condensed matter physics, for example, are effectively described
by 2+1 dimensional models.
Indeed, anyons are known to be realized as quasiparticles 
in fractional quantum Hall systems~\cite{hal, laugh}.
Further, it has been been shown that an ideal gas of electrically 
charged anyons is superconducting~\cite{chen, fetter, laughsup, anyonbook}.
At present it is unclear whether this new and rather exotic type 
of superconductivity is actually realized in nature.

Furthermore, remarkable calculations by 't Hooft
revealed a nonperturbative mechanism for baryon
decay in the standard model through instantons 
and sphalerons~\cite{thooftinst}.
Afterwards, Rubakov and Callan  discovered
the phenomenon of baryon decay catalysis induced by 
grand unified monopoles~\cite{callan, ruba}. Baryon number violating 
processes also occur in the vicinity of grand unified cosmic strings as 
has been established by Alford, March-Russell and 
Wilczek~\cite{alfbar}.

So far, we have given a (rather incomplete) enumeration of 
properties and processes that involve
the interactions between topological and ordinary excitations. 
However, the interactions between defects themselves 
can also be highly nontrivial. Here, one should not only 
think of ordinary interactions corresponding to the 
exchange of field quanta.
Consider, for instance, the case
of Alice electrodynamics which occurs if some nonabelian gauge group 
(e.g.\ $SO(3)$) is broken to the nonabelian subgroup 
$U(1) \times_{\rm s.d.} \Z_2$,
that is, the semi-direct product of the electromagnetic group $U(1)$ and 
the additional cyclic group $\Z_2$ whose nontrivial element 
reverses the sign of the electromagnetic fields~\cite{schwarz}.
This model features magnetic monopoles and in addition a magnetic $\Z_2$ 
string (the so-called Alice string) 
with the miraculous property that if a monopole 
(or an electric charge for that matter) 
is transported around the string, 
its charge will change sign.
In other words, a particle is converted into its own anti-particle.
This nonabelian analogue of the celebrated 
Aharonov-Bohm effect~\cite{ahabo} is of a topological nature.
That is, it only depends on the number of times the particle 
winds around the string and is independent of the distance 
between the particle and the string.

Similar phenomena occur in models in which a continuous gauge group
is spontaneously broken down to some {\em finite} subgroup $H$.
The topological defects supported by such a model are string-like
in three spatial dimensions and carry a magnetic flux corresponding 
to an element $h$ of the residual gauge  group $H$. 
As these string-like objects trivialize
one spatial dimension, we may just as well descend to the plane, for 
convenience.
In this arena, these defects become magnetic vortices, i.e.\
particle-like objects of characteristic size $1/M_H$ with $M_H$ the 
symmetry breaking scale. 
Besides these topological particles, 
the broken phase features matter 
charges labeled by the unitary irreducible representations $\Gamma$ 
of the residual gauge group $H$. 
Since all gauge fields are massive, there are no ordinary 
long range interactions among these  particles.
The remaining long range interactions are  
topological Aharonov-Bohm interactions. 
If the residual gauge group $H$ is 
nonabelian, for instance, the nonabelian fluxes $h \in H$ carried by the 
vortices exhibit flux metamorphosis~\cite{bais}. In the process 
of circumnavigating one vortex with another vortex their fluxes may change.
Moreover, if a charge corresponding to some representation $\Gamma$ of 
$H$ is transported around a vortex carrying the magnetic flux $h \in H$, 
it returns transformed by the matrix $\Gamma(h)$ assigned to the 
element $h$ in the representation~$\Gamma$ of $H$.

The spontaneously broken 2+1 dimensional models 
just mentioned will be the subject 
of these lecture notes. One of our aims is to show that the long distance 
physics of such a model is, in fact, governed by a Hopf algebra or 
quantum group based on the residual finite gauge 
group $H$~\cite{spm,spm1,sm,thesis}.
This algebraic framework manifestly unifies the topological
and nontopological quantum numbers as dual aspects of a single symmetry 
concept.    
The results presented here strongly suggests
that revisiting the symmetry breaking concept in general will
reveal similar underlying algebraic structures.

The outline of these notes, which are intended to be accessible to a reader
with a minimal background in field theory, quantum mechanics and finite group
theory,  is as follows. 
In chapter~\ref{physasp}, we start with a review of 
the basic physical properties of a 
planar gauge theory broken down to a finite gauge group via the Higgs 
mechanism. The main focus will be on the discrete $H$ gauge theory
describing the long distance of such a model. 
We argue that  in addition to the aforementioned
magnetic vortices and global $H$ charges 
the complete spectrum also consists of dyonic combinations 
of the two and  establish the basic topological interactions 
among these particles. In chapter~\ref{algstruct}, we then turn to the  
Hopf algebra related to this discrete $H$ gauge theory and elaborate
on the unified description this framework gives of the spin, braid and fusion
properties of the particles.
Finally,  the general formalism 
developed in the foregoing  chapters is illustrated by an explicit 
nonabelian example in chapter~\ref{exampled2b}, 
namely a planar gauge theory spontaneously broken down  
to the double dihedral gauge group $\bar{D}_2$. 
Among other things, exotic phenomena like 
Cheshire charge, Alice fluxes, nonabelian braid statistics 
and nonabelian Aharonov-Bohm scattering
are explained there.

Let us conclude this preface with some remarks on conventions.
Throughout these notes units in 
which $\hbar = c =1$ are employed.
Latin indices take the values $1,2$. Greek indices 
run from $0$ to $2$. Further, 
$x^1$ and $x^2$ denote spatial coordinates and $x^0=t$
the time coordinate. 
The signature of the three dimensional metric 
is taken as $(+,-,-)$.  Unless stated otherwise, we adopt 
Einstein's summation convention.

\chapter{Basics}
\label{physasp}

\sectiona{Introduction}

The planar gauge theories we will study in these notes 
are given by an action of the general form 
\bea                               \label{algz}
S &=& S_{\rm YMH } + S_{\rm matter}.
\eea  
The continuous gauge group $G$ of this model is assumed to be 
broken down to some finite subgroup $H$ of $G$ by means of 
the Higgs mechanism. 
That is, the Yang-Mills Higgs part $S_{\rm YMH }$ of the action features 
a Higgs field whose nonvanishing vacuum expectation values are only 
invariant under the action of $H$. Further, the matter part $S_{\rm matter}$ 
describes matter fields covariantly coupled to the gauge fields.
A discussion of the implications of 
adding a Chern-Simons term to the spontaneously broken planar 
gauge theory~(\ref{algz}) is beyond the scope of these notes.
For this, the interested reader is referred to~\cite{spm1, sm, sam, thesis}.

Since the unbroken  gauge group $H$ is finite, 
all gauge fields are massive and it seems that the 
low energy or equivalently the long distance 
physics of the model~(\ref{algz})
is trivial. This is not the case though. It is the occurrence of topological 
defects and the persistence of 
the Aharonov-Bohm effect that renders 
the long distance physics nontrivial.
Specifically, the defects supported by these models 
are (particle-like) vortices of characteristic size $1/M_H$, 
with $M_H$ the symmetry breaking scale. These vortices
carry magnetic fluxes labeled by the elements 
$h$ of the residual gauge group $H$.~\footnote{Here, we tacitly assume that 
the broken gauge group $G$ is simply connected. If $G$ is not simply 
connected and the model does not contain Dirac monopoles/instantons, 
then the vortices carry fluxes labeled by the elements of the lift 
$\bar{H}$ of $H$ into the universal covering group $\bar{G}$ of $G$.
See section~\ref{topclas} in this connection.}
In other words, the vortices introduce 
nontrivial holonomies in the locally flat gauge fields.
Consequently, if the residual gauge group ${H}$ is nonabelian,
these fluxes exhibit nontrivial topological interactions:
in the process in which one vortex circumnavigates another, the 
associated magnetic fluxes feel 
each others' holonomies and affect each other through conjugation.  
This is in a nutshell the long distance 
physics described by the Yang-Mills Higgs part 
$S_{\rm YMH}$ of the action.
The matter fields, covariantly coupled to the gauge fields
in the matter part $S_{\rm matter}$ of the action, 
form multiplets which transform irreducibly 
under the broken gauge group ${ G}$. In the 
broken phase, these branch to irreducible representations 
of the residual gauge group ${ H}$.
So, the matter fields introduce point charges in the broken  phase 
labeled by the 
unitary irreducible representations $\Gamma$ of $H$.
When such a charge encircles a magnetic 
flux $h \in H$, it undergoes an Aharonov-Bohm effect:
it  returns transformed by the matrix $\Gamma(h)$ assigned to the 
group element $h$ in the representation $\Gamma$ of $H$.

In this chapter, we establish 
the complete spectrum 
of the discrete $H$ gauge theory describing  the long distance 
physics of the spontaneously broken model~(\ref{algz}), which
besides the aforementioned  matter charges and magnetic vortices
also consists of dyons obtained by composing these charges and vortices.
In addition, we elaborate on the basic topological interactions between 
these particles. The discussion is organized as follows.
In section~\ref{braidgroups}, we start by briefly recalling 
that particle interchanges in the plane are organized by braid groups.
Section~\ref{abznz} then contains an analysis of  
a planar abelian Higgs model in which the $U(1)$ gauge group 
is spontaneously broken to the cyclic subgroup $\Z_N$.
The main emphasis will be on the $\Z_N$ gauge theory that describes the 
long distance physics of this model. 
Among other things, we show that the spectrum indeed consists of  
$\Z_N$ fluxes, $\Z_N$ charges and dyonic 
combinations of the two, establish the quantum mechanical 
Aharonov-Bohm interactions between these particles and argue that 
as a result the wave functions of the multi-particle configurations 
in this model realize nontrivial abelian representations of 
the related braid group.
Finally, the subtleties involved in the generalization to models in 
which a nonabelian gauge group $G$ is broken to a nonabelian 
finite group $H$ are dealt with in section~\ref{nonabz}.

\sectiona{Braid groups}    \label{braidgroups}

Let us consider a system of $n$ indistinguishable particles moving on a 
manifold $M$, which is assumed to be 
connected and path connected for convenience. 
The classical configuration space of this system 
is given by
\bea                                  \label{configi}
{\cal C}_n ({M}) &=& (M^{n} - D)/S_n,
\eea 
where the action of the permutation group $S_n$ on the particle 
positions is divided out to account for the indistinguishability 
of the particles. Moreover, the singular configurations $D$ 
in which two or more particles coincide are excluded.  
The configuration space~(\ref{configi}) is in 
general multiply-connected. This means that there are different 
kinematical options to  quantize this multi-particle system. 
To be precise, there is a quantization
associated to each  unitary irreducible representation (UIR) of the 
fundamental group $\pi_1 ({\cal C}_n ({M}))$.
See, for instance, the references~\cite{imma, laid, schul, schul2}.

It is easily verified that for manifolds $M$ with dimension larger then 2, 
we have the isomorphism $\pi_1({\cal C}_n ({M})) \simeq S_n$. Hence, 
the inequivalent quantizations of multi-particle systems moving on 
such manifolds are labeled by the UIR's of the permutation group $S_n$. 
There are two  1-dimensional UIR's of $S_n$. The trivial representation
naturally corresponds with Bose statistics. In this case, 
the system is quantized by a (scalar) wave function, which 
is symmetric under all permutations of the particles.
The anti-symmetric representation, on the other hand,
corresponds with Fermi statistics, i.e.\
we are dealing with a wave function which acquires a minus sign 
under odd permutations of the particles.
Finally, parastatistics is also conceivable.
In this case, the system is quantized by a multi-component
wave function which transforms as a 
higher dimensional UIR of $S_n$.

\begin{figure}[htb]    \epsfxsize=10cm
\centerline{\epsffile{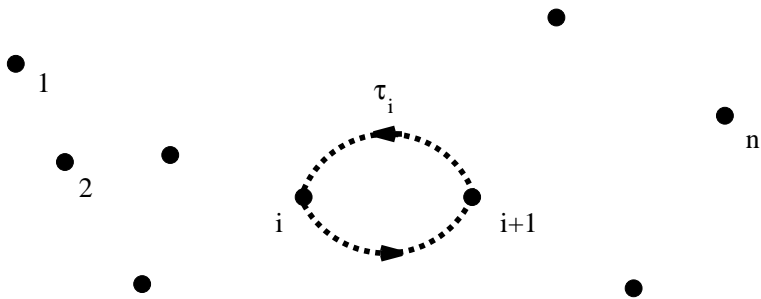}}
\caption{\sl The  braid operator $\tau_i$ establishes a counterclockwise 
interchange of the particles  $i$ and $i+1$ in a set of $n$
numbered indistinguishable particles in the plane.}
\label{touw}
\end{figure}

\begin{figure}[htb]    \epsfxsize=9.5cm
\centerline{\epsffile{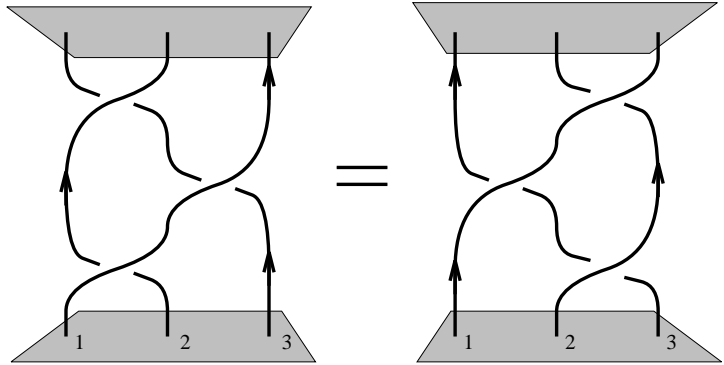}}
\caption{\sl 
Pictorial presentation of the braid relation 
$\tau_1 \tau_2 \tau_1= \tau_2 \tau_1 \tau_2$.
The particle trajectories corresponding to the composition of exchanges 
$\tau_1 \tau_2 \tau_1$ (diagram at the l.h.s.)
can be continuously deformed into the trajectories associated with 
the composition of exchanges 
$\tau_2 \tau_1 \tau_2$ (r.h.s.\ diagram).}
\label{ybes}
\end{figure}

\begin{figure}[htb]    \epsfxsize=11.5cm
\centerline{\epsffile{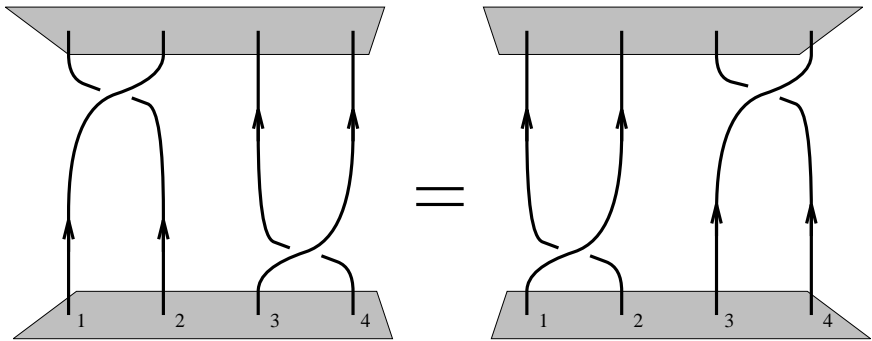}}
\caption{\sl
The braid relation $\tau_1 \tau_3 = \tau_3 \tau_1$
expresses the fact that the particle trajectories 
displayed in the l.h.s.\ diagram  can be continuously 
deformed into the trajectories in the r.h.s.\ diagram.}
\label{ybes2}
\end{figure}

It has been known for some time that quantum statistics for identical 
particles moving in the plane ($M={\bf R}^2$) can be much more exotic then 
in three or more dimensions~\cite{leinaas,wilcan}. 
The point is that the fundamental group of the associated 
configuration space  ${\cal C}_n ({\bf R}^2)$ is not given by 
the permutation group, but rather by the so-called braid 
group $B_n ({\bf R}^2)$~\cite{wu}.
In contrast with the permutation group $S_n$, 
the braid group $B_n ({\bf R}^2)$ is a nonabelian group of 
{\em infinite} order. It is 
generated by $n-1$ elements  $\tau_1, \ldots, \tau_{n-1}$, where $\tau_i$
establishes  a counterclockwise interchange 
of the particles $i$ and $i+1$ as depicted 
in figure~\ref{touw}. These generators are subject to the relations
\bea
\label{yangbax}
\ba{rcll} 
\tau_i\tau_{i+1}\tau_i &=& \tau_{i+1}\tau_i\tau_{i+1} & \qquad
 i=1,\ldots,n-2  \\
\tau_i\tau_j & = & \tau_j\tau_i & \qquad |i-j|\geq 2,
\ea
\eea  
which can be presented graphically as 
in figure~\ref{ybes} and~\ref{ybes2} respectively.
In fact, the permutation group $S_n$ ruling 
the particle exchanges in three or more dimensions,
is given by the same set of generators
with relations~(\ref{yangbax}) {\em and} the additional relations $\tau_i^2=e$
for all $i \in 1, \ldots, n-1$. 
These last relations are absent for 
$\pi_1({\cal C}_n ({\bf R}^2)) 
\simeq B_n ({\bf R}^2)$, since in the plane
a counterclockwise particle interchange $\tau_i$ ceases to 
be homotopic to the clockwise interchange $\tau_i^{-1}$.

The one dimensional UIR's of the 
braid group $B_n ({\bf R}^2)$ 
are labeled by an angular parameter $\Theta \in [0, 2\pi)$ 
and are defined  by assigning the 
same phase factor to all generators. That is, 
\bea                                \label{qstat} 
\tau_i  
&\mapsto& \exp (\im \Theta),
\eea
for all $i \in 1, \ldots, n-1$. 
The quantization of a system of $n$ identical particles in the plane
corresponding to an arbitrary but fixed $\Theta \in [0, 2\pi)$
is then given by a multi-valued (scalar) wave function that generates
the  quantum statistical phase $\exp(\im \Theta)$ upon a counterclockwise 
interchange of two adjacent particles.  
For $\Theta =0$ and $\Theta = \pi$,  we are 
dealing with bosons and fermions respectively.
The particle species related to other values 
of $\Theta$ have been called  anyons~\cite{wilcan}.
Quantum statistics deviating from conventional 
permutation statistics is known under various names in the literature,
e.g.\ fractional statistics, anyon statistics and exotic statistics.
We adopt the following nomenclature. 
An identical particle system  described by a (multi-valued) 
wave function that transforms as an one dimensional (abelian) UIR of 
the braid group $B_n ({\bf R}^2)$ ($\Theta \neq 0, \pi$)
is said to realize abelian braid statistics.
If an identical particle system is described by a multi-componenent 
wave function carrying an higher dimensional UIR of the braid group, 
then the particles are said to obey nonabelian braid statistics.

\begin{figure}    \epsfxsize=10cm
\centerline{\epsffile{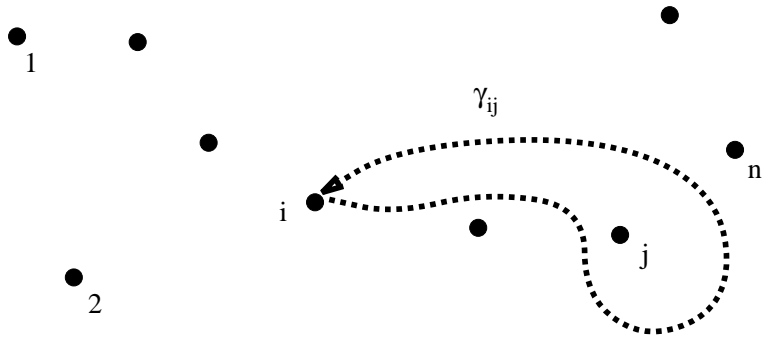}}
\caption{\sl The monodromy operator
$\gamma_{ij}$ takes particle $i$ counterclockwise
around particle $j$.}
\label{gammaije} 
\end{figure}

A system of $n$ distinguishable particles moving in the 
plane, in turn, is described by the 
non-simply connected configuration space
\bea
{\cal Q}_n ({\bf R}^{2}) &=& ({\bf R}^{2})^n - D.
\eea
The fundamental group of this configuration space is 
the so-called colored braid group $P_n({\bf R}^2)$, also known as the  
pure braid group. The colored braid group $P_n({\bf R}^2)$
is the subgroup of the ordinary braid group $B_n({\bf R}^2)$ 
generated by the monodromy operators
\bea                         \label{pbge}
\gamma_{ij} &:=& \tau_i \cdots \tau_{j-2} \tau_{j-1}^2 \tau_{j-2}^{-1}\cdots
             \tau_i^{-1}   \qquad \qquad \mbox{with} \; 1 \leq i<j \leq n.
\eea
Here, the $\tau_i$'s are the generators of $B_n({\bf R}^2)$ acting 
on the set of $n$ numbered 
distinguishable particles as displayed in figure~\ref{touw}.
It then follows from the definition~(\ref{pbge}) that 
the monodromy operator $\gamma_{ij}$ takes 
particle $i$ counterclockwise around particle $j$ as 
depicted in figure~\ref{gammaije}. The different 
UIR's of $P_n({\bf R}^2)$ now  label  
the inequivalent ways to quantize a system of $n$ 
distinguishable particles in the plane. 
Finally, a planar system that consists of a subsystem of identical particles 
of one type, a subsystem of identical particles of another type and so on, 
is of course also conceivable. 
The fundamental group of the configuration space of such a system 
is known as a partially colored braid group. 
Let the total number of particles of this system again be $n$, 
then the associated partially colored braid group 
is the subgroup of the ordinary braid group $B_n({\bf R}^2)$ 
generated by the braid operators that interchange identical 
particles and the monodromy operators acting on distinguishable 
particles. See for example~\cite{brekfa,brekke}.

To conclude, the fundamental excitations 
in planar discrete gauge theories, namely 
magnetic vortices and matter charges, are in principle bosons.
As will be argued in the next sections, 
in the first quantized description, these particles
acquire braid statistics through the Aharonov-Bohm effect.
Hence, depending on whether we are dealing with a system of 
identical particles, a system of distinguishable particles or 
a mixture, the associated multi-particle wave function
generically transforms as a nontrivial representation 
of the ordinary braid group, colored braid group
or partially colored braid group respectively.

\sectiona{$\Z_N$ gauge theory} \label{abznz}

The simplest example of a broken gauge theory
is an $U(1)$ gauge theory spontaneously broken down to the cyclic 
subgroup $\Z_N$.
This symmetry breaking scheme occurs in  an
abelian Higgs model in which the field that condenses carries
charge $Ne$, with $e$ the fundamental charge unit~\cite{krawil}.
The case $N=2$ is in fact realized in the ordinary BCS superconductor, as 
the field  that condenses in the BCS superconductor 
is that associated with the Cooper pair carrying charge $2e$.

This section is devoted to a 
discussion of such an abelian Higgs model in 2+1 dimensional space time.
We focus on the $\Z_N$ gauge theory describing the long range physics.
The organization is  as follows. 
In section~\ref{ahm}, we will start with a brief review of the 
screening mechanism for the electromagnetic fields 
of  external matter charges $q$ in the Higgs phase. 
We will argue that the external matter charges, which are multiples 
of the fundamental charge $e$ rather then multiples of the Higgs charge
$Ne$, are surrounded by screening charges provided by the Higgs condensate. 
These screening charges screen the electromagnetic fields around 
the external charges.  Thus, 
no long range Coulomb interactions persist among  the external charges.
The main point of section~\ref{mavoab} will be, however,
that the screening charges
do {\em not} screen the Aharonov-Bohm interactions between the external 
charges and the magnetic vortices, which also feature in this 
model. {\em As a consequence, 
long range Aharonov-Bohm interactions persist between the 
vortices and the external matter charges in the Higgs phase.} 
Upon circumnavigating a magnetic
vortex (carrying a flux $\phi$ which is a multiple of the 
fundamental flux unit $\frac{2\pi}{Ne}$ in this case) with 
an external charge $q$ (being a multiple of the fundamental charge unit $e$) 
the wave function of the system picks up the Aharonov-Bohm 
phase $\exp(\im q\phi)$.  
These Aharonov-Bohm phases lead to observable low energy scattering 
effects from which we conclude that the physically distinct 
superselection sectors in the Higgs phase can be labeled 
as $(a,n)$, where $a$ stands for the number of fundamental 
flux units $\frac{2\pi}{Ne}$ and $n$ for the number of fundamental 
charge units $e$. In other words, 
the spectrum of the $\Z_N$ gauge theory in the 
Higgs phase consists of pure charges $n$, pure fluxes $a$ and dyonic
combinations. Given the remaining long range Aharonov-Bohm interactions, 
these charge and flux quantum numbers are defined modulo $N$.
Having identified the spectrum and the long range interactions
as the topological Aharonov-Bohm effect, we proceed with a
closer examination of this $\Z_N$ gauge theory in section~\ref{abdyons}.
It will be argued that multi-particle systems in 
general satisfy abelian braid statistics. 
That is, the wave functions realize one dimensional 
representations of the associated braid group. In particular,
identical dyons behave as anyons.
We will also discuss the composition rules 
for the charge/flux quantum numbers
when two particles are brought together.
A key result of this section is a topological proof of the spin-statistics
connection for the particles in the spectrum. 
This proof is of a general nature and applies to all the theories
that will be discussed in these notes.

\subsection{Coulomb screening}  \label{ahm}

The planar abelian Higgs model which we will study is governed by the 
following action
\bea        \label{abhigg}
S &=& \int d \, ^3 x \;
({\cal L}_{\rm YMH} + {\cal L}_{\rm matter}) \\
{\cal L}_{\rm YMH} &=& 
-\frac{1}{4}F^{\kappa\nu} F_{\kappa\nu} 
  +({\cal D}^\kappa \Phi)^*{\cal D}_\kappa \Phi - V(|\Phi|) \label{hiks} \\
{\cal L}_{\rm matter} &=& -j^{\kappa}A_{\kappa}. \label{matcoup} 
\eea
The Higgs field $\Phi$ is assumed to carry the charge $Ne$ 
w.r.t.\ the compact $U(1)$ gauge symmetry. In the conventions 
we will adopt in these notes, this means that the covariant derivative reads 
${\cal D}_{\rho}\Phi=(\partial_{\rho}+\im Ne A_{\rho})\Phi$. 
Furthermore, the potential  
\bea                         \label{poto}
V(|\Phi|) &=& \frac{\lambda}{4}(|\Phi|^2-v^2)^2  \qquad\qquad
 \lambda, v > 0,
\eea 
endows the Higgs field with a nonvanishing vacuum  expectation value
$|\langle \Phi \rangle|=v$, which implies that the 
global continuous $U(1)$ symmetry is spontaneously broken. 
However, in this particular model the symmetry is 
not completely broken. Under global symmetry transformations 
$\Lambda(\alpha)$, with $\alpha \in [0,2\pi)$ being the $U(1)$ parameter,
the ground states transform as
\bea                     \label{reszn}
\Lambda(\alpha) \langle \Phi \rangle
&=& e^{\im N \alpha} \langle \Phi \rangle,
\eea 
since the Higgs field is assumed to carry  the charge $Ne$. Clearly,
the residual symmetry group of the ground states is the finite 
cyclic group $\Z_N$ corresponding to the 
elements  $\alpha= 2\pi k/N$ with $k \in 0,1,\ldots,N-1$.    

Further, the field equations following  from 
variation of the action~(\ref{abhigg}) 
w.r.t.\ the vector potential $A_{\kappa}$ and the Higgs field $\Phi$
are simply inferred as
\bea                            
\partial_{\nu} F^{\nu\kappa}
&=& j^\kappa+j^\kappa_H      \label{maxequ}     \\
{\cal D}_\kappa {\cal D}^\kappa \Phi^* &=& -\frac{\partial V}{\partial \Phi},
\label{hequ}
\eea   
where 
\bea       \label{Higgscur}
j^{\kappa}_H &=& \im Ne(\Phi^*{\cal D}^{\kappa}\Phi - 
({\cal D}^{\kappa}\Phi)^*\Phi),
\eea 
denotes the Higgs current.

In this section, we will only be concerned with the Higgs screening
mechanism for the electromagnetic fields induced by the matter charges 
described by the conserved matter current $j^\kappa$ in~(\ref{matcoup}). 
For convenience, we discard the dynamics of the fields that
are associated with this current and simply treat $j^\kappa$ 
as being external.
In fact, for our purposes the only important feature of 
the current $j^\kappa$ is that it allows us to introduce global 
$U(1)$ charges $q$ in the Higgs medium, which are multiples of the fundamental
charge $e$ rather then multiples of the 
Higgs charge $Ne$, so that all conceivable charge sectors can be discussed.

Let us first recall 
some of the basic dynamical features of  this model.
First of all, the complex Higgs field 
\bea
\Phi(x) &=& \rho(x)\exp (\im \sigma(x)) ,
\eea 
describes two physical degrees of freedom: the charged Goldstone boson field
$\sigma(x)$ and the physical field $\rho(x)-v$ with mass 
$M_H=v \sqrt{2\lambda}$ corresponding to the charged neutral Higgs particles.
The  Higgs mass $M_H$ sets the characteristic energy scale of this model.
At energies larger then $M_H$, the massive Higgs particles can 
be excited. At energies smaller then $M_H$ on the other hand,
the massive Higgs particles can not be excited. 
For simplicity we will restrict ourselves to 
the latter low energy regime. In that case,
the Higgs field is completely condensed, i.e.\ it acquires ground state values
everywhere
\bea                                     \label{simple}
\Phi(x) & \longmapsto & \langle \Phi(x) \rangle \; = \; v \exp (\im \sigma(x)).
\eea
The condensation of the Higgs field implies 
that in the low energy regime, the  Higgs model is 
governed by the effective action obtained from
the action~(\ref{abhigg}) by the following simplification 
\bea                               \label{efhig}
{\cal L}_{\rm YMH} & \longmapsto & -\frac{1}{4}F^{\kappa\nu} F_{\kappa\nu} 
+\frac{M_A^2}{2} \tilde A^{\kappa}\tilde A_{\kappa} \\
\tilde{A}_{\kappa} & := & A_{\kappa} + \frac{1}{Ne}\partial_{\kappa}
\sigma
\label{Atild}  \\
 M_A & := & Ne v\sqrt{2}.      \label{mal}
\eea
Thus, the  dynamics of the Higgs medium arising here 
is described by the effective field equations inferred 
from varying the effective
action w.r.t.\ the gauge field $A_\kappa$ and the Goldstone boson $\sigma$
respectively    
\bea   \label{fieldhp}                            
\partial_\nu F^{\nu\kappa} &=& 
j^\kappa+j^\kappa_{\rm scr}                         \\
\partial_\kappa  j^\kappa_{\rm scr} &=& 0,       \label{fieldhp2}
\eea
with
\bea                                \label{scrcurr}
j^\kappa_{\rm scr} &=& - M_A^2 \tilde{A}^\kappa,
\eea
the simple form the Higgs current~(\ref{Higgscur}) takes in the 
low energy regime.

It is easily verified that the field equations~(\ref{fieldhp}) 
and~(\ref{fieldhp2}) can be cast
in the following  form
\bea                     \label{kleingord}
(\partial_{\nu} \partial^{\nu} + M_A^2) \tilde{A}^\kappa &=& j^\kappa \\
\partial_{\kappa} \tilde{A}^\kappa &=& 0,
\eea
which clearly indicates that the gauge invariant 
vector field $\tilde{A}_\kappa$  has become massive.
More specifically, in this 2+1 dimensional setting 
it describes a two component massive photon field 
carrying the mass $M_A$ defined in~(\ref{mal}). 
Consequently, the electromagnetic fields around sources in the Higgs medium
decay exponentially with mass $M_A$.
Of course, the number of degrees of freedom is conserved.
We started with an unbroken 
theory with two physical degrees of freedom $\rho-v$ and $\sigma$ 
for the Higgs field and one for the massless gauge field $A_{\kappa}$.
After spontaneous symmetry breaking the Goldstone boson $\sigma$
conspires with the gauge field $A_{\kappa}$ to form a massive 
vector field $\tilde{A}_{\kappa}$ with two  degrees of freedom,
while the real scalar field $\rho$ decouples in the low energy regime.

Let us finally turn to the response of  the Higgs medium to 
the external point charges $q=ne$ (with $n \in \Z$) 
introduced by the matter current $j^\kappa$
in~(\ref{matcoup}). From~(\ref{kleingord}), we infer that 
the gauge invariant combined field 
$\tilde{A}_\kappa$  around this current 
drops off exponentially with mass $M_A$. Hence, 
the gauge field $A_\kappa$ necessarily becomes pure gauge at 
distances much larger then $1/M_A$ from these point charges,
and the electromagnetic fields generated by this current
vanish accordingly.
In other words, the electromagnetic fields generated by 
the external matter charges $q$ are completely screened by the Higgs medium.
From the field equations~(\ref{fieldhp}) and~(\ref{fieldhp2}), 
it is clear how the Higgs screening mechanism works.
The external matter current $j^{\kappa}$ induces  
a screening current~(\ref{scrcurr}) in the Higgs medium proportional to 
the vector field $\tilde{A}_\kappa$. 
This becomes most transparent upon considering
Gauss' law  in this case 
\bea                    \label{higgsgaus}
Q \; = \; \int d \, ^2 x \; \nabla \cdot {\bf E} 
\; = \; q +  q_{\rm scr} \; = \; 0,
\eea
which shows that the external point charge $q$ is surrounded
by a cloud of  screening charge density $j^0_{\rm scr}$ with
support of characteristic size $1/M_A$.
The contribution of the screening charge 
$q_{\rm scr} = \int d \, ^2x \, j^0_{\rm scr}
=- M_A^2 \int d \, ^2x \,\tilde{A}^0=-q$
to the long range Coulomb fields completely cancels 
the contribution of the external charge $q$. 
Thus, we arrive at the well-known result that long range Coulomb 
interactions between  external matter charges vanish in the Higgs phase.

It has long been believed that with the vanishing of the Coulomb interactions,
there are no long range interactions left for the external charges in the 
Higgs phase.  However, it was indicated by Krauss, Wilczek 
and Preskill~\cite{krawil,preskra} that this is not the case.
They noted that when the $U(1)$ gauge
group is not completely broken, but instead we are left with
a finite cyclic manifest gauge group $\Z_N$ in the Higgs phase,
the external matter charges may still have long range Aharonov-Bohm 
interactions with the magnetic vortices  also featuring  in this model.
These interactions are of a purely quantum mechanical nature with no classical 
analogue.
The physical mechanism behind  the survival of  Aharonov-Bohm 
interactions was subsequently uncovered in~\cite{sam}: 
the induced screening charges $q_{\rm scr}$ accompanying the matter charges
only couple to the Coulomb interactions 
and not to the Aharonov-Bohm interactions. As a result, the screening charges 
only screen the long range Coulomb interactions among the external 
matter charges,
but not the aforementioned long range Aharonov-Bohm 
interactions between the matter charges and the magnetic 
vortices. 
We will discuss this phenomenon in further detail in the next section.

\subsection{Survival of the Aharonov-Bohm effect}
\label{mavoab}

A distinguishing feature of the abelian Higgs model~(\ref{hiks}) 
is that it supports stable vortices carrying 
magnetic flux~\cite{abri, niels}. 
These are static classical solutions of the 
field equations  with finite energy and correspond
to topological defects in the Higgs condensate, which are pointlike in 
our 2+1 dimensional setting. 
Here, we will briefly review the basic properties of these 
magnetic vortices and subsequently elaborate on their long range 
Aharonov-Bohm interactions with the screened external charges.

The energy density following from the action~(\ref{hiks}) for time 
independent field configurations reads
\bea                     \label{endensz}
{\cal E} &=& \frac{1}{2} (E^i E^i +B^2) +  (Ne A_0)^2 |\Phi|^2 +
{\cal D}_i \Phi ({\cal D}_i \Phi)^*
+V(|\Phi|).
\eea
All the terms occurring 
here are obviously positive definite. 
For field configurations of finite energy these 
terms should therefore vanish separately at spatial infinity. 
The potential~(\ref{poto}) vanishes for ground states only.
Thus, the Higgs field is necessarily 
condensed~(\ref{simple}) at spatial infinity.
Of course, the Higgs condensate can still make a nontrivial 
winding in  the manifold of ground states. 
Such a winding at spatial infinity 
corresponds to a nontrivial holonomy  in the Goldstone boson field 
\bea                                      \label{goldhol}
\sigma(\theta+ 2\pi) - \sigma(\theta) &=& 2\pi a,
\eea
where $a$ is required to be an integer in order 
to leave the Higgs condensate~(\ref{simple}) 
itself single valued,
while $\theta$ denotes the polar angle. 
Requiring the fourth term in~(\ref{endensz}) to be integrable translates
into the condition
\bea        \label{atweg}
{\cal D}_i \Phi(r\rightarrow \infty) \sim 
\tilde{A}_i(r\rightarrow \infty) \; = \; 0,
\eea  
with $\tilde{A}_i$ the gauge invariant combination of the Goldstone boson
and the gauge field defined in~(\ref{Atild}).
Consequently, the nontrivial holonomy in the Goldstone boson field
has to be compensated by an holonomy in the gauge fields and the vortices
carry  magnetic flux $\phi$  quantized as 
\bea         \label{quaflux}
\phi \;= \; \oint dl^i A^i  \; = \; 
\frac{1}{Ne} \oint dl^i \partial_i \sigma  \; = \; 
\frac{2\pi a}{Ne} \qquad \mbox{with $a \in \Z$.}
\eea
To proceed, the third term 
in the energy density~(\ref{endensz}) disappears at spatial infinity 
if and only if  $A_0 (r \rightarrow \infty)=0$,
and all in all we see that the gauge field $A_\kappa$ is pure gauge 
at spatial infinity, so the first two terms vanish automatically.
To end up with a regular field configuration corresponding to a  
nontrivial winding~(\ref{goldhol}) of the Higgs condensate at 
spatial infinity,
the Higgs field $\Phi$ should obviously become zero somewhere in the plane. 
Thus the Higgs phase is necessarily destroyed in some finite region in the 
plane. A closer evaluation of the energy density~(\ref{endensz}) shows 
that the Higgs field grows monotonically from its zero value to its 
asymptotic ground state value~(\ref{simple}) at the distance  $1/M_H$, 
the so-called core size~\cite{abri, niels}.
Outside the core we are in the Higgs phase, and the physics is described
by the effective Lagrangian~({\ref{efhig}), while inside the core
the $U(1)$ symmetry is restored.
The magnetic field associated with the flux~(\ref{quaflux}) of the vortex
reaches its maximum inside the core where the gauge fields are massless.
Outside the core the gauge fields become massive and the magnetic field
drops off exponentially with the mass $M_A$. 
The core size $1/M_H$ and the penetration depth $1/M_A$ of the magnetic field
are the two length scales characterizing the magnetic vortex.
The formation of magnetic vortices depends on the ratio of these 
two scales.  An evaluation of the free energy 
(see for instance~\cite{gennes}) yields that 
magnetic vortices can be formed iff $M_H/M_A =\sqrt{\lambda}/Ne \geq 1$.
We will always assume that this inequality is satisfied, 
so that magnetic vortices may indeed appear in the Higgs medium.
In other words, we assume that we are dealing with a 
superconductor of type~II.

To summarize, there are two dually charged 
types of sources in the Higgs medium.
On the one hand, we have the vortices $\phi$ being sources for  
screened magnetic fields, and  on the other hand
the external charges $q$ being sources for screened  electric fields.
The magnetic fields of the vortices are localized within regions 
of length scale $1/M_H$ dropping off with mass $M_A$ at larger distances.
The external charges are point particles with Coulomb fields completely 
screened at distances $> 1/M_A$.
Henceforth, we will restrict our considerations to the low energy regime
(or alternatively send the Higgs mass $M_H$ and the mass $M_A$ of 
the gauge field to infinity by sending the symmetry breaking 
scale to infinity).  
This means that the distances between the sources remain much larger
then the Higgs length scale $1/M_H$. In other words, 
the electromagnetic fields
associated with the magnetic- and electric sources never overlap 
and the Coulomb interactions between
these sources vanish in the low energy regime.
Thus, from a classical point of view 
there are no long range interactions left between the sources.
From a quantum mechanical perspective, however, it is known  that 
in ordinary electromagnetism 
shielded localized magnetic fluxes can affect electric charges 
even though their mutual electromagnetic fields do not interfere.
When an electric charge $q$ encircles  a localized magnetic
flux $\phi$, 
it notices the nontrivial holonomy in the locally
flat gauge fields around the flux and in this process the wave function
picks up  a quantum phase $\exp (\im q\phi)$ in the first 
quantized description.
This is the celebrated Aharonov-Bohm effect~\cite{ahabo}, which is  
a purely quantum mechanical effect with no classical analogue.
These long range Aharonov-Bohm interactions are of a topological 
nature, i.e.\ as long as the charge never enters the region where the 
flux is localized,
the Aharonov-Bohm interactions only depend on the number of windings of 
the charge around the flux and not on the distance between the charge and the 
flux.
Due to a remarkable cancellation in the effective action~(\ref{efhig}),
the screening charges $q_{\rm scr}$ accompanying the external charges do 
{\em not} exhibit the Aharonov-Bohm effect. As a result 
the long range Aharonov-Bohm effect {\em persists} 
between the external charges $q$
and the magnetic vortices $\phi$ in the Higgs phase. We will argue this
in further detail.

\begin{figure}[tbh]    \epsfxsize=9cm
\centerline{\epsffile{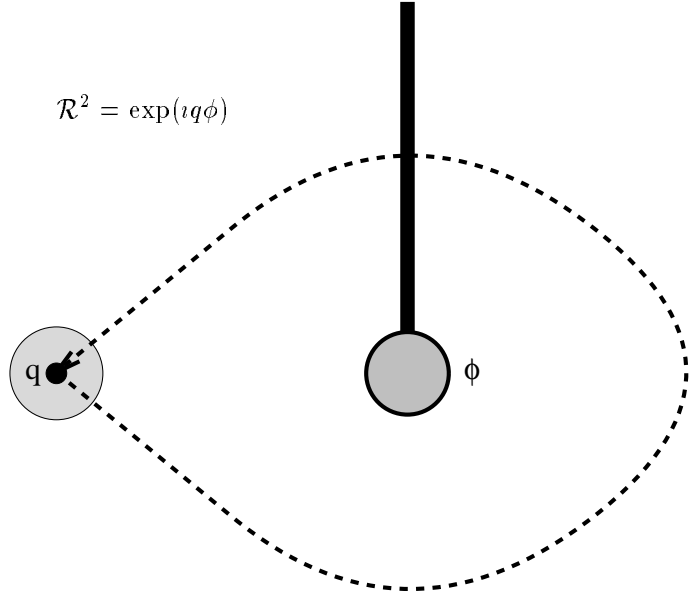}}
\caption{\sl Taking a screened
external charge $q$ around a magnetic vortex $\phi$ 
in the Higgs medium generates
the Aharonov-Bohm phase $\exp (\im q \phi)$. 
We have emphasized the extended structure of these 
sources, although this structure will not be probed 
in the low energy regime to which we confine ourselves here.
The shaded region around the external 
point charge $q$ represents the cloud of screening charge 
of characteristic size $1/M_A$.
The flux of the vortex is confined to the shaded circle bounded by the core 
at the distance $1/M_H$ from its centre. 
The string attached to the core represents the 
Dirac string of the flux, i.e.\ the  
strip in which the nontrivial parallel transport in the gauge fields 
takes place.}
\label{chargefluxbra}
\end{figure}

Consider the system depicted in figure~\ref{chargefluxbra} 
consisting of an external 
charge $q$ and a magnetic vortex $\phi$ in the Higgs medium
well separated from each other.
We have depicted
these sources as extended objects, but in the low energy regime their
extended structure will never be probed and it is legitimate
to describe these sources as point particles moving in the plane.
The magnetic vortex introduces a nontrivial holonomy~(\ref{quaflux})
in the gauge fields to which the external charge couples through  
the matter coupling~(\ref{matcoup})
\bea 
-\int \!\! d\, ^{2} x \; j^\kappa A_\kappa &=& \frac {q \phi}{2 \pi} 
\dot{\chi}_\phi({\bf y}(t)-{\bf z}(t)).
\label{int}
\eea 
Here, ${\bf y}(t)$ and ${\bf z}(t)$ respectively denote the 
worldlines of the external charge $q$ and magnetic vortex $\phi$ in the plane.
In the conventions we will use throughout these notes, 
the nontrivial parallel transport in the gauge fields
around the magnetic vortices 
takes place in a thin strip (simply called Dirac string from now)
 attached  to the core of the vortex going
off to spatial infinity in the direction of the positive vertical axis.
This situation can always be reached by a smooth gauge transformation, and
simplifies the bookkeeping for the braid processes involving more than two 
particles. The 
multi-valued function $\chi_\phi({\bf x})$ with support 
in the aforementioned
strip of parallel transport is a direct translation of this convention.
It increases from $0$ to $2\pi$ if the strip is passed from right to left.
Thus, when the external charge $q$ 
moves through this strip once in the counterclockwise
fashion indicated in figure~\ref{chargefluxbra},
the topological interaction Lagrangian~(\ref{int}) 
generates the action $q\phi$.
In the same process the screening charge $q_{\rm scr}= -q$ 
accompanying the external charge $q$ also moves 
through this strip of parallel transport. Since the screening charge 
has a sign opposite to the sign of the external charge,
it seems, at first sight, that the total topological action associated 
with encircling a flux by a screened external charge vanishes. 
This is not the case though.
The screening charge $q_{\rm scr}$ not only couples to the holonomy 
in the gauge field $A_\kappa$ around the vortex but also to the holonomy
in the Goldstone boson field $\sigma$. This follows directly from 
the effective low energy Lagrangian~(\ref{efhig}).
Let $j_{\rm scr}^\kappa$ be the screening current~(\ref{scrcurr}) 
associated with the screening charge $q_{\rm scr}$.
The interaction term in~(\ref{efhig}) couples this current to the 
massive gauge invariant  field $\tilde A_\kappa$ around the vortex:
$-j_{\rm scr}^{\kappa}\tilde{A}_{\kappa}$.
As we have seen in~(\ref{atweg}), the holonomies in the gauge field
and the Goldstone boson field are related 
at large distances from the core of the vortex, such
that $\tilde A_\kappa$ strictly vanishes.  As a consequence, 
the interaction term $-j_{\rm scr}^{\kappa}\tilde{A}_{\kappa}$ 
vanishes and indeed the matter coupling~(\ref{int}) summarizes 
all the remaining long range 
interactions in the low energy regime~\cite{sam}.

Being a  total time derivative, the topological interaction term~(\ref{int})
does not appear in the equations of motion and has no effect at
the classical level.   
In the first quantized description, however, the appearance of this term 
has far reaching consequences. This is most easily seen using 
the path integral method for quantization. In the path integral formalism,
the transition amplitude or propagator
from one point in the configuration space at some
time to another point at some later time, is given by a weighed sum 
over all the paths connecting the two points. In this sum, the paths 
are weighed by their action $\exp (\im S)$. 
If we apply this prescription to our charge/flux system, we see that 
the Lagrangian~(\ref{int}) assigns amplitudes differing by $\exp (\im q\phi)$
to paths differing by an encircling 
of the external charge $q$ around the flux $\phi$.
Thus  nontrivial interference takes place between  paths 
associated with different winding numbers of the charge around the flux.
This is the Aharonov-Bohm effect which becomes observable 
in quantum interference experiments~\cite{ahabo},
such as low energy scattering 
experiments of external charges from the magnetic vortices. 
The cross sections measured in  these
Aharonov-Bohm scattering experiments can be found in 
appendix~\ref{ahboverl}.

There are two equivalent ways to present the appearance of
the Aharonov-Bohm interactions.
In the above discussion of the path integral formalism we kept 
the topological Aharonov-Bohm interactions in the Lagrangian for this 
otherwise free charge/flux system. In this description we work
with single valued wave functions on the 
configuration space for a given  time slice
\bea
\Psi_{q \phi}({\bf y}, {\bf z}, t) &=& \Psi_q ({\bf y}, t) 
\Psi_{\phi} ({\bf z}, t) \qquad\qquad \mbox{with ${\bf y} \neq {\bf z}$}.
\label{wave}
\eea
The factorization of the wave functions follows because there 
are no interactions between the external charge and the magnetic flux
other then the topological one~(\ref{int}).
The time evolution of these wave functions is given by the 
propagator associated with the two particle  Lagrangian
\bea
L &=& \frac{1}{2} m_q \dot{{\bf y}}^2 + \frac{1}{2} m_\phi \dot{{\bf z}}^2
+\frac {q \phi}{2 \pi}
\dot{\chi}_\phi({\bf y}(t)-{\bf z}(t)).
\eea
Equivalently, we may absorb the  topological interaction~(\ref{int}) 
in the boundary condition of the wave functions and 
work with multi-valued wave functions
\bea
\tilde{\Psi}_{q \phi}({\bf y}, {\bf z}, t) &:=&
e^{\im   \frac {q \phi}{2 \pi} \chi_\phi({\bf y}-{\bf z}) } \;
\Psi_q ({\bf y}, t) \Psi_\phi ({\bf z}, t),
\label{wave2}
\eea                   
which propagate with a completely free two particle Lagrangian~\cite{wu}
(see also~\cite{forte})
\bea
\tilde{L} &=& 
\frac{1}{2} m_q \dot{{\bf y}}^2 + \frac{1}{2} m_\phi \dot{{\bf z}}^2.
\eea
We cling to the latter description from now on. That is,
we will always absorb the topological interaction terms in 
the boundary condition of the wave functions.
For later use and convenience, we set some more conventions.
We will adopt a compact Dirac notation emphasizing the internal charge/flux
quantum numbers of the particles.
In this notation, the quantum state describing a  charge or flux 
localized at some position ${\bf x}$ in the plane is presented as
\bea
|{\rm charge/flux} \rangle := |{\rm charge/flux}, {\bf x}\rangle
= |{\rm charge/flux}\rangle |{\bf x} \rangle.
\eea  
To proceed, the charges $q=ne$ will be abbreviated by the number $n$
of fundamental charge units $e$ and the fluxes $\phi$ by the number $a$  of 
fundamental flux units $\frac{2\pi}{Ne}$.
With the two particle quantum state  $|n \rangle |a\rangle$ 
we then indicate the multi-valued wave function
\bea
 | n \rangle |a \rangle &:=&  e^{\im \frac{n a}{N} 
\chi_{a}({\bf x}- {\bf y})}  \;
|n, {\bf x} \rangle  |a, {\bf y} \rangle,
\label{Diracnota}
\eea
where by convention the particle that is located  most left in the plane 
(in this case the external charge $q=ne$), appears most 
left in the tensor product.
The process of transporting the charge adiabatically around the flux 
in a counterclockwise fashion as depicted in figure~\ref{chargefluxbra}
is now summarized by the action of the monodromy
operator on this two particle state
\bea         \label{monodro}
{\cal R}^2 \; | n \rangle |a \rangle 
&=& e^{\frac{2\pi \im}{N} na} \; | n \rangle |a \rangle,
\eea
which boils down to a residual global $\Z_N$  transformation 
by the flux $a$ of the vortex on the charge $n$.

Given the remaining long range Aharonov-Bohm interactions~(\ref{monodro}) 
in the Higgs phase, the labeling of the charges 
and the fluxes by  integers is, of course, highly redundant.
Charges $n$ differing by a multiple of $N$ 
can not be distinguished. The  same holds for the fluxes $a$.
Hence, the charge and flux quantum numbers are defined modulo $N$
in the residual manifest $\Z_N$ gauge theory describing the long
distance physics of the model~(\ref{abhigg}).
Besides these pure $\Z_N$ charges and fluxes the full spectrum 
naturally consists of 
charge/flux composites or dyons produced by fusing the charges and fluxes.
We return to a detailed discussion of this spectrum and the topological
interactions it exhibits in the next section.

Let us recapitulate our results from 
a more conceptual point of view (see also~\cite{alfrev, kli, preskra} 
in this connection). In unbroken (compact) 
quantum electrodynamics, the quantized matter charges $q=ne$ (with $n \in \Z$),
corresponding to the different unitary irreducible representations
(UIR's) of the global symmetry group  $U(1)$,  
carry long range Coulomb fields.
In other words, the Hilbert space of this theory decomposes into a direct
sum of orthogonal charge superselection sectors that can be distinguished
by measuring the associated Coulomb fields at spatial infinity.   
Local observables preserve this decomposition, since they can not affect 
these long range properties of the charges. 
The charge sectors can alternatively be distinguished by their 
response to global $U(1)$  transformations, since these are  
related to physical measurements of the Coulomb fields at spatial infinity 
through Gauss' law. Let us emphasize that the states in the 
Hilbert space are of course invariant under local gauge transformations, 
i.e.\ gauge transformations with finite support, which 
become trivial at spatial infinity.

Here, we touch upon the important distinction between global 
symmetry transformations and local gauge transformations. 
Although both leave the action of the model invariant, their physical 
meaning is rather different. 
A global symmetry (independent of the coordinates)  
is a true symmetry of the theory and in particular leads to a conserved 
Noether current.
Local gauge transformations, on the other hand, correspond 
to a redundancy in the variables describing a given model and should therefore 
be modded out in the construction of the physical Hilbert space.
In the $U(1)$ gauge theory under consideration the fields that transform 
nontrivially under the global $U(1)$ symmetry are the matter fields. 
The associated Noether current $j^\kappa$ shows up in the Maxwell equations.
More specifically, the  conserved Noether charge $q=\int d\,^2 x \, j^0$, 
being the generator of the global symmetry, is identified with  
the Coulomb charge $Q=\int d\,^2 x \, \nabla \cdot {\bf E}$ 
through Gauss' law. This 
is the aforementioned relation between the global symmetry transformations
and physical Coulomb charge measurements at spatial infinity.

Although the long range Coulomb fields vanish
when this $U(1)$ gauge theory is spontaneously broken down to 
a finite cyclic group $\Z_N$, we are still able to detect
$\Z_N$ charge at arbitrary long distances through the Aharonov-Bohm effect.
In other words, there remains a relation between residual 
global symmetry transformations and physical charge measurements at 
spatial infinity.
The point is that we are left with a {\em gauged} $\Z_N$ 
symmetry in the Higgs phase, as witnessed by the appearance of 
stable magnetic fluxes in the spectrum. The magnetic fluxes introduce 
holonomies in the (locally flat) gauge fields, which take 
values in the residual manifest gauge group $\Z_N$ to 
leave the Higgs condensate single valued.  To be specific, 
the holonomy of a given flux
is classified by the group element picked up by the Wilson loop operator
\bea                \label{wilsonab}
W({\cal C}, {\bf x}_0)   &=& P \exp \, ( \im e \oint A^i dl^i) \; \in \Z_N,
\eea 
where ${\cal C}$ denotes a loop enclosing the flux starting  and 
ending at some fixed base point ${\bf x}_0$ at spatial infinity. 
The path ordering indicated by $P$ is trivial in this abelian case. 
These fluxes can be used for charge measurements in the Higgs phase
by means of the Aharonov-Bohm effect~(\ref{monodro}).  
This purely quantum mechanical effect,
boiling down to a global $\Z_N$ gauge transformation on the charge 
by the group element~(\ref{wilsonab}), is topological. 
It persists at arbitrary long ranges and therefore distinguishes 
the nontrivial $\Z_N$ charge  sectors in the Higgs phase. 
Thus the result of the Higgs mechanism 
for the charge sectors can  be summarized as follows:  
the charge superselection sectors of the original $U(1)$ gauge theory, 
which were in one-to-one correspondence with 
the UIR's of the global symmetry group $U(1)$, branch 
to UIR's of the residual  (gauged) symmetry group $\Z_N$ 
in the Higgs phase.

An important conclusion from the foregoing
 discussion is that a spontaneously broken 
$U(1)$ gauge theory in general can have distinct Higgs phases 
corresponding to different manifest cyclic gauge groups $\Z_N$. 
The simplest example is a  $U(1)$ gauge  theory 
with two Higgs fields; one carrying a charge $Ne$ and the 
other a charge $e$.  There are in principle
two possible Higgs phases  in this particular theory,
depending on whether the $\Z_N$ gauge symmetry
remains manifest or not. In the first case only  
the Higgs field with charge $Ne$ is condensed and we are left with 
nontrivial $\Z_N$ charge sectors.
In the second case the Higgs field carrying the fundamental 
charge $e$ is condensed. No charge sectors survive 
in this completely broken phase.
These two  Higgs phases, separated by a phase transition, can clearly 
be distinguished by probing the existence of $\Z_N$ charge sectors.
This is exactly the content of the nonlocal order parameter constructed 
by Preskill and Krauss~\cite{preskra} 
(see also~\cite{alflee, alfmarc, alfrev, lo, poli} in this context).  
In contrast with the Wilson loop operator 
and the 't Hooft loop operator  distinguishing the Higgs
and confining phase of a given gauge theory through the dynamics 
of electric and magnetic flux tubes~\cite{thooft, wilson},
this order parameter is of a topological nature.
To be specific, in this 2+1 dimensional setting
it amounts to evaluating the expectation value 
of  a closed electric flux tube linked with a closed magnetic flux 
loop corresponding to the worldlines of 
a minimal $\Z_N$ charge/anti-charge
pair linked with  the worldlines of a minimal $\Z_N$ magnetic 
flux/anti-flux pair. 
If the $\Z_N$ gauge symmetry is manifest, this order parameter
gives rise to the Aharonov-Bohm phase~(\ref{monodro}), whereas
it becomes trivial in the completely broken phase  with  minimal
stable flux  $\frac{2\pi}{e}$.

\subsection{Braid and fusion properties of the spectrum} \label{abdyons}

We proceed with a more thorough discussion of the topological interactions
described by the residual $\Z_N$ gauge theory featuring in the Higgs phase 
of the model~(\ref{abhigg}). 
As we have argued in the previous section, the
complete spectrum consists
of pure $\Z_N$ charges labeled by $n$, pure $\Z_N$ fluxes labeled by $a$
and dyons produced by fusing these charges and fluxes: 
\bea                              \label{compspectr}
|a \rangle \times |n \rangle  &=& 
|a, n \rangle \qquad  \qquad \mbox{with} 
\qquad a,n \in 0,1, \ldots, N-1.
\eea   
We have depicted this spectrum for a $\Z_4$ gauge theory 
in figure~\ref{z4z}.

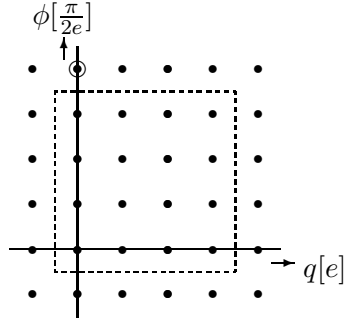
\begin{figure}[htb] 
\begin{center}
\begin{picture}(85,80)(-15,-15)
\put(-5,-5){\dashbox(40,40)[t]{}}
\put(-15,0){\line(1,0){60}}
\put(0,-15){\line(0,1){60}}
\thinlines
\multiput(-10,-10)(0,10){6}{\multiput(0,0)(10,0){5}{\circle*{2.0}}}
\multiput(0,-10)(0,10){6}{\multiput(0,0)(20,0){3}{\circle*{2.0}}}
\put(0,40){\circle{3.6}}
\put(-3,42){\vector(0,1){5}}
\put(43,-3){\vector(1,0){5}}
\put(-10,50){\small$\phi[\frac{\pi}{2e}]$}
\put(50,-6){\small$q[e]$}
\end{picture}
\vspace{0.5cm}
\caption{\sl The spectrum of a Higgs phase featuring a residual manifest
gauge group $\Z_4$  compactifies to the particles inside the dashed box.
The particles outside the box are identified with the ones inside by means 
of modulo $4$ calculus along the charge and flux axes.
The modulo $4$ calculus for the fluxes corresponds to Dirac 
monopoles/instantons, if these are present. That is, the minimal monopole
tunnels the encircled flux into the vacuum.}
\label{z4z}
\end{center}
\end{figure}

The topological interactions described by a $\Z_N$ gauge theory are 
completely governed by the Aharonov-Bohm effect~(\ref{monodro}) 
and can simply be summarized as follows
\bea      
{\cal R}^2\;|a,n\rangle |a',n'\rangle &=&
e^{\frac{2 \pi \im}{N} (n a' +n'a)}
\;|a,n\rangle |a',n'\rangle      \label{monozoco}   \\
{\cal R} \; |a,n\rangle |a,n\rangle &=&
e^{\frac{2 \pi \im}{N} n a } \;|a,n\rangle |a,n\rangle
\label{brazoco}  \\
|a,n\rangle \times |a',n'\rangle 
       &=& |[a+a'],[n+n']\rangle  \label{fusionzoco} \\
{\cal C} \, |a,n \rangle &=& |[-a], [-n ]\rangle  \label{CC}  \\
T \, |a,n\rangle &=& e^{\frac{2 \pi \im}{N} na}  \;  |a,n \rangle.
\label{modulT}      
\eea 
The expressions~(\ref{monozoco}) and~(\ref{brazoco}) sum up 
the braid properties of the particles in the spectrum~(\ref{compspectr}). 
These realize abelian representations of the braid groups 
discussed in section~\ref{braidgroups}. Of course, for distinguishable 
particles only the monodromies, as contained in the pure 
or colored braid groups  are 
relevant. 
(See the discussion concerning
relation~(\ref{pbge}) for the definition of colored braid groups).
In the present context, 
particles carrying different charge and magnetic flux are distinguishable. 
When a  given particle $|a,n\rangle$ located at some position in the plane 
is adiabatically transported around another remote particle $|a',n'\rangle$
in the counterclockwise fashion depicted in figure~\ref{gammaije}, the 
total multi-valued wave function of the system picks up the 
Aharonov-Bohm phase
displayed in~(\ref{monozoco}). In this process, the charge $n$
of the first particle moves through the Dirac string attached to the 
flux $a'$ 
of the second particle, while the charge $n'$ of the second particle 
moves through
the Dirac string of the flux $a$ of the first particle.
In short, the total 
Aharonov-Bohm effect for this monodromy 
is the composition of a global $\Z_N$ symmetry transformation on
the charge $n$ by the flux $a'$ and a global transformation on 
the charge $n'$ by the flux $a$.  
We confined ourselves to the case of two particles so far. 
The generalization  to systems containing more then two particles is 
straightforward. The quantum states describing these systems are 
tensor products of localized single particle states 
$|a,n, {\bf x}\rangle$, where we cling to the convention that the particle 
that appears most left in the plane appears most left in the tensor product.
These multi-valued wave functions carry abelian representations of the 
colored braid group: the action of the monodromy operators~(\ref{pbge}) 
on these wave functions boils down to the quantum phase     
in expression~(\ref{monozoco}).

For identical particles, i.e.\ particles carrying 
the same charge $n$ and flux $a$, the braid operation depicted
in figure~\ref{touw} becomes meaningful.  In this braid process, in which
two adjacent identical particles $|a,n\rangle$ 
located at different positions in the 
plane are exchanged in a counterclockwise way, the charge of the particle
that moves `behind' the other dyon encounters the Dirac string attached to the 
flux of the latter. The result of this exchange in the multi-valued wave 
function is the quantum statistical phase factor 
(see expression~(\ref{qstat}) of section~\ref{braidgroups})
presented in~(\ref{brazoco}).
In other words, the dyons in the spectrum of this $\Z_N$ theory 
are anyons. In fact, these charge/flux composites are very close to 
Wilczek's original proposal for anyons~\cite{wilcchfl}. 

An important aspect of this theory is that the particles in the  
spectrum~(\ref{compspectr}) satisfy the canonical 
spin-statistics connection.
The proof of this connection is of a topological nature and applies 
in general to all the models that will be considered in these notes.
The fusion rules play a role in this proof and we will discuss these first.

Fusion and braiding are intimately related. 
Bringing two particles together  is essentially a local process. 
As such, it can never affect global properties of the system.
Hence, the single particle state that arises after fusion should exhibit
the same global properties as the two particle state we started with.
In this topological theory, the global properties of a given 
configuration are determined by its braid properties with the 
different particles in the spectrum~(\ref{compspectr}). 
In the previous section, 
we had already established that the charges and fluxes 
become $\Z_N$ quantum numbers under these braid properties. 
Therefore, the complete set of fusion rules, determining
the way the charges and fluxes of a two particle state 
compose into the charge and flux of a single particle state when the pair is 
brought together, can be summarized as~(\ref{fusionzoco}). 
The rectangular brackets denote modulo $N$ calculus such that the sum 
always lies in the range $0,1,\ldots, N-1$. 

It is worthwhile  to digress  a little
on the dynamical mechanism underlying the  modulo $N$ calculus 
compactifying the flux part of the spectrum. 
This modulo calculus is induced by magnetic monopoles, when these 
are present.
The presence of magnetic monopoles can be accounted for by 
assuming that the compact $U(1)$ gauge theory~(\ref{abhigg}) 
arises from a spontaneously broken  $SO(3)$ gauge theory. The 
 monopoles we obtain in this particular model are 
the regular 't~Hooft-Polyakov monopoles~\cite{thooftmon,polyamon}.
Let us, alternatively, assume that we have singular 
Dirac monopoles~\cite{dirac} in this compact $U(1)$ gauge theory.  
In three spatial dimensions, these are  point particles carrying
magnetic charges $g$ quantized as  $\frac{2\pi}{e}$.
In the present  2+1 dimensional Minkowski setting,
they become instantons describing flux tunneling events 
$|\Delta \phi|= \frac{2\pi}{e}$.   As has been shown by 
Polyakov~\cite{polyakov}, the presence of these instantons 
in unbroken  $U(1)$ gauge theory has a striking dynamical effect.
It leads to linear confinement of electric charge. 
In the broken version of these theories, in which we are 
interested,  electric charge is screened and  the presence of 
instantons in the Higgs phase merely implies that the magnetic 
flux~(\ref{quaflux}) of the vortices is conserved modulo $N$ 
\bea                            \label{instans}
\mbox{instanton:}  & &  a \; \mapsto \; a -N.
\eea 
In other words, a flux $N$ moving in the plane (or $N$ minimal fluxes
for that matter) can disappear by ending on an instanton.
The fact that the instantons tunnel between
states that can not be distinguished 
by the braidings in this theory is nothing but 
the 2+1 dimensional space time translation of the unobservability of the  
Dirac string in three spatial dimensions.

We  turn to the connection between spin and statistics.
There are in principle two approaches to prove this deep relation, both 
having their own merits. 
One approach, originally due to Wightman~\cite{streater},
involves the axioms of local relativistic quantum field theory, and 
leads to the observation that integral spin fields commute,
while half integral spin fields anticommute.
The topological approach that we will take here was first
proposed by Finkelstein and Rubinstein~\cite{fink}. 
It does not rely upon the heavy framework
of local relativistic quantum field theory and among other things 
applies to the topological defects considered in this thesis.
The original formulation of Finkelstein and Rubinstein
was in the 3+1 dimensional context, but 
it naturally extends to 2+1 dimensional space time as we will 
discuss now~\cite{balach, sen}. See also the 
references~\cite{frohma, frogama}
for an algebraic approach.

The crucial ingredient in the topological proof of the spin-statistics 
connection for a given model is the existence of an anti-particle 
for every particle in the spectrum, such that the pair can annihilate 
into the vacuum after fusion. 
Consider the process depicted at the l.h.s. of the equality
sign in figure~\ref{spinstafig}. It describes the creation of 
two separate identical particle/anti-particle pairs from the vacuum,
a subsequent counterclockwise exchange of the particles of the 
two pairs and finally
annihilation of the pairs.
To keep track of the  writhing of the particle trajectories we 
depict them as ribbons with a white- and a dark side.
It is easily verified now that the closed ribbon associated with the 
process just explained  can be continuously deformed into the ribbon at the 
r.h.s., which corresponds to a counterclockwise 
rotation of the particle over an angle of
$2\pi$ around its own centre. In other words, the effect of interchanging
two identical particles in a consistent 
quantum description should be the same as the effect 
of rotating one particle over an angle of $2\pi$ around its centre. 
The effect of this rotation in the wave function is the spin factor
$\exp (2\pi \im s)$ with $s$ the spin of the particle, which in contrast with
three spatial dimensions may be any real number in two spatial dimensions.
Therefore, the result of exchanging  the two identical particles
necessarily boils down to a quantum statistical phase 
factor $\exp (\im \Theta)$ in the wave function being
the same as the spin factor
\bea                          \label{spistath}
\exp (\im \Theta) &=& \exp (2\pi \im s).
\eea
This is the canonical spin-statistics connection. Actually, a further
consistency condition  can be inferred from this ribbon argument. The writhing 
in the particle trajectory can be continuously deformed to a writhing 
with the same orientation in the anti-particle trajectory. Hence, the 
anti-particle necessarily carries the same spin and statistics 
as the particle.

\begin{figure}[htb]    \epsfxsize=13cm
\centerline{\epsffile{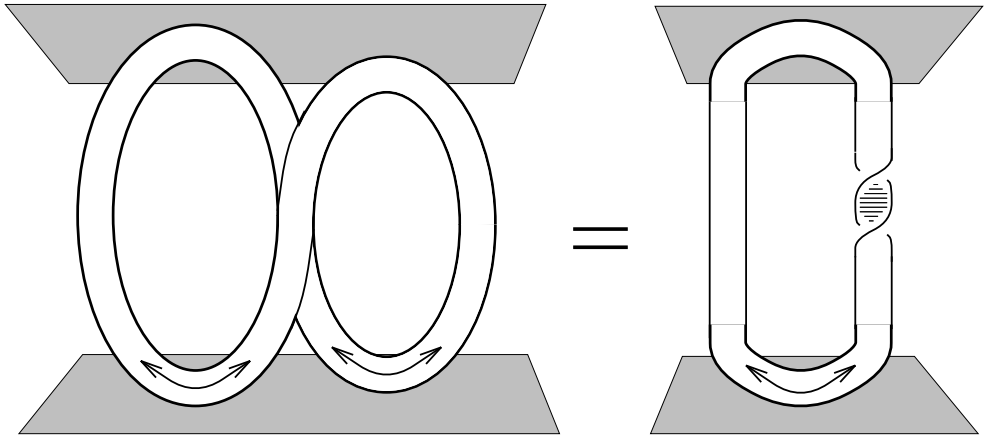}}
\caption{\sl Canonical spin-statistics connection. The trajectories 
describing a counterclockwise interchange of two particles in
separate particle/anti-particle pairs (the 8 laying on its back)
can be continuously 
deformed into a single pair in which the particle undergoes a 
counterclockwise rotation
over an angle of $2\pi$ around its own centre (the 0 with a twisted leg).}
\label{spinstafig}
\end{figure}

Sure enough the topological proof of the canonical spin-statistics connection
applies to  the $\Z_N$ gauge theory at hand.
First of all, we can naturally assign an anti-particle to every 
particle in the spectrum~(\ref{compspectr}) through  the
charge conjugation operator~(\ref{CC}). Under charge conjugation
the charge and flux of the particles in the spectrum reverse sign and
amalgamating a particle with its charge conjugated partner yields
the quantum numbers of the vacuum as follows from the fusion 
rules~(\ref{fusionzoco}). Thus the basic assertion for the 
above ribbon argument is satisfied. 
From the quantum statistical phase factor~(\ref{brazoco}) assigned
to the particles and~(\ref{spistath}), we then conclude that the particles
carry spin. Specifically, under rotation over $2\pi$ the single particle
states should give rise to the spin factors displayed in~(\ref{modulT}). 
In fact, these spin factors can be interpreted as the Aharonov-Bohm phase 
generated when the charge of a given dyon rotates around its own flux.
Of course, a small separation between the charge and the flux of the dyon 
is required for this interpretation.
Also, note that the particles and their anti-particles indeed 
carry the same spin and statistics, as follows immediately from the 
invariance of the Aharonov-Bohm effect under charge conjugation.

Having established a complete classification of the topological 
interactions described by a $\Z_N$ gauge theory, we conclude with 
some remarks on the Aharonov-Bohm scattering  experiments 
by which these interactions can be probed.
(A concise discussion of these purely quantum mechanical  
experiments can be found in appendix~\ref{ahboverl} of 
chapter~\ref{exampled2b}).
It is the monodromy effect~(\ref{monozoco}) that 
is measured in these two particle elastic scattering experiments.
To be explicit, the symmetric cross section for scattering a
particle~$|a,n\rangle$ from a particle~$|a',n'\rangle$ is given by
\bea                        \label{ZNab}
\frac{{\rm d} \sigma}{{\rm d} \theta} &=&
\frac{\sin^2 \left( \frac{\pi}{N}(na' +n'a ) \right)}{2\pi p 
\sin^2 (\theta/2)},
\eea    
with  $p$ the relative momentum of the two particles and $\theta$ 
the scattering angle. 
A subtlety arises in scattering experiments involving two 
identical particles, however. Quantum statistics enters the scene:
exchange processes between the scatterer and the 
projectile have to be taken into account~\cite{anyonbook, preslo}.
This leads to the following cross section for Aharonov-Bohm 
scattering of two identical particles $|a,n\rangle$
\bea                        \label{ZNid}
\frac{{\rm d} \sigma}{{\rm d} \theta} &=&
\frac{\sin^2 (\frac{2\pi na}{N})}{2\pi p \sin^2 (\theta/2)} \; + \;
\frac{\sin^2 (\frac{2\pi na}{N})}{2\pi p \cos^2 (\theta/2)},
\eea     
where the second term summarizes the effect of the 
extra exchange contribution to the direct scattering amplitude.

\sectiona{Nonabelian discrete gauge theories} \label{nonabz}

The generalization of the foregoing analysis to spontaneously broken models
in which we are left with a {\em nonabelian} finite gauge group ${ H}$ 
involves some essentially new features. 
In this introductory section, we 
will establish the complete flux/charge spectrum of such a  nonabelian 
discrete ${ H}$ gauge theory and discuss 
the basic topological interactions among the different flux/charge composites.
The outline is as follows. Section~\ref{topclas} contains
a general discussion on the topological classification of stable magnetic 
vortices and the subtle role magnetic monopoles play in this classification. 
In section~\ref{fluxmetamorphosis}, we  subsequently
review the properties of the nonabelian magnetic vortices that occur
when the residual symmetry group $H$ is nonabelian. 
The most important one being that these  vortices 
exhibit a nonabelian Aharonov-Bohm effect.
To be specific, the fluxes of the vortices, which are labeled by 
the group elements of ${ H}$, affect each other through conjugation 
when they move around each other~\cite{bais}.
Under the residual global symmetry group ${ H}$ the magnetic fluxes 
transform by conjugation as well, and the conclusion is that the vortices 
are organized in degenerate multiplets, corresponding to
the different conjugacy classes of ${ H}$.  
These classical properties will then be elevated
into the first quantized description in which the magnetic 
vortices are treated  
as point particles moving in the plane. 
In section~\ref{inclumatter}, we finally 
turn to the matter charges that may occur 
in these Higgs phases and their
Aharonov-Bohm interactions with the magnetic vortices.
As has been pointed out in~\cite{almawil,preskra}, these matter charges
are labeled by the different UIR's $\Gamma$ of the residual global symmetry 
group ${ H}$ and when such a charge encircles a nonabelian
vortex it picks up a global symmetry transformation by the matrix
$\Gamma(h)$ associated with the 
flux $h$ of the vortex in the representation $\Gamma$.
To conclude, we elaborate on the subtleties~\cite{spm} 
involved in the description
of dyonic combinations of the nonabelian magnetic fluxes and the matter
charges $\Gamma$.

\subsection{Classification of stable magnetic vortices}
\label{topclas}

Let us start by briefly specifying the spontaneously broken gauge 
theories  in which we are left with a nonabelian discrete  gauge theory.
In this case, we are dealing with a model governed by a 
Yang-Mills Higgs action of the form
\bea
S_{\rm YMH} &=& \int d \, ^3 x \;  
(-\frac{1}{4}F^{a \; \kappa\nu} F_{\kappa\nu}^a
  +({\cal D}^\kappa \Phi)^\dagger \cdot {\cal D}_\kappa \Phi - 
V(\Phi) )     \label{nonhiks}.
\eea
Here, the Higgs field $\Phi$ 
transforms according to some higher dimensional representation 
of a continuous nonabelian gauge group ${ G}$, 
the superscript $a$ naturally labels the generators of the Lie algebra
of $G$
and the potential $V(\Phi)$ 
gives rise to a degenerate set of ground states 
$\langle \Phi \rangle \neq 0 $ which are only invariant under 
the action of a finite nonabelian subgroup ${ H}$ of ${ G}$.
For simplicity, we make two assumptions. 
First of all, we assume that this Higgs potential is normalized such
that $V(\Phi) \geq 0$ and equals zero for the ground states 
$\langle \Phi \rangle $. More importantly, we assume 
that all ground states can be reached from any given one by global 
${ G}$ transformations. This last assumption  
implies that the ground state manifold 
becomes isomorphic to the coset ${ G}/{ H}$.
(Renormalizable examples of potentials doing the job for 
${ G}\simeq SO(3)$ and ${ H}$ some of its point groups 
can be found in~\cite{ovrut}). In the following, we will only be concerned
with the low energy regime of this theory,  so that the massive 
gauge bosons can be ignored.

The  topologically stable vortices that can be formed in 
the spontaneously broken 
gauge theory~(\ref{nonhiks})
correspond to noncontractible maps
from the circle at spatial infinity (starting and ending at 
a fixed base point ${\bf x}_0$) 
into the ground state  manifold ${ G}/{ H}$. 
Different vortices are related to noncontractible maps 
that can not be continuously deformed into each other.
In short, the different vortices are labeled by the elements 
of the fundamental group $\pi_1$ of ${ G}/{ H}$ based at the 
particular ground state $\langle \Phi_0 \rangle$ the Higgs 
field takes at the base point ${\bf x}_0$ in the plane.
(Standard references on the use of homotopy groups 
in the classification of topological defects 
are~\cite{cola,mermin,presbook,trebin}. See also~\cite{poen} for an early 
discussion on the occurrence of nonabelian fundamental groups 
in models with a spontaneously broken {\em global} symmetry).

The content of the fundamental group $\pi_1 ({ G}/{ H})$  of the ground
state manifold for a specific spontaneously broken model~(\ref{nonhiks})
can be inferred from the exact sequence
\bea              \label{exactseq}
0 \simeq \pi_1 (H) \rightarrow \pi_1 (G) \rightarrow \pi_1 (G/H) \rightarrow
\pi_0 (H) \rightarrow \pi_0 (G) \simeq 0,
\eea
where the first isomorphism  follows from the fact that $H$ is discrete.
For convenience, we restrict our considerations to  continuous
Lie groups $G$ that are path connected, which 
accounts for the last isomorphism.
If  $G$ is simply connected as well, 
i.e.\ $\pi_1 ({ G}) \simeq 0$, 
then the exact sequence~(\ref{exactseq}) yields the isomorphism 
\bea   \label{flulaord}
\pi_1({ G}/{ H}) &\simeq& { H},
\eea 
where we used the result $\pi_0 (H) \simeq H$, which holds 
for finite $H$.
Thus, the different magnetic vortices in this case
are in one-to-one correspondence
with the group elements $h$ of the residual symmetry group ${ H}$. 
When ${ G}$ is {\em not} simply connected, however,  
this is not a complete classification. This can be seen 
by the following simple argument. 
Let $\bar{ G}$ denote the universal covering
group of ${ G}$ and $\bar{ H}$ the corresponding lift 
of ${ H}$ into $\bar{ G}$.
We then have $G/H = \bar{ G}/\bar{ H}$ and in particular
$\pi_1({ G}/{ H}) \simeq \pi_1(\bar{ G}/\bar{ H})$.
Since the  universal covering group of $G$ is by definition simply connected,
that is, $\pi_1(\bar{G}) \simeq 0$, we obtain the following isomorphism
from the exact sequence~(\ref{exactseq}) for the lifted groups  
$\bar{G}$ and $\bar{H}$
\bea                                  \label{barh}
\pi_1({ G}/{ H}) \; \simeq \; \pi_1(\bar{ G}/\bar{ H}) \;\simeq \;
 \bar{ H}.
\eea   
Hence, for a non-simply connected broken gauge 
group $G$, the different stable
magnetic vortices are labeled by the elements of $\bar{H}$
rather then $H$ itself. 

It should be emphasized that the extension~(\ref{barh}) 
of the magnetic vortex spectrum is  
based on the tacit assumption that there are no 
Dirac monopoles featuring in this model. In any theory with a   
non-simply connected gauge group $G$, however, we have the freedom 
to introduce singular Dirac monopoles `by hand'~\cite{later,cola}. 
The  magnetic charges of these 
monopoles are characterized by the elements of the fundamental 
group $\pi_1(G)$, which is abelian for continuous Lie groups $G$.
The exact sequence~(\ref{exactseq}) for the present spontaneously 
broken model now implies the identification
\bea
\pi_1(G) &\simeq& \mbox{Ker} \left(\pi_1(G/H) \rightarrow \pi_0(H)\right) \\
         &\simeq& \mbox{Ker} \left(\bar{H} \rightarrow H \right). \nn 
\eea
In other words, the magnetic charges of the Dirac monopoles 
are in one-to-one correspondence with the
nontrivial elements of $\pi_1(G/H)\simeq \bar{H}$ 
associated with the trivial element  in $\pi_0(H) \simeq H$.
The physical interpretation of this formula is as follows. In the 
2+1 dimensional Minkowsky setting, in which we are interested,
the Dirac monopoles become instantons describing
tunneling events between magnetic vortices $\bar{h} \in \bar{H}$
differing by the elements of $\pi_1(G)$. 
Here, the decay or tunneling time will naturally depend exponentially
on the actual mass of the monopoles.
The important conclusion is  that in the presence of 
these Dirac monopoles the magnetic
fluxes $\bar{h} \in \bar{H}$ are conserved
modulo the elements of $\pi_1(G)$ and the proper labeling of the stable 
magnetic vortices boils down to the elements of the residual symmetry 
group~$H$ itself
\bea         \label{fluxspcompi}
\bar{H}/\pi_1(G) &\simeq& H.
\eea   
To proceed, the introduction of Dirac monopoles has a bearing on the matter 
content of the model as well. The only matter fields allowed in the theory
with monopoles are those that transform according to an ordinary 
representation of $G$. Matter fields carrying a faithful 
representation of the universal covering group 
$\bar{G}$ are excluded. This means that the 
matter charges appearing in the broken phase 
correspond to ordinary representations of $H$, while faithful 
representations of the lift $\bar{H}$ do not occur. As a result, 
the fluxes $\bar{h} \in \bar{H}$ related by tunneling events induced
by  the Dirac monopoles can not be distinguished through long 
range Aharonov-Bohm experiments with the available matter charges, 
which is consistent with the fact that the
stable magnetic fluxes are labeled by elements of $H$ 
rather then $\bar{H}$ in this case.

The whole discussion can now be summarized as follows. First of all,
if a simply connected  gauge group $G$ is spontaneously broken down 
to a finite subgroup $H$, we are left with a discrete $H$ gauge theory
in the low energy regime. 
The magnetic fluxes are labeled by the elements  of $H$, whereas the 
different electric charges correspond to the full set of UIR's of $H$.
When we are dealing with a non-simply connected gauge group $G$
broken down to a finite subgroup $H$, 
there are two possibilities depending on whether we allow 
for Dirac monopoles/instantons in the theory or not.
In case Dirac monopoles are ruled out, 
we obtain a discrete $\bar{H}$ gauge theory.
The stable fluxes are labeled by the elements of $\bar{H}$ and 
the different charges by the UIR's of $\bar{H}$.
If the model features singular Dirac monopoles, on the other hand,
then the stable fluxes simply correspond to the elements of the group $H$  
itself, while the allowed matter charges constitute UIR's of $H$.
In other words, we are left with a discrete $H$ gauge theory under these 
circumstances.

Let us illustrate these general considerations by some explicit examples.
First we return to the model discussed in section~\ref{abznz}, in which
the non-simply connected gauge group $G \simeq U(1)$ is spontaneously 
broken down to the finite cyclic group $H \simeq \Z_N$.
The topological classification~(\ref{barh}) for this particular model gives
\beas
\pi_1(U(1)/\Z_N) \; \simeq \; \pi_1({\bf R}/\Z_N \times \Z) \; \simeq 
\; \Z_N \times \Z \; \simeq \; \Z.
\eeas
Thus, in the absence of Dirac monopoles, 
the different stable vortices are labeled by the
integers in accordance with~(\ref{quaflux}), where we found  
that the  magnetic fluxes associated with these vortices are quantized as
$\phi = \frac{2\pi a}{Ne}$ with $a\in \Z$. In principle, we are dealing with
a discrete $\Z$ gauge theory now
and the complete magnetic flux spectrum could be distinguished 
by means of long range Aharonov-Bohm experiments with 
electric charges $q$ being fractions of the fundamental unit $e$,
which correspond to the UIR's of $\Z$.
Of course, this observation is rather academic in this context,
since free charges carrying fractions of the fundamental charge unit $e$
have never been observed. With matter charges $q$ being multiples of $e$, the 
low energy theory then boils down to a $\Z_N$ gauge theory, although
the topologically stable magnetic vortices in the broken phase are 
labeled by the integers $a$.
The Dirac monopoles/instantons 
that can be introduced in this theory correspond to the elements 
of $\pi_1(U(1)) \simeq \Z$. The presence of these monopoles, which
carry magnetic charge $g=\frac{2\pi m}{e}$ with $m \in \Z$, imply that 
the magnetic flux $a$ of the vortices is  conserved modulo $N$, 
as we have seen explicitly in~(\ref{instans}). 
In other words, the proper 
labeling of the stable magnetic fluxes is by the elements
of $\Z_N \times \Z /\Z \simeq \Z_N$, as indicated by~(\ref{fluxspcompi}).
Moreover, electric charge is necessarily quantized in 
multiples of the fundamental charge unit $e$ now, 
so that the tunneling events induced by the instantons 
are unobservable at long distances.
The unavoidable conclusion then becomes that 
in the presence of Dirac monopoles,
we are left with a $\Z_N$ gauge theory 
in the low energy regime of this spontaneously broken model, in complete
accordance with the general discussion of the foregoing paragraphs.

When a gauge theory at some intermediate stage of symmetry breaking exhibits 
regular 't~Hooft-Polyakov monopoles, their effect on the stable magnetic 
vortex 
classification is automatically taken care of, as it should because the 
monopoles can not be left out in such a  theory.
Consider, for example, a model in which the non-simply connected
gauge group $G \simeq SO(3)$ is initially 
broken down to $H_1 \simeq U(1)$ and subsequently to $H_2 \simeq \Z_N$
\bea      \label{sbh}
     SO(3) \; \longrightarrow \; U(1) \; \longrightarrow \; \Z_N.
\eea
The first stage of symmetry breaking is accompanied by the appearance 
of regular 't~Hooft-Polyakov monopoles~\cite{thooftmon,polyamon} carrying
magnetic charges characterized by the  
elements of the second homotopy group 
$\pi_2  (SO(3)/U(1)) \simeq \Z$. A simple exact sequence
argument shows 
\bea
\pi_2  (SO(3)/U(1)) & \simeq & \mbox{Ker} \left(\pi_1(U(1)) \rightarrow 
\pi_1(SO(3)) \right) \\
 & \simeq &  \mbox{Ker} \left(\Z \rightarrow \Z_2 \right).   \nn 
\eea
Hence, the magnetic charges of the regular monopoles correspond to 
the elements of $\pi_1 (U(1))$
associated with the trivial element 
of $\pi_1(SO(3))$, that is, the even elements of $\pi_1(U(1))$. 
In short, the regular monopoles carry 
magnetic charge $g=\frac{4\pi m}{e}$ with $ m \in \Z$.
To proceed, the residual topologically stable magnetic vortices emerging
after the second symmetry breaking are labeled by the elements of
$\bar{H}_2 \simeq \Z_{2N}$, which follows
from~(\ref{barh}) 
\beas
\pi_1 (SO(3)/\Z_N) \; \simeq \; \pi_1 (SU(2)/\Z_{2N}) \; \simeq \; \Z_{2N}.
\eeas
As in the previous example, the magnetic fluxes carried by these vortices are 
quantized as $\phi=\frac{2\pi a}{Ne}$, while the presence of the regular
't Hooft-Polyakov monopoles now causes the fluxes $a$ to be conserved modulo
$2N$. 
The tunneling or decay time will depend on the 
mass of the regular monopoles, that is, 
the energy scale associated with the first symmetry breaking 
in the hierarchy~(\ref{sbh}).
Here it is assumed that the original $SO(3)$ gauge theory
does not feature Dirac monopoles ($g=\frac{2\pi m}{e}$, with $m=0,1$) 
corresponding to the elements of $\pi_1 (SO(3))\simeq \Z_2$.
This means that additional  matter fields carrying  
faithful (half integral spin) representations of the universal 
covering group $SU(2)$ are allowed in this model, which  leads to 
half integral charges $q=\frac{ne}{2}$ with $n\in \Z$ in the $U(1)$ phase.
In the final Higgs phase, the half integral charges 
$q$ and the quantized magnetic
fluxes $\phi$ then span the complete spectrum of the associated 
discrete $\Z_{2N}$ gauge theory. 

Let us now, instead, suppose that the original $SO(3)$ gauge 
theory contains Dirac monopoles.  
The complete monopole spectrum arising after the first symmetry breaking
in~(\ref{sbh}) 
then consists of the  magnetic charges $g=\frac{2 \pi m}{e}$ 
with $m \in \Z$, which implies that magnetic flux $a$ 
is conserved modulo $N$ in the final Higgs phase.
This observation is in complete agreement 
with~(\ref{fluxspcompi}), 
which states that the proper magnetic flux labeling is by the elements
of $\Z_{2N}/\Z_2 \simeq \Z_N$ under these circumstances.
In addition, the incorporation of Dirac monopoles 
rules out  matter fields which carry 
faithful representations of the universal covering group 
$SU(2)$. Hence, only integral electric charges are conceivable
($q=ne$ with $n\in \Z$) and all in all we end up with a discrete 
$\Z_N$ gauge theory in the Higgs phase.
This last situation can alternatively be 
implemented by embedding this spontaneously broken $SO(3)$ 
gauge theory in an $SU(3)$ gauge theory.
In other words,  the symmetry breaking hierarchy is extended to 
\bea    \label{sbh2}
SU(3) \; \longrightarrow \; SO(3) \; \longrightarrow \; U(1) 
\; \longrightarrow \; \Z_N.
\eea
The singular Dirac monopoles in the $SO(3)$ phase then turn into
regular 't Hooft-Polyakov monopoles 
\beas
\pi_2(SU(3)/SO(3)) \; \simeq \; \pi_1(SO(3)) \; \simeq \; \Z_2.
\eeas
The unavoidable presence of these monopoles automatically imply that
the magnetic flux $a$ of the vortices in the final Higgs phase 
is conserved modulo $N$. To be specific,  a magnetic flux $a=N$ 
can decay by ending on a regular monopole in this model, where the 
decay time will again depend on the mass of the monopole or equivalently
on the energy scale associated with the first symmetry breaking 
in~(\ref{sbh2}). 
The existence of such a dynamical decay process is 
implicitly taken care of in the 
classification~(\ref{flulaord}), which indicates
that the stable magnetic fluxes are indeed labeled by the elements of
$\pi_1 (SU(3)/\Z_N)\simeq \Z_N$.

To conclude, in the above examples we restricted ourselves to the case 
where we are left with an abelian finite gauge group in the Higgs phase. 
Of course, the discussion extends to nonabelian finite groups as well. 
The more general picture then becomes as follows.
If the non-simply connected gauge 
group $G \simeq SO(3)$ is spontaneously broken to some (possibly nonabelian)
finite  subgroup $H \subset SO(3)$,
then the topologically stable magnetic fluxes correspond to the elements of 
the lift $\bar{H} \subset SU(2) \simeq \bar{G}$.  In the Higgs phase, we are 
then left with a discrete $\bar{H}$ gauge theory.
If we have embedded $SO(3)$
in $SU(3)$ (or alternatively introduced the conceivable $\Z_2$
Dirac monopoles), on the other hand, 
then the topologically stable magnetic fluxes 
correspond to the elements of $H$ itself and we end up with a discrete $H$
gauge theory.

\subsection{Flux metamorphosis}    \label{fluxmetamorphosis}

Henceforth, we assume  that the spontaneously broken 
gauge group $G$ in our model~(\ref{nonhiks})
is simply connected, for convenience. Hence, the stable magnetic 
vortices are  labeled by the elements of the nonabelian residual symmetry
group $H$, as indicated by the isomorphism~(\ref{flulaord}). 

We start with a discussion of the classical field configuration 
associated with a single static nonabelian  vortex in the plane.
In principle, we are dealing with  an extended object with a finite core 
size proportional to the inverse of the 
symmetry breaking scale $M_H$. In the low energy regime, however,
we can neglect this finite core size and we will idealize the 
vortex as a point singularity in the plane. 
For finite energy, the associated static classical field 
configuration then satisfies the equations
$V(\Phi)=0$, $F^{\kappa \nu}=0$, 
${\cal D}_i \Phi=0$ and $A_0=0$ outside the core.
These equations imply that  the Higgs field takes ground state values 
$\langle \Phi \rangle$ and the Lie algebra valued vector 
potential $A_\kappa$ is pure gauge so that all nontrivial 
curvature $F^{\kappa \nu}$ is localized inside the core. 
To be explicit, a path (and gauge) dependent 
solution w.r.t.\  an arbitrary but fixed 
ground state $\langle \Phi_0 \rangle$ 
at an arbitrary but fixed base point ${\bf x}_0$ can be presented as
\bea
\langle \Phi(\bf x) \rangle &=& W({\bf x}, {\bf x}_0, \gamma) 
\langle \Phi_0 \rangle ,
\eea 
where the  untraced path ordered 
Wilson line integral
\bea
W({\bf x}, {\bf x}_0, \gamma) &=& 
P \exp ( \im e \int_{{\bf x}_0}^{\bf x} A^i dl^i),
\eea
is evaluated along an oriented path $\gamma$ (avoiding the 
singularity where the vortex is located) 
from the  base point to some other 
point ${\bf x}$ in the plane. 
Here, we merely used the fact that the relation
${\cal D}_i \langle \Phi \rangle=0$ identifies the parallel transport
in the Goldstone boson fields  with that in the gauge fields,
as we have argued in full detail for the abelian case 
in section~\ref{mavoab}.
Now in order to keep the Higgs field
single valued, the magnetic flux of the vortex, picked up by 
the  Wilson line integral along a counterclockwise 
closed loop ${\cal C}$, which starts and ends at the base point 
and encloses the core, 
necessarily takes values in the subgroup  ${ H}_0$ of 
${ G}$ that leaves  the ground state  
$\langle \Phi_0 \rangle$ at the base point invariant, i.e.\
\bea                \label{paraflux}
W({\cal C}, {\bf x}_0 )   
&=& P \exp ( \im e \oint A^i dl^i) \; = \;  h \in { H}_0.
\eea   
The untraced Wilson loop operator~(\ref{paraflux}) completely classifies
the long range properties of the vortex solution.
It is invariant under a continuous deformation of the loop ${\cal C}$ that 
keeps the base point fixed and avoids the core of the vortex.
Moreover, it is invariant under continuous gauge transformations 
that leave the ground state $\langle \Phi_0 \rangle$
at the base point invariant. 
As in the abelian case, we fix this residual 
gauge freedom by sending all  nontrivial parallel transport 
into a narrow wedge or Dirac string from the core of the vortex to 
spatial infinity  as depicted in figure~\ref{sinvo}. 
It should be emphasized that our gauge fixing procedure for 
these vortex solutions involves two physically irrelevant choices.
First of all, we have chosen a fixed ground state $\langle \Phi_0 \rangle$
at the base point ${\bf x}_0$. This choice merely determines the embedding 
of the  residual symmetry group in ${ G}$ to be the stability group
${ H}_0$ of $\langle \Phi_0 \rangle$.
A different choice for this ground state gives rise to a 
different embedding of the residual symmetry group, 
but will eventually lead to an  unitarily equivalent 
quantum description of the discrete ${ H}$ gauge theory in the
Higgs phase.
For convenience, we subsequently fix the remaining gauge freedom by sending 
all nontrivial transport around the vortices to a small wedge.
Of course, physical phenomena will not depend on this choice.
In fact, an equivalent formulation of the low energy theory,
without fixing this residual gauge freedom for the vortices, 
can also be given, see for example~\cite{bucher}.

\begin{figure}[tbh]    \epsfxsize=3cm
\centerline{\epsffile{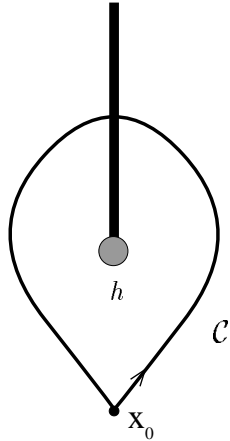}}
\caption{\sl  Single vortex solution. We have fixed the gauge freedom by 
sending all nontrivial parallel transport around the core in the Dirac 
string attached to the core. Thus, outside the core,
the Higgs field takes the same 
ground state value $\langle \Phi_0 \rangle$ everywhere
except for the region where the Dirac string is localized. Here it makes a 
noncontractible winding in the ground state manifold. 
This winding corresponds to a holonomy in the gauge field classified 
by the result of the untraced Wilson loop operator
$W({\cal C}, {\bf x}_0) = h \in { H}_0$, 
which picks up the nonabelian  magnetic flux located inside the core.}
\label{sinvo}
\end{figure}

In the gauge fixed prescription described above, 
we are still able to perform 
global symmetry transformations  $g \in { H}_0$  
on the vortex solutions that leave  
the ground state $\langle \Phi_0 \rangle $ invariant.
These transformations affect the 
field configuration of the vortex in the following way 
\bea
\Phi ({\bf x})      &\longmapsto&  g \; \Phi ({\bf x}) \\
A_\kappa  ({\bf x}) &\longmapsto&  g \; A_\kappa  ({\bf x}) \; g^{-1}.
\eea
As an immediate  consequence, we then obtain 
\bea 
W({\cal C}, {\bf x}_0 ) &\longmapsto& g \; W({\cal C}, {\bf x}_0 ) \; g^{-1},
\eea 
which shows that the flux of the vortex becomes conjugated
$h \mapsto ghg^{-1}$ under a residual global symmetry 
transformation $g \in { H}_0$. 
The conclusion is that  the nonabelian vortex solutions 
are in fact organized in degenerate multiplets under 
the residual global symmetry transformations ${ H}_0$, 
namely the different conjugacy classes of ${ H}_0$ denoted as  $^A C$,
where $A$ labels a particular conjugacy class. 
For convenience, we will refer to the stability group
of $\langle \Phi_0 \rangle $ as ${ H}$ from now on.

The different vortex solutions in a given conjugacy class $^A C$ of ${ H}$, 
being related by internal global symmetry transformations 
that leave the action~(\ref{nonhiks}) invariant, clearly 
carry the same external quantum numbers, that is, the  total energy of the 
configuration, the coresize etc. 
These solutions only differ by their internal magnetic flux quantum number.
This internal degeneracy becomes relevant in  
adiabatic interchange processes of  remote vortices in the plane.
Consider, for instance, the configuration of 
two remote vortices as presented in figure~\ref{metamo}. 
In the depicted adiabatic counterclockwise interchange of these vortices,
the  vortex initially carrying  the magnetic flux $h_2$
moves through the Dirac string attached to
the other vortex. As a result, its flux picks up a global symmetry 
transformation by the flux $h_1$ of the latter, i.e.\ 
$h_2 \mapsto h_1 h_2 h^{-1}_1$, such that the total flux 
of the configuration is conserved. 
This classical nonabelian Aharonov-Bohm effect 
appearing for noncommuting fluxes, which
has been called flux metamorphosis~\cite{bais}, 
leads to physical observable phenomena. Suppose, for example,
that  the magnetic flux $h_2$ was a member of a flux/anti-flux pair 
$(h_2,h_2^{-1})$ created from the vacuum. When $h_2$ encircles 
$h_1$, it returns as the flux $h_1 h_2 h^{-1}_1$ and will not
be  able to annihilate the flux $h_2^{-1}$ anymore.
Upon rejoining the pair we now obtain the stable flux
$h_1 h_2 h^{-1}_1 h_2$.  
Moreover, at the quantum level, 
flux metamorphosis leads to nontrivial Aharonov-Bohm scattering 
between nonabelian vortices as we will argue in more detail later on.

Residual global symmetry transformations naturally 
leave the aforementioned  observable Aharonov-Bohm effect for nonabelian 
vortices invariant. This simply follows from the fact that these 
transformations commute with this nonabelian Aharonov-Bohm effect.
To be precise,
a residual global symmetry transformation $g \in { H}$ on the two vortex 
configuration in figure~\ref{metamo}, for example,
affects the flux of both vortices through conjugation by the group 
element $g$, and it is easily verified that it makes no difference whether
such a transformation is performed before the interchange is started or 
after the interchange is completed. 
The extension of these classical 
considerations to configurations of more 
then two vortices in the plane is straightforward. Braid processes,
in which the fluxes of the vortices affect each other by conjugation, 
conserve the total flux of the configuration. The residual global symmetry
transformations $g \in { H}$ of the low energy regime,
which act by an overall conjugation of the fluxes 
of the vortices in the configuration by $g$, 
commute with these  braid processes.

\begin{figure}[tbh]    \epsfxsize=\textwidth
\centerline{\epsffile{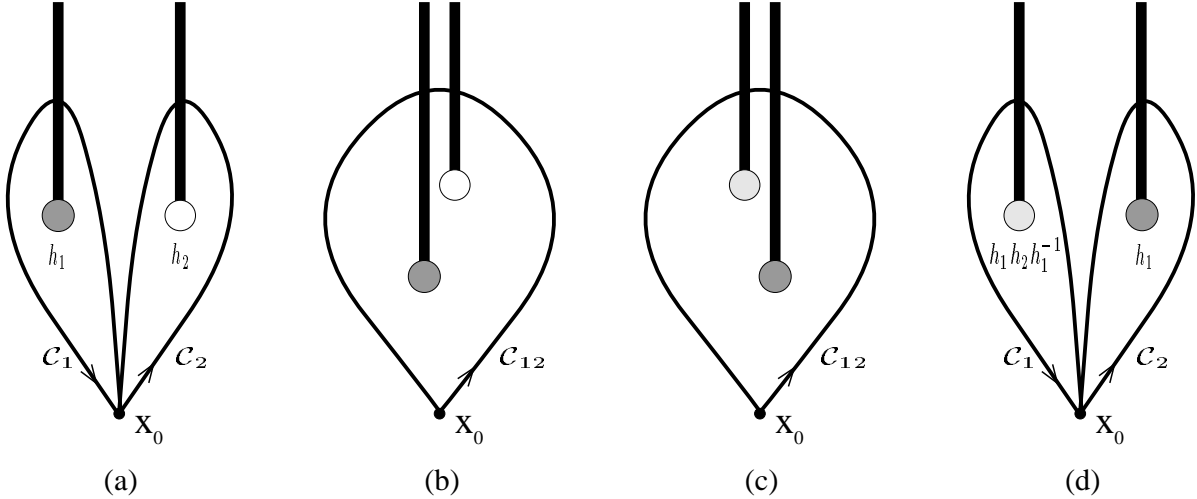}}
\caption{\sl  Flux metamorphosis. We start off with a classical configuration
of  two patched vortex solutions, as visualized in figure~(a). The vortices 
are initially assumed to carry the fluxes 
$W({\cal C}_{1}, {\bf x}_0 ) = h_1$ and 
$W({\cal C}_{2}, {\bf x}_0 ) = h_2$. The total flux of this 
configuration is picked up by the Wilson line integral along the loop
${\cal C}_{12}$ encircling both vortices as depicted in figure~(b): 
$W({\cal C}_{12}, {\bf x}_0 )
= W({\cal C}_{1} \circ {\cal C}_{2}, {\bf x}_0 )= 
W({\cal C}_{1}, {\bf x}_0 ) \cdot W({\cal C}_{2}, {\bf x}_0 )= h_1 h_2$.
Now suppose that the two vortices are interchanged in the counterclockwise 
fashion depicted in figures~(b)--(d). In this process  vortex~2 moves through
the Dirac string attached to vortex~1 and as a result its flux 
will be affected $h_2 \mapsto h_2'$. Vortex~1, on the other hand,
never meets any nontrivial parallel transport in the gauge 
fields and its flux remains the same. 
Since this local braid process should not be able to change the 
global properties of this system, i.e.\ the total flux, we have
$h_1 h_2= W({\cal C}_{12}, {\bf x}_0 )= W({\cal C}_{1}, {\bf x}_0 ) \cdot 
W({\cal C}_{2}, {\bf x}_0 )= h_2'h_1$. 
Thus the flux of vortex 2 becomes conjugated
$h_2'=h_1 h_2 h^{-1}_1$ by the flux of vortex~1 in this braid process.}
\label{metamo}
\end{figure}

As in the abelian case discussed in the previous sections, we wish 
to treat these nonabelian vortices as point particles in the first quantized 
description.  The degeneracy of these vortices under the residual 
global symmetry group ${ H}$ then indicates that 
we have to assign a finite dimensional  internal 
Hilbert space $V^A$ to these particles, which is 
spanned by the different fluxes in a given conjugacy 
class $^A C$ of ${ H}$ and endowed with the standard 
inner product~\cite{spm}
\bea
\langle h'| h \rangle &=& \delta_{h',h} \qquad \qquad \forall
\,  h,h'\in\, ^A C.
\eea    
Under the residual global symmetry transformations 
the flux eigenstates in this internal Hilbert space 
$V^A$ are affected through conjugation
\bea                         \label{fluxconjump}
g \in { H} : \qquad   |h \rangle &\longmapsto &
|g h g^{-1} \rangle.
\eea
In general, the particle can be in a 
normalized linear combination of the different 
flux eigenstates in the internal Hilbert space $V^A$. 
The residual global symmetry transformations~(\ref{fluxconjump}) 
act linearly on such states. 
Of course, the conjugated action of the residual symmetry group is in general 
reducible and, at first sight, it seems that we have to decompose 
this internal Hilbert space into the different irreducible components. 
This is not the case as we will see in more detail later on (see the 
discussion concerning relation~(\ref{scatamplo})). 
The point is that we can independently perform physical 
flux measurements by means of quantum interference experiments with 
electric charges. These measurements project 
out a particular flux eigenstate.
Clearly, these flux measurements do not commute with the residual global 
symmetry transformations and under their combined action the internal 
Hilbert spaces $V^A$ associated with the different conjugacy classes 
$^A C$ form irreducible representations.

The complete quantum state of these particles consists of an internal flux
part and an external part. The quantum state describing a single particle
in the flux eigenstate $|h_1 \rangle \in V^{A_1}$ at a fixed 
position ${\bf y}$ in the plane, for instance, is the formal tensor product 
$|h_1, {\bf y} \rangle = |h_1 \rangle |{\bf y} \rangle$. 
To proceed, the initial configuration depicted in 
figure~\ref{metamo} is described by 
the multi-valued two particle quantum state 
$|h_1, {\bf y} \rangle|h_2, {\bf z} \rangle$,
where again by convention the particle located most left in the plane 
appears most left in the tensor product.
The result of an adiabatic counterclockwise interchange 
of the two particles can now be summarized by the action of the 
braid operator
\bea                   \label{fluxmetam}
{\cal R} \; |h_1, {\bf y} \rangle|h_2, {\bf z} \rangle 
&=& |h_1 h_2 h^{-1}_1, {\bf y} \rangle|h_1, {\bf z}\rangle,
\eea
which acts linearly on linear combinations of these flux eigenstates.
What we usually measure in quantum interference experiments, however, 
is the effect in the internal wave function of a monodromy of 
the two particles 
\bea               \label{flumonodr}
{\cal R}^2  \; |h_1, {\bf y} \rangle|h_2, {\bf z} \rangle
&=& |(h_1 h_2) h_1 (h_1 h_2)^{-1}, {\bf y} \rangle|
h_1 h_2 h^{-1}_1, {\bf z}\rangle.
\eea 
This nonabelian Aharonov-Bohm effect can be probed either through 
a double slit experiment~\cite{colem,preslo} or through an Aharonov-Bohm 
scattering experiment as discussed in appendix~\ref{ahboverl} of
chapter~\ref{exampled2b}.
In the first case, we keep one particle fixed between the two slits, whereas
the other particle comes in as a plane wave. The geometry of the Aharonov-Bohm 
scattering experiment, depicted in figure~\ref{abverpres}
of appendix~\ref{ahboverl} 
is more or less similar. The interference pattern in both experiments
is determined by the internal transition amplitude
\bea             \label{transamp}
\langle u_2 |\langle u_1 | \; {\cal R}^2 \; | u_1 \rangle |u_2 \rangle,
\eea 
where $| u_1 \rangle$ and  $|u_2 \rangle$ respectively 
denote the properly normalized internal flux states of the two particles, 
which are generally linear combinations of the flux eigenstates in 
the corresponding internal Hilbert spaces $V^{A_1}$ and $V^{A_2}$.
The topological interference amplitudes~(\ref{transamp}) 
summarize all the physical 
observables for vortex configurations in the low energy regime to which we 
confine ourselves here.
As we have argued before, the  residual global symmetry transformations
affect internal multi-vortex states  through an overall conjugation  
\bea                         \label{fluxconjump2}
g \in { H} : \qquad   
|h_1 \rangle|h_2 \rangle & \longmapsto &
|g h_1 g^{-1} \rangle|g h_2 g^{-1} \rangle,
\eea
which commutes with the braid operator and therefore
leave the interference amplitudes~(\ref{transamp}) invariant.

\subsection{Including matter}  \label{inclumatter}

Let us now suppose that the total model is 
of the actual form
\bea                          \label{totmod}
S &=& S_{\rm YMH} + S_{\rm matter},
\eea
where  $S_{\rm YMH}$ denotes the action for the nonabelian 
Higgs model given in~(\ref{nonhiks}) and the action 
$ S_{\rm matter}$ describes additional matter fields minimally coupled  
to the gauge fields. 
In principle, these matter fields correspond 
to multiplets which transform 
irreducibly under the spontaneously broken symmetry group ${ G}$. 
Under the residual symmetry group 
${ H}$ in the Higgs phase, however, these representations
will become reducible  and  branch to  
UIR's $\Gamma$ of ${ H}$. Henceforth, it is assumed that 
the matter content of the model is such that all UIR's 
$\Gamma$ of ${ H}$ are indeed realized. We will 
treat the different charges $\Gamma$, appearing 
in the Higgs phase in this way~\cite{almawil,preskra}, as point particles. 
In the first quantized description, these point charges 
then carry an internal Hilbert space, namely  
the representation space associated with $\Gamma$.
Let us now consider a configuration of a nonabelian vortex in 
a flux eigenstate $|h \rangle$ at some fixed position in the plane 
and  a remote charge $\Gamma$ in a normalized internal charge 
state $|v\rangle$ fixed at another position. 
When the charge encircles the vortex in a counterclockwise fashion,
it meets the Dirac string and picks up a global symmetry 
transformation by the flux of the vortex
\bea                   \label{charmetam}
{\cal R}^2 \; |h, {\bf y} \rangle|v, {\bf z} \rangle 
&=& |h, {\bf y} \rangle|\Gamma(h) \, v, {\bf z} \rangle.
\eea 
Here, $\Gamma(h)$ is the matrix assigned to the group element $h$ in the 
representation  $\Gamma$. Note, that this Aharonov-Bohm effect boils down
to the abelian one given in~(\ref{monodro}) in case the residual gauge 
group $H \simeq \Z_N$.
Further, the residual global symmetry transformations on the two particle 
configuration 
\bea
g \in { H} : \qquad |h, {\bf y} \rangle|v, {\bf z} \rangle
&\longmapsto&
|ghg^{-1}, {\bf y} \rangle |\Gamma (g) \, v, {\bf z} \rangle,
\eea 
again commutes with the monodromy operation~(\ref{charmetam}). Thus, the 
interference amplitudes
\bea               \label{scatamplo}
\langle v| \langle h| \; {\cal R}^2 \; |h \rangle | v \rangle &=&
\langle h|h \rangle \langle v | \Gamma (h) \, v  \rangle \; = \; 
\langle v | \Gamma (h) \, v  \rangle,
\eea                          
measured in either double slit or Aharonov-Bohm scattering experiments 
involving these particles are invariant under the residual 
global symmetry transformations. As alluded to before,
these interference experiments can be used to measure the flux of 
a given vortex~\cite{colem,alfrev,vohgs,preslo}. To that end, we  
place the vortex between the two slits
(or alternatively use it as the scatterer in an Aharonov-Bohm 
scattering experiment)
and evaluate the interference pattern for  an incident beam of charges 
$\Gamma$ in the same internal state $| v \rangle$. In this way, we determine 
the interference amplitude~(\ref{scatamplo}). Upon repeating this experiment 
a couple of times with different internal states 
for the incident charge $\Gamma$, we can determine all matrix elements
of $\Gamma (h)$ and hence, iff $\Gamma$ corresponds to a faithful UIR of 
${ H}$, the group element $h$ itself. In a similar fashion, we may
determine the charge $\Gamma$ of a given particle 
and, moreover, its internal quantum state $|v \rangle$. In this case,
we put the unknown charge between the double slit
(or use it as the scatterer in an Aharonov-Bohm scattering experiment), 
measure the interference pattern for an incident beam of vortices
in the same flux eigenstate $|h\rangle$ and again repeat this experiment 
for all $h \in { H}$.

At this point, we have established the purely magnetic flux and the purely 
electric charge superselection sectors of the discrete $H$ gauge theory 
describing the long distance physics of the model~(\ref{totmod}).
The different magnetic sectors are labeled by the conjugacy classes  $^A C$
of  the residual gauge group ${ H}$, whereas the different electric 
charge sectors correspond to the different UIR's $\Gamma$  of $H$. 
The complete spectrum of this discrete gauge theory 
also contains dyonic combinations of these sectors.  The relevant remark
in this context is that we have not yet completely exhausted the action 
of the residual global symmetry transformations on the internal magnetic
flux quantum numbers.  As we have seen in~(\ref{fluxconjump}),
the residual global ${ H}$  transformations 
affect the magnetic fluxes through conjugation. 
The transformations that slip through this conjugation may 
in principle be implemented on an additional internal charge degree of freedom 
assigned to these fluxes~\cite{spm}.
More specifically, the global symmetry transformations that leave 
a given flux $|h\rangle$ invariant are those that 
commute with this flux, i.e.\ the group elements in the centralizer 
$^h N \subset { H}$. The internal charges that we can assign to this flux
correspond to the different UIR's $\alpha$ of the group $^h N$. Hence, 
the inequivalent dyons that can be formed in the composition of 
a global ${ H}$ charge $\Gamma$ with a magnetic flux 
$|h\rangle$ correspond to the different irreducible
components of the subgroup $^h N$ of ${ H}$ contained in the representation
$\Gamma$.
Two remarks are pertinent now.
First of all, the centralizers of different fluxes in a given 
conjugacy class $^A C$ are isomorphic. Secondly, the full set of the residual 
global ${ H}$ symmetry transformations relate the fluxes in a given 
conjugacy class carrying unitary equivalent centralizer charge 
representations. In other words, the different dyonic sectors are labeled
by $(\, ^A \! C, \alpha \, )$, where $^A \! C$ runs over the different 
conjugacy classes of ${ H}$ and $\alpha$ over the different nontrivial 
UIR's of the associated centralizer. The explicit transformation properties
of these dyons under the full global symmetry group ${ H}$ involve 
some conventions, 
which will be  discussed in 
in the next chapter, where we will identify the Hopf algebra 
related to a discrete $H$ gauge theory.

The  physical observation behind the formal construction of 
the dyonic sectors~\cite{spm} described above, is that we can, in fact, 
only measure the transformation 
properties of the charge of a given 
flux/charge composite under the centralizer of 
the flux of this composite, see also~\cite{preslo}. 
A similar phenomenon occurs in the 3+1 dimensional
setting for monopoles carrying a nonabelian magnetic charge, 
where it is known as the global color problem~\cite{balglob,nelson,nelsonc}. 
To illustrate this phenomenon,
we suppose that we have a composite of a pure flux $|h\rangle$
and a pure global ${ H}$ charge $\Gamma$ in some internal 
state $|v\rangle$. Thus, the complete internal state of the composite becomes
$|h, v\rangle$.
As we have argued before, the charge of a given object can be determined 
through double slit or Aharonov-Bohm scattering experiments involving 
beams of vortices in the same internal flux state $|h'\rangle$ and repeating 
these experiments for all $h' \in { H}$.
The interference amplitudes measured in 
this particular case are of the form 
\bea                \label{topobstr}
\langle h,v| \langle h'| \; {\cal R}^2 \; | h' \rangle  | h,v \rangle 
&=&   \langle h,v| h' h h'^{-1}, \Gamma (h') \, v \rangle
\langle h'| (h'h) h'(h'h)^{-1} \rangle \\
&=& \langle v |\Gamma (h') \, v \rangle \: \delta_{h, h' h h'^{-1}} \; , \nn 
\eea 
where we used~(\ref{flumonodr}) and~(\ref{charmetam}). As a result of  the 
flux metamorphosis~(\ref{flumonodr}), the interference term is only 
nonzero for experiments involving fluxes $h'$ that commute with the 
flux of the composite, i.e.\ $h' \in$$\, ^h N$. Hence, we are only 
able to detect the response of the charge $\Gamma$ of the composite 
to global symmetry transformations in $^h N$. 
This topological obstruction
is usually summarized with the 
statement~\cite{alice,balglob2,preskra,schwarz} 
that in the background of a single vortex $h$, the only `realizable' 
global symmetry transformations are those taking values 
in the centralizer $^h N$.

Let us close this section with a summary of the main conclusions. 
First of all, the complete spectrum  of the nonabelian 
discrete ${ H}$ gauge theory describing the long distance physics of 
the spontaneously broken model~(\ref{totmod}) can be 
presented as   
\bea                                            \label{conencen}
(\, ^A \! C, \alpha \, ) ,
\eea
where $^A \! C$ runs over the conjugacy classes of ${ H}$ and $\alpha$
denotes the different UIR's of the centralizer associated to a specific
conjugacy class $^A \! C$.  The purely magnetic sectors correspond to trivial
centralizer representations and are labeled by the different  nontrivial 
conjugacy classes. The pure charge sectors, on the other hand,
correspond to the trivial conjugacy class 
(with centralizer the full group ${ H}$) and are labeled by the different 
nontrivial UIR's of the residual symmetry group ${ H}$. 
The other sectors describe the dyons in this theory. 
Note that the sectors~(\ref{conencen}) boil down to the sectors of the 
spectrum~(\ref{compspectr}) in case ${ H} \simeq \Z_N$.

The residual long range interactions
between the particles in the spectrum~(\ref{conencen}) 
of a discrete $H$ gauge theory are topological Aharonov-Bohm interactions.
In a counterclockwise braid process involving two given particles, the 
internal quantum state of the particle that moves through
the Dirac string attached to the flux of the other particle picks 
up a global symmetry transformation by this flux. 
This (in general nonabelian) Aharonov-Bohm effect conserves the total flux 
of the system and moreover commutes with the residual 
global ${ H}$ transformations, which act simultaneously 
on the internal quantum states 
of all the particles in the system. The last property ensures that the 
physical observables for a given system, which are all 
related to this Aharonov-Bohm effect, are invariant
under global ${H}$ transformations.

An exhaustive treatment of the spin, braid and fusion properties 
of the particles in the spectrum~(\ref{conencen}) of a (nonabelian)
discrete  $H$ gauge theory involves the Hopf algebra $D(H)$ related to
a discrete ${ H}$ gauge theory, 
which will be discussed in the next chapter. 
For notational
simplicity, we will 
omit explicit mentioning of the external degrees of freedom of the particles 
in the following. In our considerations, we usually work with  
position eigenstates for the particles unless we are discussing double slit-
or Aharonov-Bohm scattering experiments in which the incoming projectiles 
are in momentum eigenstates.

\chapter{Algebraic structure} \label{algstruct}

It is by now well-established that 
there are deep connections between two dimensional rational conformal field 
theory, three dimensional topological field theory and quantum groups or
Hopf algebras. See for instance~\cite{alvarez1,alvarez2,witten} and references
therein.  
Discrete $H$ gauge theories, being examples of 
three dimensional topological field theories, naturally fit in this 
general scheme. As has been argued in~\cite{spm}, see also the 
references~\cite{spm1,sm}, the algebraic 
structure related to a discrete $H$ gauge theory 
is the quasitriangular Hopf algebra  $D(H)$   being the result 
of applying Drinfeld's
quantum double construction~\cite{drin,drinfeld} to 
the abelian algebra ${\cal F}(H)$ of functions on the finite 
group ${ H}$.~\footnote{For a thorough treatment of Hopf algebras in general 
and related issues, the interested reader is referred to 
the excellent book by Shnider and Sternberg~\cite{shnider}.}
Considered as a vector space, we then have 
$D(H) = {\cal F}(H) \ot {\bf C}[H]$, where ${\bf C}[H]$ denotes 
the group algebra  over the complex numbers ${\bf C}$.
Loosely speaking, the elements spanning the Hopf algebra $D(H)$
signal the flux of the particles~(\ref{conencen}) in the spectrum of the 
related discrete $H$ gauge theory and implement
the residual global symmetry transformations.
Under this action the particles form irreducible representations.
Moreover, the algebra  $D({ H})$  provides an unified description of 
the spin, braid and fusion properties of the particles.
Henceforth, we will simply refer to the  algebra $D({ H})$ 
as the quantum double. 
This name,  inspired by its mathematical construction, also 
summarizes nicely the physical content of a 
Higgs phase with a residual finite gauge group $H$.
The topological interactions between the particles 
are of a quantum mechanical nature, whereas the 
spectrum~(\ref{conencen}) exhibits an electric/magnetic self-dual 
(or double) structure.

In fact, the quantum double  $D({ H})$ was first proposed
by Dijkgraaf, Pasquier and Roche~\cite{dpr}. They identified it 
as the Hopf algebra associated with certain holomorphic orbifolds 
of rational conformal field theories~\cite{dvvv} 
and the related three 
dimensional topological field theories with finite gauge 
group $H$ as introduced by Dijkgraaf and Witten~\cite{diwi}. 
The new insight that emerged in~\cite{spm,spm1,sm}
was that such a topological field theory finds 
a natural realization as the residual discrete $H$ gauge theory 
describing the long range physics of gauge theories in which some 
continuous gauge group $G$ is spontaneously 
broken down to a finite subgroup $H$.

Here, we review the notion of the quantum double $D(H)$ and elaborate on the 
unified description this framework gives of the spin, braid and 
fusion properties of the topological and ordinary particles in the spectrum
of a  discrete $H$ gauge theory.

\sectiona{Quantum double}   \label{thequdo}

As has been  argued in section~\ref{inclumatter}, we are basically left 
with two physical operations on the particles~(\ref{conencen}) 
in the spectrum of a discrete ${ H}$ gauge theory. 
We can independently measure their magnetic flux and their 
electric charge through quantum 
interference experiments.
The magnetic flux of a particle is given by a group element $h \in { H}$, 
while the charge forms  an unitary irreducible representation of 
the centralizer $^h N$ of the flux $h \in { H}$ carried by the particle. 
Flux measurements then correspond to operators ${\rm P}_h$
projecting out a particular flux $h$,
while the charge of a given particle can be detected through 
its transformation properties under the residual global  
symmetry transformations $g \in ^h N \subset H$ that commute with the 
flux $h$ of the particle.

The operators ${\rm P}_h$ projecting out the 
flux $h \in { H}$ of a given quantum state   
naturally realize the projector  algebra
\bea           \label{multi}
{\rm P}_h {\rm P}_{h'} &=& \delta_{h,h'} \; {\rm P}_h, 
\eea 
with $\delta_{h,h'}$ the kronecker delta function for the group elements
$h,h' \in { H}$.   As we have seen in relation~(\ref{fluxconjump}),
global symmetry transformations $g \in { H}$ 
affect the fluxes through  conjugation. This implies that 
the flux projection operators and global symmetry  transformations 
for a nonabelian finite gauge group $H$ do not commute
\bea               \label{multip}
g \, {\rm P}_h  &=& {\rm P}_{ghg^{-1}} \, g. 
\eea
The combination 
of global symmetry transformations followed by flux measurements
\bea                          \label{lusien}
\{ {\rm P}_h \, g \}_{h,g\in { H}},  
\eea
generate  the quantum double $D({ H})= {\cal F}(H) \ot {\bf C}[H]$ 
and the multiplication~(\ref{multi}) and~(\ref{multip}) 
of these elements can be recapitulated 
as~\footnote{In \cite{spm,spm1,sm,dpr} 
the elements of the quantum double were denoted by $\hook{h}{g}$.
For notational simplicity, we use the presentation ${\rm P}_h \, g$  in these
notes.} 
\bea
{\rm P}_h \, g \cdot {\rm P}_{h'} \, g' &=& 
\delta_{h,g h'g^{-1}} \; {\rm P}_h \, gg' .
\label{algeb}
\eea

The different particles~(\ref{conencen}) in the spectrum of the associated 
discrete ${ H}$ gauge theory constitute the complete set of inequivalent 
irreducible representations of the quantum double  $D({ H})$. 
To make explicit the irreducible action of the quantum double on these 
particles, 
we have to develop some further notation. To start with,
we will label the group elements in the different conjugacy classes of 
${ H}$ as
\bea
^A\!C &=& \{^A\!h_1,\;^A\!h_2, \ldots,\,^A\!h_k\}.
\eea 
Let $^A\!N \subset { H}$ be the centralizer of the group element 
$^A\!h_1$ and 
$\{^A\!x_1,\,^A\!x_2,\ldots,\,^A\!x_k\}$ a set of representatives for the 
equivalence classes of ${ H}/^A\!N$, 
such that $^A\!h_i=\,^A\!x_i\,^A\!h_1\,^A\!x_i^{-1}$. 
For convenience, we will always take $^A\!x_1=e$, with $e$ the unit
element in ${ H}$. 
To proceed, the  basis vectors  of the 
unitary irreducible representation ${\alpha}$  of the centralizer 
$^A\!N$ will be denoted by $^{\alpha}\!v_j$.  With these conventions, the 
internal Hilbert space $V^A_{\alpha}$ is spanned by the quantum states
\bea                 \label{quantum states}
\{|\,^A\!h_i,\,^{\alpha}\!v_j\rangle\}_{i=1,\ldots,k}^{j=1,\ldots, 
\mbox{\scriptsize dim}{ \, \alpha}}. 
\eea
The combined action of a global symmetry  transformation $g \in { H}$ 
followed by a flux projection operation ${\rm P}_{h}$
on these internal flux/charge eigenstates spanning the  
Hilbert space $V^A_{\alpha}$ can then be presented as~\cite{dpr} 
\bea \label{13zo}                                              
\Pi^A_{\alpha}(\, {\rm P}_{h} \, {g}\, ) 
\; |\,^A\!h_i,\,^{\alpha}\!v_j \rangle &=&
\delta_{h,g\,^A\!h_i \, g^{-1}}\;\; |\, g\,^A\!h_i \, g^{-1},
\,{\alpha}(\tilde{g})_{mj}\,^{\alpha}v_m \rangle,
\eea
with
\bea   \label{centdef}
\tilde{g} &:=& \,^A\!x_k^{-1}\, g \,\, ^A\!x_i,
\eea
and  $\,^A\!x_k$  defined through $\,^A\!h_k := g\,^A\!h_i \, g^{-1}$.
It is easily verified that this element $\tilde{g}$ constructed from $g$ 
and the flux $^A\!h_i$ indeed commutes with $^A\!h_1$ and therefore can 
be implemented on the centralizer charge.  Two remarks are pertinent now.
First of all, there is of course arbitrariness involved in the 
ordering of the elements in the conjugacy classes and the choice of 
the representatives $\,^A\!x_k$ for the equivalence classes of the coset
${ H}/^A\!N$. However, different choices lead to unitarily equivalent 
representations of the quantum double. Secondly,
note that~(\ref{13zo}) is exactly the action anticipated in 
section~\ref{nonabz}.  
The flux $^A\!h_i$ of the associated particle is conjugated by the 
global symmetry transformation $g \in { H}$, 
while the part of $g$ that slips
through this conjugation is implemented on the 
centralizer charge of the particle.
The operator ${\rm P}_{h}$ subsequently projects out the flux $h$.

We will now argue that the  
flux/charge eigenstates~(\ref{quantum states}) 
spanning the internal Hilbert space $V^A_{\alpha}$  carry
the same spin, i.e.\ a counterclockwise
rotation over an angle of $2\pi$ gives rise to 
the same spin factor for all quantum states in $V^A_{\alpha}$.
As in our discussion of the (abelian) $\Z_N$ gauge  theory 
in section~\ref{abdyons}, we assume 
a small seperation between the centralizer charge and the flux of the 
particles.
In the aforementioned rotation, the centralizer charge of
the particle then moves through the Dirac string attached to its flux  
and as a result picks up a transformation by this flux. 
The element in the quantum double that implements  this effect 
on the internal quantum states~(\ref{quantum states})
is the central element 
\bea   \label{centraal}
\sum_h \; {\rm P}_h \,  h.
\eea 
It signals the flux of the internal quantum state and implements this flux
on the  centralizer charge
\bea
\Pi^A_{\alpha}(\, \sum_h \; {\rm P}_h \, h \,) \; 
|\,^A\!h_i,\,^{\alpha}\!v_j\rangle &=& 
|\,^A\!h_i,\,{\alpha}(^A\!h_1)_{mj}\,^{\alpha}v_m \rangle,
\eea
which boils down to the same matrix ${\alpha}(^A\!h_1)$
for all fluxes $^A\!h_i$  in $^A \! C$.
Here, we used~(\ref{13zo}) and~(\ref{centdef}).
Since $^A\!h_1$ by definition commutes with all the elements in the 
centralizer $^A\!N$, it follows from  Schur's lemma  that it 
is proportional to the unit matrix in the irreducible representation $\alpha$
\bea                 \label{spin!}
{\alpha}(^A\!h_1) &=& e^{2\pi \im s_{(A,\alpha)}} \,  {\bf 1}_\alpha.
\eea
This proves our claim. The conclusion is that there is an overall spin value 
$s_{(A,\alpha)}$ assigned to the sector $(\, ^A \! C, \alpha \, )$.
Note that the only sectors carrying a nontrivial spin are the dyonic sectors
corresponding to nontrivial conjugacy classes paired with 
nontrivial centralizer charges.

The  internal Hilbert space  describing a system of two particles 
$(\, ^A C,\alpha \,)$ and  $(\,^B C,\beta\,)$ 
is the tensor product $V_{\alpha}^A \ot V_{\beta}^B$.
The extension of the action of the quantum double
$D({ H})$ on the single particle states~(\ref{13zo})
to the two particle states in $V_{\alpha}^A \ot 
V_{\beta}^B$ is given  by the comultiplication
\bea
 \Delta(\,{\rm P}_h \, g \,) &=& 
\sum_{h' \cdot h''=h} {\rm P}_{h'} \, g \ot {\rm P}_{h''} \, {g},
\label{coalgeb}
\eea
which is an algebra morphism  from $D({ H})$ to 
$D({ H}) \ot D({ H})$. To be concrete, 
the tensor product representation of $D({ H})$ carried by 
the two particle internal Hilbert space $V_{\alpha}^A \ot V_{\beta}^B$  
is defined as  
$\Pi^A_{\alpha} \ot \Pi^B_{\beta}(\Delta (\, {\rm P}_h \, g \,)  )$.
The action~(\ref{coalgeb}) of the quantum double on the internal
two particle quantum states in $V_{\alpha}^A \ot V_{\beta}^B$
can be summarized as follows. In accordance with our observations 
in section~\ref{fluxmetamorphosis} and~\ref{inclumatter}, the residual 
global symmetry transformations $g \in { H}$ 
affect the internal quantum states of the two particles separately. 
The projection operator ${\rm P}_h$ subsequently projects out the 
total flux of the two particle quantum state, i.e.\ the product of the 
two fluxes. 
Hence, the action~(\ref{coalgeb}) of the quantum 
double determines the global properties of a given two particle
quantum state, which are conserved under the local process of fusing 
the two particles. 
It should be mentioned now that
the tensor product representation 
$(\Pi^A_{\alpha} \ot \Pi^B_{\beta}, V_{\alpha}^A \ot V_{\beta}^B)$ 
of $D({ H})$ is in general reducible 
and can be decomposed into a direct sum of irreducible representations
$(\Pi^C_{\gamma}, V^C_\gamma)$.
The different single particle states that can be obtained by 
the aforementioned fusion process are the states in the different internal 
Hilbert spaces $V^C_\gamma$ that occur in this decomposition.
We will return to an  elaborate discussion of the fusion rules 
in section~\ref{amalgz}.

An important  property of the 
comultiplication~(\ref{coalgeb}) is that it is coassociative, i.e. 
\bea \label{coas}
({\rm id} \ot \Delta)\, \Delta(\, {\rm P}_h\, g\,) = 
(\Delta \ot {\rm id}) \, \Delta(\, {\rm P}_h \, g \,) = 
\sum_{h' \cdot h''\cdot h'''=h} {\rm P}_{h'} \, g \ot {\rm P}_{h''} \, {g}
\ot {\rm P}_{h'''} \, {g}.
\eea     
This means that the action of the quantum double $D(H)$ 
on the three particle internal Hilbert space 
$V_\alpha^A \ot V_\beta^B \ot V_\gamma^C$
defined either through 
$({\rm id} \ot \Delta)\, \Delta$ or through 
$(\Delta \ot {\rm id}) \, \Delta$ is the same. 
Extending the action of the quantum double to systems containing  an 
arbitrary number of particles is now straightforward:
the global symmetry transformations
$g \in { H}$ are implemented on all the particles separately, 
while the operator ${\rm P}_h$ projects out the total flux of the system.

\begin{figure}[htb]    
\epsfxsize=14cm
\centerline{\epsffile{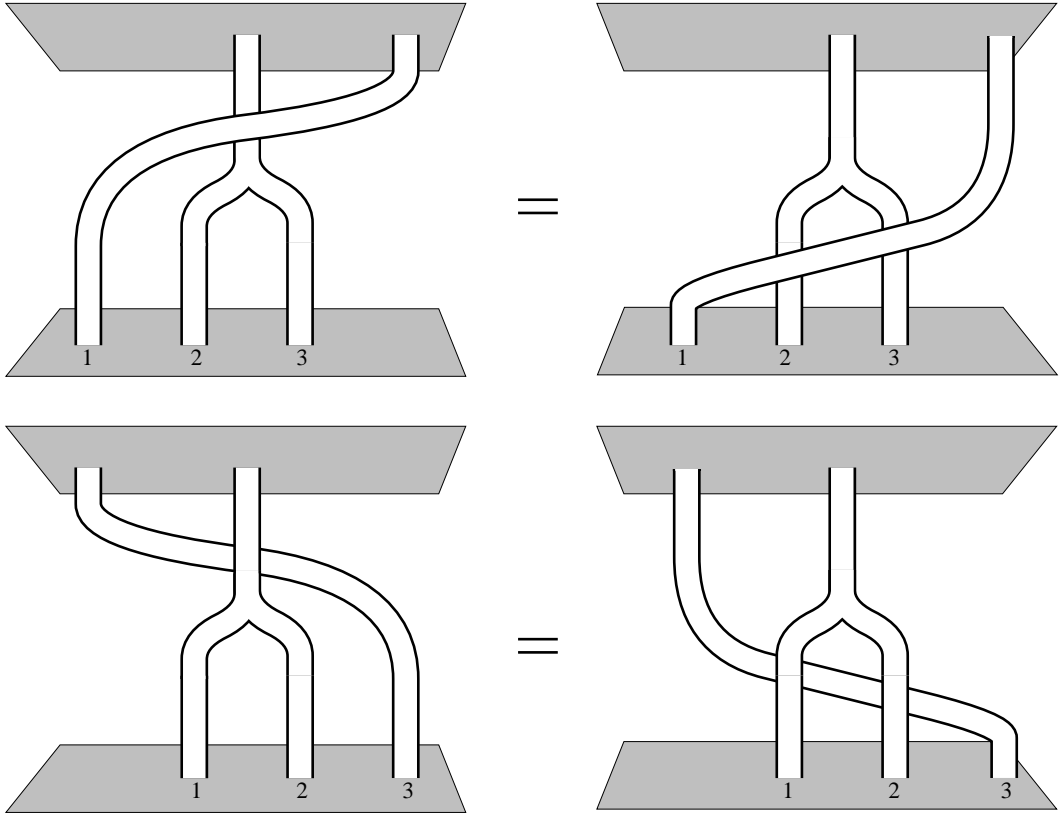}}
\caption{\sl  Compatibility of fusion
and braiding as expressed by the quasitriangularity conditions.
It  makes no difference whether a third particle  braids with two 
particles separately or with the composite that arises after fusing these two 
particles. The ribbons represent the trajectories of the particles.}
\label{qu1zon}
\end{figure}

The braid operation is formally implemented by the universal $R$-matrix, 
which is an element of $D({ H}) \ot D({ H})$
\bea
R &=&\sum_{h, g}\, {\rm P}_g \ot {\rm P}_h \, {g} .    \label{cruijff}
\eea
The $R$ matrix acts on a two particle state as a global symmetry 
transformation on the  second particle by the flux of the first particle.
The physical braid operator ${\cal R}$ that 
effectuates a counterclockwise interchange of  
the two particles  is  defined as the action of this $R$ matrix 
followed by a permutation $\sigma$  of the two particles
\bea \label{jordi}
{\cal R}_{\alpha\beta}^{AB} &:=& 
\sigma\circ(\Pi_{\alpha}^A\otimes\Pi_{\beta}^B)(\, R \, ),
\eea
To be explicit, on the two particle charge flux eigenstate
$|\,^A\!h_i,\,^{\alpha}\!v_j \rangle |\,^B\!h_m,\,^{\beta}\!v_n \rangle
\in  V^A_{\alpha}\ot V^B_{\beta}$, we have
\bea                 \label{braidact}
{\cal R} \; |\,^A\!h_i,\,^{\alpha}\!v_j \rangle 
|\,^B\!h_m,\,^{\beta}\!v_n \rangle
&=& 
|\,^A\!h_i \, ^B\!h_m \,^A\!h_i^{-1},\,\beta(\,^A\!\tilde{h}_i\,)_{ln} \,
^{\beta}\!v_l \rangle
|\,^A\!h_i,\,^{\alpha}\!v_j \rangle,
\eea  
where the element  $ ^A\!\tilde{h}_i$ is defined as in~(\ref{centdef}). 
Note that the expression~(\ref{braidact}), which
summarizes the braid operation on all conceivable two particle states 
in this theory, contains the braid  effects
established
in section~\ref{fluxmetamorphosis} and~\ref{inclumatter}, 
namely flux metamorphosis
for two pure magnetic fluxes~(\ref{fluxmetam}) and the Aharonov-Bohm
effect for a pure magnetic flux with a pure charge~(\ref{charmetam}).

It is now easily verified  that the
braid operator defined in~(\ref{braidact}) and 
the comultiplication given by~(\ref{coalgeb}) 
satisfy the quasitriangularity conditions 
\bea                              \label{grcommo}
{\cal R}\: \Delta(\, {\rm P}_h\, g\,) 
&=& \Delta(\, {\rm P}_h\, g\,) \:{\cal R}  \\
({\rm id} \ot \Delta)({\cal R}) &=& 
{\cal R}_2 \; {\cal R}_1
\label{triazo2}       \\
(\Delta \ot {\rm id})({\cal R}) &=& {\cal R}_1 \;  {\cal R}_{2}.
\label{triazo1}
\eea    
Here, the  braid operators ${\cal R}_1$ and ${\cal R}_2$ respectively 
act as ${\cal R} \ot {\bf 1}$ and ${\bf 1} \ot {\cal R}$ 
on three particle states in the internal Hilbert space
$V_\alpha^A \ot V_\beta^B \ot V_\gamma^C$.
The relation~(\ref{grcommo}) expresses the fact that the braid operator
commutes with the global symmetry transformations $g \in { H}$
and conserves the total magnetic flux of the configuration as
measured by ${\rm P}_h$.
In addition, the quasitriangularity 
conditions~(\ref{triazo2}) and~(\ref{triazo1}), which can 
be presented graphically as in figure~\ref{qu1zon},  imply  
consistency between braiding and fusing.
From the complete set of quasitriangularity conditions, it follows that the 
braid operator satisfies the  Yang-Baxter equation
\bea                    \label{yobo}
{\cal R}_1 \;  {\cal R}_{2}\; {\cal R}_1 &=&
{\cal R}_2 \; {\cal R}_1\;  {\cal R}_{2}.
\eea
Thus the braid operators~(\ref{braidact}) define representations 
of the braid groups discussed in section~\ref{braidgroups}. 
These unitary representations are in general reducible.  
So the internal Hilbert space describing a  multi-particle  system 
in general splits up into a direct sum of irreducible subspaces 
under the action of the braid group.   
The braid properties of the system
depend on the particular irreducible subspace.
If the dimension of the irreducible
representation is one, we are dealing with abelian braid 
statistics or ordinary anyons. If the dimension is larger
then one, we are dealing with nonabelian braid statistics, i.e.\
the nonabelian generalization of anyons.
Note that the latter higher dimensional irreducible representations only 
occur for systems consisting  of more than two particles, 
because the braid group for two particles is abelian. 

To conclude, the internal Hilbert space describing 
a multi-particle system in a discrete $H$ gauge theory
carries a representation of the internal symmetry
algebra $D({ H})$ and a braid group representation. 
Both representations are in general reducible. The quasitriangularity 
condition~(\ref{grcommo}) implies (see for instance~\cite{alvarez1,alvarez2})
that the action of the associated braid 
operators commutes with the action of the elements of $D({ H})$.
Hence, the multi-dyon internal Hilbert space can in fact be 
decomposed into a direct sum of irreducible subspaces under 
the direct product action of $D({ H})$ and the braid group. 
We discuss this in further detail in the next two sections. 
We first introduce the notion of truncated braid groups.

\sectiona{Truncated braid groups}  \label{trunckbr}

We turn to a closer examination of the braid group representations 
that occur in discrete ${ H}$ gauge theories. 
An important observation  in this respect is that 
the braid operator~(\ref{braidact}) is of finite order:
\bea                        \label{finR}
{\cal R}^m &=& {\bf 1} \ot {\bf 1},
\eea
with ${\bf 1}$ the identity operator and $m$ some integer depending on the 
specific particles on which the braid operator acts. 
In other words, we can assign a finite number  $m$ to any two particle 
internal Hilbert space $V_\alpha^A \ot V_\beta^B$, such that the effect
of $m$ braidings is trivial for all states in this internal Hilbert space.
This result, which can be traced back directly to the 
finite order of ${ H}$, 
implies that the multi-particle configurations appearing in a discrete ${ H}$ 
gauge theory actually realize representations of factor groups
of the braid groups discussed in section~\ref{braidgroups}.
Consider, for instance, a system consisting of $n$ 
indistinguishable particles. 
Hence, all particles carry the same internal 
Hilbert space  $V^A_{\alpha}$ and  the $n$ particle   
internal Hilbert space describing this system is  the 
tensor product space $(V^A_{\alpha})^{\ot n}$. 
The abstract generator $\tau_i$, which establishes a counterclockwise 
interchange of the two adjacent particles $i$ and $i+1$, acts on
this internal Hilbert space by means of  the operator
\bea                                 \label{brare}
\tau_i &\longmapsto& {\cal R}_i  \, ,
\eea
with
\bea                          \label{ridef}
{\cal R}_i & := & {\bf 1}^{\ot (i-1)} \ot {\cal R} \ot {\bf 1}^{\ot (n-i-1)}.
\eea
That is, the generator 
$\tau_i$ acts as~(\ref{braidact}) on the $i^{\rm th}$ and 
$(i+1)^{\rm th}$ entry in the tensor product space $(V^A_{\alpha})^{\ot n}$.
As follows from~(\ref{yobo}) and~(\ref{finR}),   
the homomorphism~(\ref{brare}) furnishes a representation of the 
braid group 
\bea
\label{eqy}
\ba{rcll}
\tau_i\tau_{i+1}\tau_i &=& \tau_{i+1}\tau_i\tau_{i+1} &
\qquad i=1,\ldots,n-2  \\
\tau_i\tau_j &=& \tau_j\tau_i & \qquad |i-j|\geq 2,
\ea
\eea
 with  the {\em extra} relation
\bea
\tau_i^m &=& e  \;\;\;\;\;\;\;\;\;\; i=1, \ldots , n-1,
\label{truncate}
\eea
where $e$ denotes the unit element or trivial braid.
For obvious reasons, we will call the factor groups
with defining relations~(\ref{eqy}) and the additional 
relation~(\ref{truncate}) {\em truncated} 
braid groups  $B(n,m)$. Here $n$ naturally stands 
for the number of particles and $m$ for the order of 
the generators $\tau_i$.

The observation  of the previous paragraph
naturally extends to a system containing 
$n$ distinguishable particles, i.e.\ 
the particles carry different internal Hilbert spaces or `colors' now.
The group that governs the monodromy properties of such a system 
is the truncated version $P(n,m)$ of the colored braid group $P_n({\bf R}^2)$
defined in section~\ref{braidgroups}. To be specific, the truncated 
colored braid 
group $P(n,m)$ is the subgroup of $B(n,m)$ generated by the elements
\bea                         
\gamma_{ij} &=& \tau_i \cdots \tau_{j-2}\; \tau_{j-1}^2 \; 
\tau_{j-2}^{-1}\cdots
\tau_i^{-1}   \qquad \qquad   1 \leq i<j \leq n,
\eea
with the extra relation~(\ref{truncate}) incorporated. 
Thus the generators of the pure braid group satisfy 
\bea     \label{puretru}           
\gamma_{ij}^{m/2} &=& e,  
\eea                                              
from which it is clear that the colored braid group $P(n,m)$ is,
in fact, only  
defined for even $m$.
The representation of the colored braid group $P(n,m)$ realized by a system 
of $n$ different particles in a discrete $H$ gauge theory
then  becomes  
\bea                                 \label{purarepo}
\gamma_{ij} &\longmapsto& {\cal R}_i \cdots {\cal R}_{j-1} \; {\cal R}_j^2 \;
{\cal R}_{j-1}^{-1} \cdots {\cal R}_i^{-1},
\eea
where the operators ${\cal R}_i$ defined by expression~(\ref{ridef}) 
now act on the tensor product space $ V^{A_{1}}_{\alpha_{1}} \ot \cdots \ot 
V^{A_{n}}_{\alpha_{n}}$ of $n$ different internal Hilbert spaces
$V^{A_{l}}_{\alpha_{l}}$ with $l \in 1,2,\ldots,n$.

Finally, a `mixture' of the above systems is of course also possible, that is,
a system containing a subsystem consisting of $n_1$ particles with `color' 
$V^{A_{1}}_{\alpha_{1}}$, a subsystem of $n_2$ particles carrying the 
different   
`color' $V^{A_{2}}_{\alpha_{2}}$ and so on.
Such a system  realizes a
representation of a truncated partially colored braid 
group (see section~\ref{braidgroups} and the references given there
for the definition of ordinary partially colored braid groups).
Let $n=n_1+n_2+\ldots$ again be the total number of particles in the system.
The truncated partially colored braid group associated with this system
then becomes the subgroup of some truncated braid group
$B(n,m)$, generated by 
the braid operations on  particles with the same `color' and the monodromy
operations on particles carrying different `color'.

The appearance of truncated rather than ordinary braid groups facilitates
the decomposition of a given multi-particle internal Hilbert space into 
irreducible subspaces under the braid/monodromy  operations. 
The point is that the representation theory of  
ordinary braid groups is rather complicated due to their infinite order.
The extra relation~(\ref{truncate}) for
truncated braid groups $B(n,m)$, however, causes these to become
finite for various values of the labels $n$ and $m$,  which 
leads to identifications  with well-known groups of 
finite order~\cite{pema}. It is instructive to consider
some of these cases explicitly.
The truncated braid group $B(2,m)$ for two indistinguishable 
particles, for instance, has 
only one generator $\tau$, which satisfies  $\tau^m = e$. 
Thus, we obtain the isomorphism
\bea
B(2,m) &\simeq& \Z_m.
\label{eq:rela}
\eea
For $m=2$, the relations~(\ref{eqy}) and~(\ref{truncate}) are the 
defining relations of the permutation group $S_n$ on $n$ strands
\bea
B(n,2) &\simeq& S_n.
\label{eq:rela1}
\eea
A less trivial example is the nonabelian truncated braid group $B(3,3)$ for 3
indistinguishable particles.
By explicit construction  from the defining relations~(\ref{eqy}) 
and~(\ref{truncate}), we arrive at the identification   
\bea     
B(3,3) &\simeq& \bar{T},
\eea
with $\bar{T}$ the lift of the tetrahedral group into $SU(2)$.
The structure of the truncated braid group $B(3,4)$ and its subgroup 
$P(3,4)$, which for example occur in a $\bar{D}_2$ gauge theory 
(see section~\ref{d2bqst}), can be found in appendix~\ref{trubra}.

To our knowledge,
truncated braid groups have not been studied in the literature so far 
and a complete classification is not available. 
An interesting group theoretical question in this context is
whether the truncated braid groups are of finite order for all 
values of the labels $n$ and $m$.

\sectiona{Fusion, spin, braid statistics and all that $\dots$}   
\label{amalgz}

Let $(\Pi^A_{\alpha}, V_\alpha^A)$ and
$(\Pi^B_{\beta}, V_\beta^B)$  be  two irreducible 
representations of the quantum double $D({ H})$ as defined in~(\ref{13zo}).
The tensor product representation 
$(\Pi^A_{\alpha} \ot \Pi^B_{\beta},V_\alpha^A \ot V_\beta^B)$, constructed
by means of the comultiplication~(\ref{coalgeb}), 
need not be irreducible. In general, it  gives rise to a
decomposition
\bea               \label{pietk}
\Pi^A_{\al}\otimes\Pi^B_{\beta}& = & \bigoplus_{C,\gamma}
N^{AB\gamma}_{\alpha\beta C} \; \Pi^C_{\gamma}, 
\eea
where $N^{AB\gamma}_{\alpha\beta C}$  stands for  the 
multiplicity of the irreducible representation 
$(\Pi^C_{\gamma}, V^C_{\gamma})$.
From  the orthogonality relation for the characters of the irreducible 
representations of $D({ H})$, we infer~\cite{dpr}
\bea          \label{Ncoe}
N^{AB\gamma}_{\alpha\beta C} &=& \frac{1}{|{ H}|} \sum_{h,g}  \;
            \mbox{tr}  \left( \Pi^A_{\alpha} \ot \Pi^B_{\beta}
                      (\Delta (\, {\rm P}_h \, g \,)  ) \right)   \;
            \mbox{tr}  \left( \Pi^C_{\gamma} (\, {\rm P}_h \, g \,) \right)^*, 
\eea
where $|{ H}|$ denotes the order of the group ${ H}$ and $*$ indicates
complex conjugation.
The fusion rule~(\ref{pietk}) 
now determines which particles 
$(\,^C C,\gamma)$ can be formed in the composition 
of  two given particles $(\, ^A C,\alpha \,)$ and  $(\,^B C,\beta\,)$,
or if read backwards, gives the decay channels of the particle
$(\,^C C,\gamma\,)$.

The fusion algebra, spanned by the elements
$\Pi^A_{\alpha}$ with multiplication rule~(\ref{pietk}), is 
commutative and associative and can therefore be diagonalized.
The matrix  implementing this diagonalization is the so-called
modular $S$ matrix~\cite{ver0} 
\bea                                \label{fusionz}   
S^{AB}_{\alpha\beta} &:=& \frac{1}{|{ H}|} \, \mbox{tr} \; {\cal 
R}^{-2 \; AB}_{\; \; \; \; \; \alpha\beta}       \\
&=&
\frac{1}{|{ H}|} \, \sum_{\stackrel{\,^A\!h_i\in\,^A\!C\,,^B\!h_j\in\,
^B\!C}{[\,^A\!h_i,\,^B\!h_j]=e}} 
\mbox{tr} \left(\alpha (\,^A\!x_i^{-1}\,^B\!h_j\,^A\!x_i) \right)^*
\; \mbox{tr} \left( \beta(\,^B\!x_j^{-1}\,^A\!h_i\, ^B\!x_j) \right)^*, \nn
\eea
The modular $S$ matrix~(\ref{fusionz})  
contains all information concerning 
the fusion algebra defined in~(\ref{pietk}).
In particular, the multiplicities~(\ref{Ncoe}) can be expressed in 
terms of the modular $S$ matrix by means of Verlinde's formula~\cite{ver0}
\bea      \label{verlindez}
N^{AB\gamma}_{\alpha\beta C}&=&\sum_{D,\delta}\frac{
S^{AD}_{\alpha\delta}S^{BD}_{\beta\delta}
(S^{*})^{CD}_{\gamma\delta}}{S^{eD}_{0\delta}}.
\eea 
Whereas the modular $S$ matrix is determined through the monodromy operator
following from~(\ref{braidact}),
the modular matrix $T$ contains the spin 
factors~(\ref{spin!}) assigned to the particles in the spectrum of 
a discrete $H$ gauge theory 
\bea                                         \label{modutz}
T^{AB}_{\alpha\beta} &:=& 
\delta_{\alpha,\beta} \, \delta^{A,B} \; \exp(2\pi \im s_{(A,\alpha)})
\; = \;
\delta_{\alpha,\beta} \, \delta^{A,B} 
\frac{1}{d_\alpha}   \, \mbox{tr} \left(  \alpha(^A\!h_1) \right),
\eea
where $d_\alpha$ stands for the dimension of the centralizer charge 
representation $\alpha$ of the particle $(\, ^A C, \alpha \,)$. 
The  matrices~(\ref{fusionz}) and~(\ref{modutz}) now 
realize an unitary representation of the modular group $SL(2,\Z)$
with the following  relations~\cite{dvvv}
\begin{eqnfourarray}
{\cal C} &=&(ST)^3 \; = \; S^2,      &         \label{charconj} \\
S^* &=& {\cal C} S \; = \;S^{-1}, & $\qquad S^t \;=\; S, $        \label{sun}      \\
T^* &=&T^{-1}, & $\qquad T^t \;=\; T.$                   \label{tun}
\end{eqnfourarray}  
The  relations~(\ref{sun}) and~(\ref{tun}) express the fact that 
the matrices~(\ref{fusionz}) and~(\ref{modutz}) 
are symmetric and unitary. To proceed, the matrix ${\cal C}$ defined 
in~(\ref{charconj}) represents the charge conjugation operator, which
assigns an unique anti-partner 
${\cal C} \, (\, ^A C,\alpha \,)=(\, ^{\bar{A}} C,\bar{\alpha} \,)$
to each particle  $(\, ^A C,\alpha \,)$ in the spectrum,
such that the vacuum channel occurs in the fusion rule~(\ref{pietk}) 
for the particle/anti-particle pairs.
Also, note that the complete set of relations imply that 
the charge conjugation matrix ${\cal C}$ commutes with 
the modular matrix $T$, which implies
that a given particle carries the same spin as its anti-partner.

Having determined the fusion rules and the associated modular algebra,
we turn to the issue of braid statistics and the fate of the spin
statistics connection in  nonabelian discrete $H$ gauge theories.
Let us emphasize from the outset that much of what follows has 
been established elsewhere in a more general setting.
See~\cite{alvarez1,alvarez2,moseco} and the references therein 
for the conformal field theory point 
of view and~\cite{frohma,frogama,witten} for the related 
2+1 dimensional space time perspective.

We first discuss a system consisting of two distinguishable 
particles $(\, ^A C,\alpha \,)$ and  $(\,^B C,\beta\,)$. The associated 
two particle internal Hilbert space $V_\alpha^A \ot V_\beta^B$ carries 
a representation of the abelian truncated colored braid group $P(2,m)$ with 
$m/2 \in \Z$ the order of the monodromy matrix ${\cal R}^2$ for this 
particular two-particle system.
This representation decomposes into a direct sum 
of one dimensional irreducible subspaces, each being labeled by the 
associated eigenvalue of the monodromy matrix ${\cal R}^2$. 
Recall from section~\ref{thequdo}, that the monodromy operation
commutes with the action of the quantum double. 
This implies that the decomposition~(\ref{pietk}) 
simultaneously diagonalizes
the monodromy matrix. To be specific, the two particle 
total flux/charge eigenstates spanning a given 
fusion channel $V_\gamma^C$ all carry the same monodromy eigenvalue,
which in addition can be shown to satisfy 
the generalized spin-statistics connection~\cite{dpr}
\bea             \label{gekspist}
K^{ABC}_{\alpha\beta\gamma} \;
{\cal R}^2 &=& 
e^{2\pi \im(s_{(C,\gamma)}-s_{(A,\alpha)}-s_{(B,\beta)})} \;\;
K^{ABC}_{\alpha\beta\gamma}.
\eea
Here,  $K^{ABC}_{\alpha\beta\gamma}$ stands for the projection on  
the irreducible component $V_\gamma^C$  of  $V_\alpha^A \ot V_\beta^B$.  
In other words, the monodromy operation on a two particle state in a given
fusion channel is the same as a clockwise rotation over an angle of $2\pi$ 
of the two particles separately accompanied by a counterclockwise
rotation over an angle 
of $2\pi$ of the single particle state emerging after fusion. 
This is consistent with the observation that these two processes can be 
continuously deformed into each other, see the associated
ribbon diagrams depicted in figure~\ref{kanaal}.
The discussion can now be summarized by the statement that 
the total internal Hilbert space $V_\alpha^A \ot V_\beta^B$ decomposes 
into the following  direct sum of irreducible representations of the direct 
product  $D({ H}) \times P(2,m)$ 
\bea
 \bigoplus_{C,\gamma} N^{AB\gamma}_{\alpha\beta C} \; 
(\Pi^C_{\gamma}, \Lambda_{C-A-B} ),
\eea 
where $\Lambda_{C-A-B}$ denotes the one dimensional 
irreducible representation of $P(2,m)$ in which the monodromy 
generator $\gamma_{12}$ acts as~(\ref{gekspist}).

\begin{figure}[htb]    \epsfxsize=11cm
\centerline{\epsffile{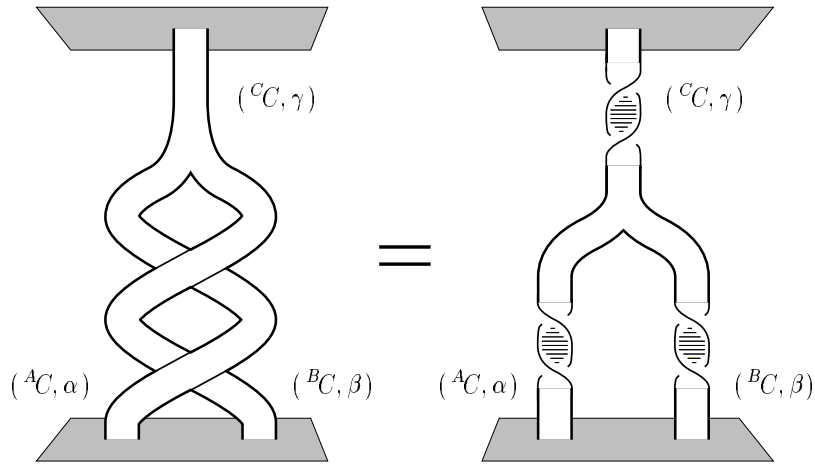}}
\caption{\sl Generalized spin-statistics connection. 
The displayed ribbon diagrams are homotopic as can 
be checked with the pair of pants you are presently wearing. 
This means that
a monodromy of two particles in a given fusion channel followed  
by fusion of the pair  can be continuously deformed into the process
describing a rotation over an angle of $-2\pi$ of 
the two particles seperately followed by fusion of the pair and 
a final rotation over an angle of $2\pi$ of the composite. }
\label{kanaal}
\end{figure}

The analysis for a configuration 
of two indistinguishable particles $(\, ^A C,\alpha \,)$
is analogous. The total internal Hilbert space 
$V_\alpha^A \ot V_\alpha^A$ decomposes
into one dimensional irreducible subspaces under the action 
of the truncated braid group $B(2,m)$ with $m$ the order 
of the braid operator ${\cal R}$, which depends on the system
under consideration. 
By the same argument 
as before, the two particle total flux/charge eigenstates spanning 
a given fusion channel $V_\gamma^C$ all 
carry  the same one dimensional representation
of $B(2,m)$.  The quantum statistical parameter assigned to this channel
now satisfies the square root version of the generalized spin-statistics 
connection~(\ref{gekspist})
\bea             \label{gespietst}
K^{AAC}_{\alpha\alpha\gamma} \;
{\cal R} &=& \epsilon \;
e^{ \pi \im(s_{(C,\gamma)}-2 s_{(A,\alpha)})} \; \;
K^{AAC}_{\alpha\alpha\gamma},
\eea
with $\epsilon$ a sign depending on whether the 
fusion channel $V_\gamma^C$ appears in a symmetric or an
anti-symmetric fashion~\cite{alvarez1,alvarez2}. 
In other words, the internal space Hilbert space for a system of two 
indistinguishable particles $(\, ^A C,\alpha \,)$ breaks up into the following
irreducible representations of the direct product $D(H) \times B(2,m)$
\bea                           \label{nil!}
 \bigoplus_{C,\gamma} N^{AA\gamma}_{\alpha\alpha C} \; 
(\Pi^C_{\gamma}, \Lambda_{C-2A} ),
\eea 
with $\Lambda_{C-2A}$ the one dimensional representation 
of the truncated braid group $B(2,m)$ defined in~(\ref{gespietst}).

The result~(\ref{gespietst}) is actually rather surprising.
It states that  indistinguishable particle systems
in a nonabelian discrete $H$ gauge theory quite
generally violate the canonical spin-statistics connection~(\ref{spistath}).
More accurately, in a nonabelian discrete gauge theory
we are dealing with the generalized connection~(\ref{gespietst}), which 
incorporates the canonical one. In fact, the canonical spin-statistics
connection is retrieved in some particular channels occurring in~(\ref{nil!}),
as we will argue now.
Let us first emphasize that the basic assertions for the ribbon proof 
depicted in figure~\ref{spinstafig} are naturally  
satisfied in the nonabelian setting as well. For every particle 
$(\, ^A C,\alpha \,)$ in the spectrum there exists an 
anti-particle $(\, ^{\bar{A}} C,\bar{\alpha} \,)$ such that 
under the proper composition the pair acquires the quantum numbers of the 
vacuum and may decay. Moreover, every particle carries the same spin as its
anti-partner, as indicated by the fact that the 
charge conjugation operator ${\cal C}$ commutes with the modular matrix $T$. 
It should be noted now that the ribbon proof in figure~\ref{spinstafig}
actually {\em only} applies to states in which
the particles that propagate along the exchanged ribbons are 
in strictly  identical internal states. Otherwise the ribbons can not be
closed. Indeed, we find that the action~(\ref{braidact}) of the braid operator 
on two particles in identical internal flux/charge eigenstates
\bea                 \label{braidactid}
{\cal R} \; |\,^A\!h_i,\,^{\alpha}\!v_j \rangle 
|\,^A\!h_i,\,^{\alpha}\!v_j \rangle
&=& 
|\,^A\!h_i,\,\alpha\,(\,^A\! h_1\,)_{mj} \,
^{\alpha}\!v_j \rangle
|\,^A\!h_i,\,^{\alpha}\!v_j \rangle ,
\eea  
boils down to the diagonal matrix~(\ref{spin!}) and therefore to the 
same spin factor~(\ref{spist}) for all  $i,j$
\bea                             \label{spist}
\exp(\im \Theta_{(A,\alpha)}) &=& \exp(2\pi \im s_{(A,\alpha)}).
\eea
The conclusion is that the canonical spin-statistics connection is 
restored in the fusion channels spanned by linear combinations 
of the states~(\ref{braidactid}) in which the particles are in strictly 
identical internal flux/charge eigenstates. 
The quantum statistical parameter~(\ref{gespietst}) assigned to 
these channels reduces to the spin factor~(\ref{spist}). Thus the 
effect of a counterclockwise interchange 
of the two particles in the states in these channels is the same 
as the effect of rotating one of the particles over an angle of $2\pi$.
To conclude, the closed ribbon proof does not apply to the other channels
and we are left with the more involved connection~(\ref{gespietst}) following 
from the open ribbon argument displayed in figure~\ref{kanaal}.

Finally, higher dimensional irreducible braid group representations 
are conceivable for a system which consists of more than two particles. 
The occurrence of such 
representations simply means that the generators of the braid group can 
not be diagonalized simultaneously. What happens in this situation is that
under the full set of braid operations, the system jumps between isotypical 
fusion channels, i.e.\ fusion channels of the same type or `color'.
Let us make this
statement more precise. To keep the discussion general, we
do not specify the nature of the particles in the system.
Depending  on whether the system consists of 
distinguishable particles, indistinguishable particles or some `mixture', 
we are dealing
with a truncated braid group, a colored braid group or a 
partially colored braid group respectively.
The internal Hilbert for such a system again decomposes  
into a direct sum of irreducible subspaces (or fusion channels) under 
the action of the quantum double $D(H)$.  
Given the fact that the action of the associated  
braid group commutes with that of the quantum double, we are left 
with two possibilities.
First of all, there will in general be some 
fusion channels separately  being  invariant under
the action of the full  braid group.
As in the two particle systems discussed before, 
the total flux/charge eigenstates spanning such a 
fusion channel, say $V_\gamma^C$, carry the same one 
dimensional irreducible representation $\Lambda_{ab}$ of the braid group. 
That is, these states realize abelian braid statistics with the same
quantum statistical parameter.
The fusion channel $V_\gamma^C$ then carries the irreducible representation 
$(\Pi_\gamma^C, \Lambda_{ab})$ of the direct product of the 
quantum double and the braid group.
In addition, it is also feasible 
that states carrying the {\em same} total flux and charge 
in {\em different} (isotypical) fusion channels 
are mixed under the action of 
the full  braid group.  In that case, we are dealing with 
a higher dimensional irreducible representation of the 
truncated braid group or nonabelian braid statistics. 
Note that nonabelian braid statistics is conceivable, 
if and only if some fusion channel, say $V_\delta^D$, occurs more 
then once  in the decomposition of the Hilbert space
under the action of the quantum double. 
Only then there are some orthogonal states with 
the same total flux and charge available to span an higher dimensional 
irreducible representation of the braid group.
The number $n$ of fusion channels $V_\delta^D$ 
related by the action of the braid operators
now constitutes the dimension of the irreducible representation 
$\Lambda_{nonab}$ of the braid group and the multiplicity of
this representation 
is the dimension $d$ of the fusion channel $V_\delta^D$. 
To conclude, the direct sum of these $n$ fusion 
channels $V_\delta^D$ carries an $n \cdot d$ dimensional 
irredicible representation $(\Pi_\delta^D, \Lambda_{nonab})$
of the direct product of the quantum double and the braid group.

\chapter{$\bar{D}_2$  gauge theory}   \label{exampled2b}

In this last chapter, we will illustrate the foregoing 
general considerations with one of the simplest nonabelian 
discrete ${ H}$ gauge theories,
namely that with finite gauge group the double dihedral group
${ H} \simeq \bar{D}_2$. See also the references~\cite{spm,spm1,sm}
in this connection. 
The plan is as follows. In section~\ref{chesd2b}, we establish 
the spectrum of a $\bar{D}_2$ gauge theory, 
the spin factors assigned to the particles
and the fusion rules. 
Here, we also 
elaborate on a feature special for nonabelian discrete ${ H}$ 
gauge theories: a pair of nonabelian magnetic 
fluxes can carry charges that are not 
localized on any of the two fluxes nor anywhere else. 
Among other things, we will show that these so-called 
Cheshire charges  a nonabelian flux pair may carry, can be excited 
by monodromy  processes with other particles in the spectrum.
In section~\ref{abd2b}, we  treat the 
(nonabelian) cross sections measured in Aharonov-Bohm scattering 
experiments involving the particles in a $\bar{D}_2$ gauge theory.
Further, the issue of (nonabelian) braid statistics 
realized by the multi-particle configurations in this theory 
will be dealt with in 
section~\ref{d2bqst}. We have also included two appendices.
Appendix~\ref{ahboverl}  contains a concise review of the
Aharonov-Bohm scattering experiment focussing on the 
cross sections appearing in (non)abelian discrete $H$ gauge theories.
Finally, in appendix~\ref{trubra}, we  give the group structure of 
two particular truncated braid groups which enter the analysis  
in section~\ref{d2bqst}.

\sectiona{Alice in physics}     \label{chesd2b}

A $\bar{D}_2$  gauge theory may, for instance, 
arise as `the long distance remnant' of a Higgs model of the 
form~(\ref{totmod}) in which 
the gauge group  $G \simeq SU(2)$ is spontaneously broken 
down to the double dihedral 
group $H \simeq \bar{D}_2 \subset SU(2)$.
Since $SU(2)$ is simply connected, the fundamental group 
$\pi_1(SU(2)/\bar{D}_2)$ coincides with the residual symmetry group
$\bar{D}_2$. Hence, the stable magnetic fluxes in this broken 
theory are indeed labeled by the group elements of $\bar{D}_2$.
See the discussion concerning 
the isomorphism~(\ref{flulaord}) in section~\ref{topclas}.
In the following, we will not dwell any further on the explicit
details of this or other possible
embeddings in broken gauge theories and simply focus on the
features of the $\bar{D}_2$ gauge theory itself. 
We start with a discussion of the spectrum.

\begin{table}[htb] 
\begin{center}
\begin{tabular}[t]{lcl} \hline
$ \mbox{Conjugacy class}                      $ & & $ \qquad 
\mbox{Centralizer}       $  \\       \hl \\[-4mm] 
$ e=\{ e \}                                      $& &$ \qquad \bar{D}_2$\\ 
$ \bar{e}=\{\bar{e}\}                           $& &$ \qquad \bar{D}_2$\\ 
$ X_1=\{X_1,\bar{X}_1\}\ $& &$ 
\qquad \Z_4\simeq \{e,X_1,\bar{e},\bar{X}_1\} $\\ 
$ X_2=\{X_2,\bar{X}_2\} $& &$ 
\qquad \Z_4\simeq \{e,X_2,\bar{e},\bar{X}_2\} $\\ 
$ X_3=\{X_3,\bar{X}_3\} $& &$ 
\qquad \Z_4 \simeq \{e,X_3,\bar{e},\bar{X}_3\}$\\[1mm]
\hl
\end{tabular}
\end{center} 
\caption{\sl Conjugacy classes of the double dihedral group 
$\bar{D}_2$ together with their centralizers.}  
\label{tabcond2}  
\end{table}
\begin{table}[htb] 
\begin{center}
\begin{tabular}[t]{lrrrrr}     \hline   \\[-4mm]
$\str \bar{D}_2\qquad$&$ e $&$ \bar{e}$&$ X_1  $&$ X_2  $&$ X_3  $\\ 
\hl \\[-4mm]
$ 1   \str$&$ 1 $&$ 1      $&$ 1  $&$ 1  $&$ 1  $\\ 
$ J_1 \str$&$ 1 $&$ 1      $&$ 1  $&$-1  $&$-1  $\\ 
$ J_2 \str$&$ 1 $&$ 1      $&$-1  $&$ 1  $&$-1  $\\ 
$ J_3 \str$&$ 1 $&$ 1      $&$-1  $&$-1  $&$ 1  $\\ 
$ \chi \str$&$ 2 $&$ -2     $&$ 0  $&$ 0  $&$ 0  $\\[1mm]   \hl
\end{tabular} \qquad \qquad
\begin{tabular}[t]{lrrrrr} \hline \\[-4mm]
$\str \Z_4\qquad$&$ e $&$ X_a      $&$ \bar{e} $&$ \bar{X}_a $\\ 
\hl  \\[-4mm]
$   {\Gamma^0}\str$&$ 1 $&$ 1       $&$ 1  $&$ 1     $\\ 
$   {\Gamma^1}\str$&$ 1 $&$ \imath  $&$-1  $&$-\imath$\\ 
$   {\Gamma^2}\str$&$ 1 $&$ -1      $&$ 1  $&$-1     $\\ 
$   {\Gamma^3} \str$&$ 1 $&$ -\imath $&$-1  $&$\imath $\\[1mm]   \hl
\end{tabular}
\end{center}
\caption{\sl Character tables of $\bar{D}_2$ and $\Z_4$.}
\label{chartabd2}
\end{table}

The double dihedral group $\bar{D}_2$ is a group of order~$8$ 
with a nontrivial centre of order~$2$. The magnetic fluxes 
associated with its group elements are organized in the 
conjugacy classes exhibited in  table~\ref{tabcond2}.
There are five conjugacy classes which we will 
denote as $e,\bar{e},X_1,X_2$ and $X_3$.   
The conjugacy class $e$ naturally corresponds to the trivial 
magnetic flux sector, while
the conjugacy class $\bar{e}$ consists of the nontrivial centre element. 
The conjugacy classes  $X_1,X_2$ and $X_3$ all contain two commuting 
elements of order~$4$. In other words, a $\bar{D}_2$ gauge theory
features four nontrivial purely magnetic 
flux sectors: one singlet flux $\bar{e}$ 
and three different doublet fluxes $X_1,X_2$ and $X_3$. 
The purely electric charge sectors, on the other hand, correspond to 
the UIR's of $\bar{D}_2$.
From the character table displayed in table~\ref{chartabd2}, 
we infer that there are four nontrivial pure charges in the spectrum:
three singlet charges $J_1,J_2,J_3$ and one doublet charge $\chi$.  
The magnetic fluxes $X_a$ and $\bar{X}_a$ (with $a \in 1,2,3$)
act on the doublet charge $\chi$ 
as $\im \sigma_a$ and $-\im \sigma_a$, respectively,
where the symbol $\sigma_a$ denotes the Pauli matrices.
Let us now turn to the dyonic sectors. These are constructed by 
assigning a nontrivial centralizer representation to the nontrivial
fluxes. The centralizers associated with the different flux sectors 
can be found in table~\ref{tabcond2}. The flux $\bar{e}$ 
obviously commutes with the full group $\bar{D}_2$, while the centralizer
of the other flux sectors is the cyclic group $\Z_4$.  
Hence, we arrive at thirteen different dyons:
three singlet dyons and one doublet dyon
associated with the flux $\bar{e}$ and nine doublets dyons 
associated with the fluxes $X_1,X_2$ and $X_3$ paired with 
nontrivial $\Z_4$ representations.
All in all, the spectrum of this theory features 22 particles, which 
will be labeled as
\bea   \label{spectred2}       \ba{rclrcl}
1        &:=& (e,\, 1 )
& \qquad \qquad \bar{1}        &:=& (\bar{e},\, 1 )       \\
J_a      &:=& (e,\, J_a )  
& \qquad \qquad \bar{J}_a      &:=& (\bar{e},\, J_a)       \\
\chi     &:=& (e,\, \chi )     
& \qquad \qquad \bar{\chi}     &:=& (\bar{e},\, \chi)     \\
\sigma_a^+&:=& (X_a,\, {\Gamma^0})
& \qquad \qquad \sigma_a^-&:=& (X_a,\, {\Gamma^2})  \\ 
\tau_a^+&:=& (X_a,\, {\Gamma^1})
& \qquad \qquad \tau_a^-&:=& (X_a,\,{\Gamma^3}),
\ea
\eea 
for convenience.
Note that the square of the dimensions of the internal Hilbert spaces
carried by these particles indeed 
add up to the order of the quantum double $D(\bar{D}_2)$: 
$8 \cdot 1^2 +14 \cdot 2^2 = 8^2$.

As has been  argued in section~\ref{amalgz}, 
the topological interactions described 
by a discrete ${ H}$ gauge theory are encoded in the 
associated modular matrices $S$ and $T$. 
The modular $T$ matrix~(\ref{modutz})
contains the spin factors assigned to the different particles. 
With relation~(\ref{spin!}) and table~\ref{chartabd2}, 
we easily infer the following spin factors
for the particles 
in the spectrum~(\ref{spectred2}) of  
a $\bar{D}_2$ gauge theory 
\bea   \label{spind2}  \ba{cc} 
\mbox{particle}       & \qquad   \exp (2\pi \im  s)    \\
1,J_a             & \qquad    1   \\
\bar{1},\bar{J}_a & \qquad    1  \\
\chi , \bar{\chi} & \qquad    1, -1 \\
\sigma_a^{\pm}    & \qquad    \pm 1  \\
\tau_a^{\pm}      & \qquad    \pm \im.
\ea
\eea  
The modular $S$ matrix~(\ref{fusionz}), on the other hand, 
is determined  by the monodromy matrix following from~(\ref{braidact}).
A lengthy but straightforward calculation shows that  
the modular $S$ matrix for a $\bar{D}_2$ gauge theory 
takes the form displayed in table~\ref{modsd2}. We proceed by enumerating
the fusion rules following from plugging this modular $S$ matrix in
Verlinde's formula~(\ref{verlindez}).

\begin{table}[hbt]
\begin{center}
\begin{tabular}{crrrrrrrrrrr}      \hline \\[-4mm]
$S$& &$1 $& $\bar{1} $&$J_a $&$\bar{J}_a $&$\chi $&$\bar{\chi} 
$&$\sigma^+_a$ &$\sigma^-_a $&$\tau^+_a $&$\tau^-_a $\\ \hl
\\[-4mm]  
$1$& &$1 $&$1 $&$1 $&$1 $&$2 $&$2 $&$2 $&$2 $&$2 $&$2 $\\ 
$\bar{1}$& &$1 $&$1 $&$1 $&$1 $&$-2 $&$-2 $&$2 $&$2 $&$-2 $
&$-2 $\\ 
$J_b$& &$1 $&$1 $&$1 $&$1 $&$2 $&$2 $&$2\ep_{ab} $
&$2\ep_{ab} $&$2\ep_{ab} 
$&$2\ep_{ab} $\\ 
$\bar{J}_b$& &$1 $&$1 $&$1 $&$1 $
&$-2 $&$-2 $&$2\ep_{ab} $&$2\ep_{ab} $&$-2\ep_{ab} 
$&$-2\ep_{ab} $\\ 
$\chi$& &$2$&$-2$&$2$&$-2$&$4 $&$-4 $&$0 $&$0 $&$0 $&$0 $\\ 
$\bar{\chi}$& &$2$&$-2$&$2$&$-2$&$-4$&$4  $&$0 $&$0 $&$0 
$&$0 $\\ 
$\sigma^+_b$& &$2 $&$2 $&$2\ep_{ab} $&$2\ep_{ab} $&$0 $&$0  
$&$4\delta_{ab} $&$-4\delta_{ab} $&$0 $&$0 $\\ 
$\sigma^-_b$& &$2 $&$2 $&$2\ep_{ab} $&$2\ep_{ab} $&$0 $&$0  
$&$-4\delta_{ab}     $&$4\delta_{ab} $&$0 $&$0 $\\ 
$\tau^+_b$& &$2 $&$-2$&$2\ep_{ab} $&$-2\ep_{ab} $&$0 $&$0 $&$0 
$&$0 $&$-4\delta_{ab} $&$4\delta_{ab} $\\ 
$\tau^-_b$& &$2 $&$-2$&$2\ep_{ab} $&$-2\ep_{ab} $&$0 $&$0 $&$0 
$&$0 $&$4\delta_{ab} $&$-4\delta_{ab} $\\[1mm]   \hline
\end{tabular}
\end{center}
\caption{\sl Modular S-matrix of the quantum double 
$D(\bar{D}_2)$ up to an overall factor
$\frac{1}{8}$. We defined $\epsilon_{ab}:= 1$ iff $a=b$ 
and $\epsilon_{ab}:=-1$ iff $a\neq b$.}
\label{modsd2}
\end{table}

The fusion rules for the purely electric charges are, of course, 
dictated by the representation ring of $\bar{D}_2$
\bea                \label{puch}
J_a \times J_a=1, \qquad J_a \times J_b = J_c, \qquad J_a \times \chi=\chi,
\qquad \chi \times \chi = 1+ \sum_a J_a.
\eea
Here, the subscripts $a,b$ and $c$ take the values $1,2$ or $3$
and where by convention $a \neq b$, $a \neq c$ and $b \neq c$.
The latter convention will be used throughout the following.
To continue, the dyons associated with the flux $\bar{1}$ 
are obtained by simply
composing this flux with the purely electric charges
\bea                           \label{zendy}
J_a \times \bar{1}= \bar{J}_a, \qquad \chi \times \bar{1} = \bar{\chi}.
\eea 
In a similar fashion, we construct the other dyons 
\bea                                \label{ooknog}
J_a \times \sigma^+_a = \sigma^+_a, \qquad 
J_b \times \sigma^+_a= \sigma^-_a, \qquad 
\chi \times \sigma^+_a = \tau^+_a + \tau^-_a.
\eea
We now have produced all the constituents of the spectrum
as given in~(\ref{spectred2}). 
Recall from section~\ref{amalgz} that the fusion algebra is commutative and 
associative. This implies that the full set of fusion rules is, in fact,
completely determined by a minimal subset. 
Bearing this in mind, amalgamation involving the 
flux  $\bar{1}$ is unambiguously prescribed by~(\ref{zendy}) and
\bea                          \label{centflufu}
\bar{1} \times \bar{1} = 1, \qquad
\bar{1} \times \sigma^{\pm}_a = \sigma^{\pm}_a, \qquad 
\bar{1} \times \tau^{\pm}_a = \tau^{\mp}_a. 
\eea
The complete set of fusion rules is then fixed by the previous ones 
together with
\bea
J_a \times \tau^{\pm}_a  = \tau^{\pm}_a, \qquad
J_b \times \tau^{\pm}_a  = \tau^{\mp}_a, \qquad
\chi \times \tau^{\pm}_a = \sigma^+_a + \sigma^-_a,
\eea
and 
\bea
\sigma^{s}_a \times \sigma^{s}_a &=&1+J_a+\bar{1}+\bar{J}_a 
\label{remark} \\
\sigma^{s}_a \times \sigma^{s}_b &=& \sigma^+_c+\sigma^-_c 
\label{p3hop} \\
\sigma^{s}_a \times \tau^{s}_a &=& \chi+\bar{\chi} 
\label{p3hop1}   \\
\sigma^{s}_a \times \tau^{s}_b  &=& \tau^+_c + \tau^-_c \\
\tau^{s}_a \times \tau^{s}_a &=& 1+J_a+\bar{J}_b+\bar{J}_c \label{spreek}\\
\tau^{s}_a \times \tau^{s}_b &=& \sigma^+_c+\sigma^-_c \, ,
\eea
with $s \in +,-$.

A few remarks concerning this fusion algebra 
are pertinent. First of all,
it is easily verified that the class algebra of $\bar{D}_2$ is 
respected as an overall selection rule. 
The class multiplication in the fusion rule~(\ref{remark}), for instance,
reads $X_a * X_a = 2e+2\bar{e}$. 
The appearance of the class algebra
naturally expresses magnetic flux conservation:
in establishing the fusion rule,  all fluxes in the consecutive 
conjugacy classes are multiplied out.
Further, the modular $S$ matrix as given in table~\ref{modsd2}
is real and therefore equal to its inverse as follows directly from
relation~(\ref{sun}).
Consequently, the charge conjugation operator ${\cal C}$ is trivial,
i.e.\ it acts on the spectrum~(\ref{spectred2}) as the unit 
matrix ${\cal C}=S^2={\bf 1}$.
Hence, all particles in this $\bar{D}_2$ gauge theory feature as 
their own anti-particle. Only two similar particles are able to annihilate, 
as witnessed  by the occurrence of the vacuum channel $1$ 
in the fusion rule for two similar particles.

At first sight, the message of the fusion rule~(\ref{remark}) is actually
rather remarkable. It seems that the fusion of two pure fluxes $\sigma_a^+$ 
may give rise to electric charge creation. 
One could start wondering about electric charge conservation at this point. 
Electric charge is conserved though. Before fusion 
this charge was present in the form of so-called nonlocalizable Cheshire 
charge~\cite{alice,spm,sm,preskra}, i.e.\ the nontrivial representation 
of the global symmetry group $\bar{D}_2$ carried by the flux pair.
This becomes clear upon writing the fusion rule~(\ref{remark})
in terms of the two particle flux states corresponding to
the different channels:
\bea
\frac{1}{\sqrt{2}}
\{ |\bar{X}_a \rangle |X_a \rangle 
\; + \; |X_a \rangle  |\bar{X}_a \rangle \}
&\longmapsto& 1 \label{corvac}   \\
\frac{1}{\sqrt{2}}
\{ |\bar{X}_a \rangle |X_a \rangle
\; - \; |X_a \rangle  |\bar{X}_a \rangle \}
&\longmapsto& J_a  \label{corstat} \\
\frac{1}{\sqrt{2}}
\{ |X_a \rangle |X_a \rangle
\; + \; |\bar{X}_a \rangle  |\bar{X}_a \rangle \}
&\longmapsto& \bar{1} \label{corfluch} \\
\frac{1}{\sqrt{2}}
\{ |X_a \rangle |X_a \rangle
\; - \; |\bar{X}_a \rangle  |\bar{X}_a \rangle \}
&\longmapsto& \bar{J}_a.
\eea
The identification of the two particle flux states at the l.h.s.\
with the single
particle states at the r.h.s.\ is established by  the action~(\ref{coalgeb}) 
of the quantum double $D(\bar{D}_2)$ on the two particle states. 
On the one hand, we can perform global $\bar{D}_2$
symmetry transformations from which we learn the total  
charge carried by the flux pair.
As indicated by the 
comultiplication~(\ref{coalgeb}), the global $\bar{D}_2$ transformations 
act through overall conjugation. 
The total flux of the pair, on the other hand, is formally obtained
by applying the flux projection operators~(\ref{multi}) and is nothing 
but the product of the two fluxes of the pair. 
Since $X_a \cdot \bar{X}_a = e$, 
the total flux of the two particle state
in~(\ref{corvac}), for instance, vanishes. Moreover, it is 
easily verified that this state is invariant under global 
under global  $\bar{D}_2$ transformations. Thus it corresponds to the vacuum 
channel $1$. In a similar fashion, we obtain the identification of
the other two particle states with the single particle states. 
Note that  these two particle 
quantum states describing the flux pairs are nonseparable.
The two fluxes are correlated: by measuring the flux of 
one particle of the pair, we 
instantaneously fix the flux of the other.  This is 
the famous  Einstein-Podolsky-Rosen (EPR) paradox~\cite{epr}.
Just as in the notorious
experiment with two spin $1/2$ particles in the singlet state,
it is no longer possible to make a flux measurement on one particle 
without affecting the other instantaneously.
The Cheshire charge carried by the flux pair depends on the symmetry 
properties of these nonseparable quantum states.
The symmetric quantum states correspond to the trivial charge $1$, whereas 
the anti-symmetric quantum states carry the nontrivial charge $J_a$.
It is clear that the charge  $J_a$ 
can not be localized on any of the fluxes  
nor anywhere else. It is a property of the pair 
and only becomes localized when the fluxes are 
brought together in a fusion process.  It is this elusive 
nature, reminiscent of the 
smile of the Cheshire cat in Alice's adventures in 
wonderland~\cite{carroll}, 
that formed  the motivation to call such a charge Cheshire 
charge~\cite{colem, preskra}.

\begin{figure}[h] 
\epsfysize=8cm
\centerline{\epsffile{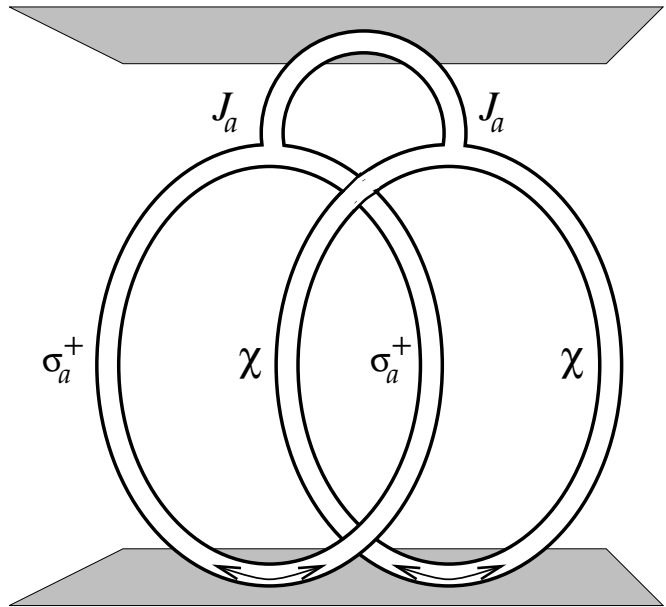}}
\caption{\sl  A  charge/anti-charge pair $\chi$ and a 
flux/anti-flux pair $\sigma_a^+$ are created from the vacuum
at a certain time slice. 
The ribbons represent the worldlines of the particles.
After the charge $\chi$  has encircled the flux $\sigma_a^+$, 
both particle/anti-particle pairs carry the nonlocalizable
Cheshire charge $J_a$. These Cheshire charges become  localized 
upon rejoining the members
of the pairs. Subsequently, the two charges 
$J_a$ annihilate each other.}
\label{ccatd2} 
\end{figure}

The Cheshire charge $J_a$ of the flux pair can be excited 
by encircling one flux in the pair by the 
doublet charge $\chi$~\cite{colem,sm,preskra}.
Here, we draw on a further analogy with 
Alice's adventures.
The magnetic fluxes 
$X_a$ and $\bar{X}_a$ act
by means of $\im \sigma_a$ and $-\im \sigma_a$ respectively on  
the doublet charge $\chi$, where the symbol $ \sigma_a$ denotes 
the Pauli matrices.
This means that  when a charge $\chi$ with its orientation down 
is adiabatically transported around,  for example, the flux $X_2$, 
it returns with its orientation up:
\bea                              \label{vlaflip}
{\cal R}^2 \; |X_2 \rangle | \left( \ba{c} 0 \\ 1  \ea \right) \rangle
&=&
|X_2 \rangle | \left( \ba{c} 1 \\ 0  \ea \right) \rangle ,
\eea
as follows from~(\ref{braidact}). In terms of Alice's adventures:
the charge has gone through the looking-glass. 
For this reason the flux $X_2$ has been called 
an Alice flux~\cite{alice,preskra,schwarz}.  
The other fluxes $X_a,\bar{X}_a$ affect the doublet charge $\chi$ 
in a similar way. Let us
now consider the process depicted in 
figure~\ref{ccatd2}.   We start with the creation of a 
charge/anti-charge pair $\chi$ and a flux/anti-flux pair $\sigma_a^+$ 
from the vacuum. Thus both pairs do not carry 
Cheshire charge at this stage. They are in the vacuum channel of the 
corresponding fusion rules~(\ref{puch}) and~(\ref{remark}).
Next, one member of the charge pair encircles a flux in the flux pair.
The flip of the charge orientation~(\ref{vlaflip})  
leads to an exchange of the internal  quantum numbers of the pairs: 
both pairs carry Cheshire charge $J_a$  after this process, i.e.\
both pairs are in the $J_a$ channel of the associated fusion rules.
The global charge of the configuration is conserved. Both charges   
$J_a$  can be annihilated  by bringing them together
as follows from the fusion rule $J_a \times J_a = 1$ given in~(\ref{puch}).
These phenomena can be made explicit  by  writing this process 
in terms of the corresponding correlated  internal quantum states: 
\bea   \label{uitzd2}
1 & \longmapsto & \frac{1}{2} 
\{ |\bar{X}_2 \rangle |X_2 \rangle 
\; + \; |X_2 \rangle  |\bar{X}_2 \rangle \} 
\{ | \left( \ba{c} 1 \\ 0  \ea \right) \rangle
| \left( \ba{c} 0 \\ 1  \ea \right) \rangle  \; - \;
| \left( \ba{c} 0 \\ 1  \ea \right) \rangle
| \left( \ba{c} 1 \\ 0  \ea \right) \rangle  \}  \nn \\
  & \stackrel{{\bf 1} \ot {\cal R}^2 \ot {\bf 1}}{\longmapsto} &
\frac{1}{2}      
\{ |\bar{X}_2 \rangle |X_2 \rangle
\; - \; |X_2 \rangle  |\bar{X}_2 \rangle \} 
\{ | \left( \ba{c} 0 \\ 1  \ea \right) \rangle
| \left( \ba{c} 0 \\ 1  \ea \right) \rangle  \; + \;
| \left( \ba{c} 1 \\ 0  \ea \right) \rangle
| \left( \ba{c} 1 \\ 0  \ea \right) \rangle  \}  \nn \\
& \longmapsto & |J_2 \rangle  |J_2 \rangle \nn \\
  & \longmapsto & 1.            
\eea 
Here, we used~(\ref{braidact}) and the fact that the fluxes 
$X_2$ and $\bar{X}_2$  act by means of the matrices
$\im \sigma_2$ and $-\im \sigma_2$ (with $\sigma_2$ the second 
Pauli matrix) 
respectively on the charge $\chi$.
After the charge has encircled the flux, the flux pair is in the
anti-symmetric quantum state~(\ref{corstat}) with Cheshire charge $J_2$,
while the same observation holds for the  quantum state of  
the charge pair. Before fusion the charge pair was in the anti-symmetric
vacuum representation $1$, while the state that emerges after the monodromy 
carries the Cheshire charge $J_2$.
For convenience, we restricted ourselves to  the flux pair $\sigma_2^+$ here.
The argument for the other flux pairs is completely similar though.

The foregoing  discussion naturally extends to the exchange of
magnetic quantum numbers in monodromy processes involving 
noncommuting fluxes, that is, the occurence 
of flux metamorphosis~(\ref{flumonodr}). 
If we replace the doublet charge $\chi$ pair in figure~\ref{ccatd2}
by a flux pair $\sigma_b^+$ (with $a \neq b$) starting off in the 
vacuum channel~(\ref{corvac}), both  flux pairs end up in the 
nontrivial flux channel~(\ref{corfluch}) after the monodromy
and both pairs now carry the total flux $\bar{1}$. 
Upon fusing the members of the pairs,
these `Cheshire' fluxes 
become localized and subsequently 
annihilate each other according to their fusion rule
given in~(\ref{centflufu}).

Let us close this section by briefly summarizing the profound role 
the fusion rules play as overall selection rules in the flux/charge 
exchange processes among the particles. It is natural
to confine our considerations to multi-particle systems which are
overall in the vacuum sector denoted by $1$, 
i.e.\ multi-particle systems for which the total flux and charge 
vanishes.
Hence, in the $\bar{D}_2$ gauge theory under consideration the 
particles necessarily appear in pairs of the same species,
as we have seen explicitly in the example of 
such a system in figure~\ref{ccatd2}.
The fusion rules then classify the different 
total fluxes and Cheshire charges these
pairs can carry and determine the flux/charge exchanges that may occur
in monodromy processes involving particles in different pairs.

\sectiona{Scattering doublet charges off Alice fluxes}   \label{abd2b}

The Aharonov-Bohm interactions among the particles in the 
spectrum of a nonabelian discrete $H$ gauge theory 
roughly fall into two classes. 
First of all, there are the interactions in which no internal 
flux/charge quantum numbers are exchanged between the particles. 
In other words, we are dealing with two particles for which
the monodromy  matrix following by taking the square of the 
braid matrix~(\ref{braidact})
is diagonal in the two particle flux/charge eigenbasis
with possibly different 
Aharonov-Bohm phases as diagonal elements.  The cross sections 
measured in Aharonov-Bohm scattering experiments with two 
such particles  simply follow from 
the well-known cross section
derived by Aharonov and Bohm~\cite{ahabo}. 
See relation~(\ref{ababcs}) of appendix~\ref{ahboverl}. 
The  more interesting
Aharonov-Bohm interactions are those in which internal 
flux/charge quantum numbers are exchanged between two particles when these 
encircle each other.
In that case, we are dealing with two particles for which 
the monodromy matrix is off diagonal in the
two particle flux/charge  eigenbasis.
The cross sections~\cite{spm,preslo,ver1} measured in 
Aharonov-Bohm scattering experiments
involving two such particles are briefly reviewed in appendix~\ref{ahboverl}. 
In this section, we will focus on a nontrivial example
in the $\bar{D}_2$ gauge theory at hand, namely 
an Aharonov-Bohm scattering experiment in which a doublet charge $\chi$ 
scatters from an Alice flux $\sigma_2^+$.

The total 
internal Hilbert space associated with the two particle system consisting 
of a pure doublet charge $\chi$ together with
a pure doublet flux $\sigma_2^+$ is four dimensional. 
We define the following natural flux/charge eigenbasis  in this internal 
Hilbert space
\bea        \label{cfbasis}     \ba{rclrcl}
e_1 &=& 
|X_2 \rangle| \left( \ba{c} 1 \\ 0  \ea \right) \, \rangle
\; := \; |\uparrow \, \rangle| \uparrow \, \rangle
             \\
e_2 
&=& |X_2  \rangle | \left( \ba{c} 0 \\ 1  \ea \right) \, \rangle 
\; := \; |\uparrow \, \rangle |\downarrow  \, \rangle        \\
e_3 
&=& |\bar{X}_2 \rangle| \left( \ba{c} 1 \\ 0  \ea \right) \, \rangle 
\; := \; |\downarrow \, \rangle | \uparrow \, \rangle       \\
e_4 
&=& |\bar{X}_2 \rangle| \left( \ba{c} 0 \\ 1  \ea \right) \, \rangle 
\; := \; |\downarrow \, \rangle |\downarrow \, \rangle \, .
\ea    
\eea
As has been mentioned before, 
the fluxes ${X}_2$ and $\bar{X}_2$ respectively 
are represented by the matrices 
$\im\sigma_2$ and $-\im\sigma_2$ (with $\sigma_2$ the second Pauli matrix)
in the doublet charge representation $\chi$.
From~(\ref{braidact}), we then infer that the monodromy matrix 
takes the following block diagonal form
in this basis
\bea                                   \label{allesd2}
{\cal R}^2 &=&  \left( \ba{rrrr} 0 & 1 & 0 & 0 \\ -1 & 0 &  0 & 0   \\
                                 0 & 0 & 0 & -1 \\ 
                                 0 & 0 & 1 & 0 \ea \right),
\eea  
which reflects the phenomenon discussed in the previous section:
the orientation of the charge $\chi$ is flipped, when
it is transported either around the Alice flux ${X}_2$ or 
around $\bar{X}_2$.

Let us now consider the Aharonov-Bohm scattering experiment in 
which the doublet charge 
$\chi$ scatters from the Alice flux $\sigma_2^+$.
We  assume that we are measuring with  a detector that only gives a signal 
when a scattered charge  $\chi$ enters the device with a specific orientation 
(either $\uparrow$ or $\downarrow$). Here we may, for instance,
think of an apparatus in which we have captured the associated 
anti-particle. This is the charge with opposite orientation, as we 
have seen in~(\ref{uitzd2}).
If the orientation of the scattered charge entering the device matches
that of the anti-particle, the pair annihilates 
and we assume that the apparatus somehow gives a signal when such an
annihilation process occurs.
The cross section measured with such a detector 
involves the matrix elements
of  the scattering matrix 
\bea                     \label{scatmd2}
{\cal R}^{-\theta/\pi}({\bf 1} - {\cal R}^{2}) 
&=&     
\sqrt{2} \, e^{-\im \theta/2} 
\left( \ba{rrrr} \cos{\frac{\pi - \theta}{4}} & \sin{\frac{\pi - \theta}{4}} 
&  0 & 0 \\ 
                -\sin{\frac{\pi - \theta}{4}}  & \cos{\frac{\pi - \theta}{4}} 
&  0 & 0 \\
   0 & 0 & \cos{\frac{\pi - \theta}{4}} & -\sin{\frac{\pi - \theta}{4}} \\
   0 & 0  & \sin{\frac{\pi - \theta}{4}} & \cos{\frac{\pi - \theta}{4}}
        \ea \right), \qquad \qquad
\eea
for the flux/charge eigenstates~(\ref{cfbasis}). 
This scattering matrix is determined using the prescription~(\ref{exclur})
in the monodromy eigenbasis in which the above monodromy 
matrix~(\ref{allesd2})  is diagonal,
and subsequently transforming back to the flux/charge 
eigenbasis~(\ref{cfbasis}).
Now suppose that the scatterer is 
in a particular flux eigenstate, while 
the projectile that  comes in is a charge with a specific orientation and
the detector is only sensitive for scattered charges with this 
specific orientation.                              
Under these circumstances, the two particle in and out state are the same, 
$|\, \mbox{in}\rangle = |\, \mbox{out}\rangle $, and equal to one 
of the flux/charge eigenstates in~(\ref{cfbasis}). 
In other words, we are measuring 
the scattering amplitudes on the diagonal of the 
scattering matrix~(\ref{scatmd2}). 
Note that the formal sum of the out state $|\mbox{out} \rangle$ 
over a complete basis of flux eigenstates for the scatterer, 
as indicated in appendix~\ref{ahboverl},
boils down to one term here, 
namely the flux eigenstate of the scatterer in the 
in state $|\mbox{in} \rangle$. The other flux eigenstate
does {\em not} contribute. The corresponding matrix element vanishes,
because the flux of the scatterer is not
affected when it is encircled by  the charge $\chi$.
From equation~(\ref{exclus}) of appendix~\ref{trubra}, 
we finally obtain the following  
exclusive cross section for this scattering experiment
\bea              \label{c+d2}
\frac{{\rm d} \sigma_+}{{\rm d} \theta}   &=&
\frac{1+\sin{(\theta/2)}}{8 \pi p \sin^2{(\theta/2)}}   \; .
\eea
The charge flip cross section, in turn, is measured by a detector
which only signals scattered charges with an orientation opposite 
to the orientation of the charge of the projectile.
In that case, the state $|\mbox{in} \rangle$ is again
one of the flux/charge eigenstates in~(\ref{cfbasis}), while
the $|\mbox{out} \rangle$ state we measure is the same as the 
in state, but with the orientation of the charge flipped. 
Thus we are now measuring the off diagonal matrix elements of the scattering
matrix~(\ref{scatmd2}). 
In a similar fashion as before, we find the following form for the 
charge flip cross section
\bea              \label{c-d2}
\frac{{\rm d} \sigma_-}{{\rm d} \theta} &=& 
\frac{1-\sin{(\theta/2)}}{8 \pi p \sin^2{(\theta/2)}}  \;.
\eea
The exclusive cross sections~(\ref{c+d2}) and~(\ref{c-d2}), 
which are the same as 
derived for scattering of electric charges from Alice fluxes in Alice 
electrodynamics by Lo and Preskill~\cite{preslo}, are clearly multi-valued
\bea 
\frac{{\rm d} \sigma_{\pm}}{{\rm d} \theta} \; (\theta + 2\pi) &=& 
\frac{{\rm d} \sigma_{\mp}}{{\rm d} \theta} \;  (\theta).
\eea
This merely reflects the fact that a detector only signalling  
charges $\chi$ with their orientation up, becomes a detector only
signalling charges with orientation down (and vice versa), when
it is transported in a counterclockwise way
over an angle $2\pi$ around the scatterer.
Specifically, in this parallel transport the anti-particle in our 
detector feels the holonomy in the gauge fields associated with
the flux of the scatterer and returns with its orientation flipped.
As a consequence, the device becomes sensitive for the opposite
charge orientation after this parallel transport.

Verlinde's detector does not suffer from this multi-valuedness.
It does not discriminate  between the orientations 
of the scattered charge, and gives a signal whenever a charge $\chi$ 
enters the device. This detector measures the total or 
inclusive cross section, i.e.\ both branches of the multi-valued 
cross section~(\ref{c+d2}) (or~(\ref{c-d2}) for that matter). 
To be specific, the exclusive cross sections~(\ref{c+d2})
and~(\ref{c-d2})
combine as  follows 
\bea
\frac{{\rm d} \sigma}{{\rm d}\theta} \; = \;
\frac{{\rm d}\sigma_-}{{\rm d}\theta}  + 
\frac{{\rm d}\sigma_+}{{\rm d}\theta} \; = \; 
 \frac{1}{4 \pi p \sin^2{(\theta/2)}} \; .
\eea
into Verlinde's single valued inclusive cross section~(\ref{Aharonov}) 
for this scattering experiment.

To conclude, the above analysis is easily extended to Aharonov-Bohm scattering
experiments involving other particles in the spectrum~(\ref{spectred2}) 
of this $\bar{D}_2$ gauge theory. It should be stressed, however, that 
a crucial ingredient in the derivation of the {\em multi-valued} exclusive
cross sections~(\ref{c+d2}) and~(\ref{c-d2}) 
is that the monodromy matrix~(\ref{allesd2}) is off diagonal
and has imaginary eigenvalues  $\pm \im$. In the other cases, where
the monodromy matrices are diagonal or off diagonal 
with eigenvalues $\pm 1$, as it for instance 
appears for scattering two noncommuting
fluxes $\sigma_a^+$ and $\sigma_b^+$ from each other, 
we arrive at {\em single valued} exclusive cross sections.

\sectiona{Nonabelian braid statistics}  \label{d2bqst}

We finally turn to the issue of nonabelian braid statistics. 
As we have argued 
in section~\ref{trunckbr}, the braidings and monodromies 
for multi-particle configurations appearing in discrete
${ H}$ gauge theories are governed by truncated rather then ordinary
braid groups. 
To be precise, the total internal Hilbert space for a given multi-particle
system carries a representation of some truncated braid group,
which in general decomposes into a direct sum of irreducible representations.
In this section, we identify the truncated braid groups ruling in this 
particular $\bar{D}_2$ gauge theory and elaborate on the aforementioned
decomposition. We first consider the indistinguishable particle configurations 
in this model.

It can easily be verified that the  braid operators  acting on a 
configuration which  only  consists of the pure singlet charges 
$J_a$ (with $a \in 1,2,3$)
are of order one. 
The same holds for the singlet dyons $\bar{1}$ and $\bar{J}_a$.
In other words, 
these particles behave as ordinary bosons, in accordance with the trivial
spin factors~(\ref{spind2}) assigned to them.
To proceed, the braid operators acting on a system of $n$
doublet charges $\chi$ are of order two and therefore realize 
a (higher dimensional) representation of the permutation group $S_n$. 
The same observation appears for the doublet dyons $\bar{\chi}$ 
and $\sigma^{\pm}_a$. The total internal Hilbert spaces for these 
indistinguishable particle systems can then be decomposed into a direct sum 
of  subspaces, each carrying an irreducible representation
of the permutation group $S_n$. The one dimensional representations that  
appear in this decomposition naturally correspond either 
to Bose or Fermi statistics, while
the higher dimensional representations describe parastatistics.
Finally, braid statistics occurs for a system consisting of $n$ 
dyons $\tau_a^{\pm}$. The braid operators that act on such a system are 
of order four. Hence, the associated internal Hilbert space splits up 
into a direct sum of 
irreducible representations  of the truncated braid group $B(n,4)$. 
The one dimensional representations
that occur in this decomposition realize abelian braid statistics
(anyons), 
whereas the higher dimensional representations  correspond to nonabelian 
braid statistics (nonabelian anyons). 
We  will illustrate these features with two 
representative examples. We first examine a system 
containing two dyons $\tau_1^{+}$. The irreducible 
braid group representations available for this system are one dimensional,
since the truncated braid group  $B(2,4)$ for two particles is abelian. 
We then turn to the more interesting system consisting of 
three dyons $\tau_1^{+}$. In that case, we are dealing with nonabelian braid
statistics. The associated total internal Hilbert space
breaks up into four 1-dimensional irreducible subspaces and two 
2-dimensional irreducible subspaces 
under the action of the nonabelian truncated braid group
$B(3,4)$.

We start by setting some conventions. First of all, the 
two fluxes in the conjugacy class associated with the 
dyon $\tau_1^{+}$ are ordered as indicated in table~\ref{tabcond2}, i.e.\
\beas
^1 h_1 &=& X_1   \\
^1 h_2 &=& \bar{X}_1,
\eeas 
while we take the following coset representatives appearing 
in the definition~(\ref{13zo}) of the centralizer charge
\beas
^1 x_1 &=& e   \\
^1 x_2 &=& X_2.
\eeas
To lighten the  notation a bit,  we furthermore 
use the following  abbreviation for  the internal flux/charge 
eigenstates of the dyon $\tau_1^{+}$
\beas
|\uparrow \,\rangle &:=& |X_1, ^{1} v\rangle \\
|\downarrow \,\rangle &:=& |\bar{X}_1, ^{1} v\rangle.
\eeas 

Let us now consider a system consisting of two dyons $\tau^{+}_1$.
Under the action of the quantum  double  $D(\bar{D}_2)$,
the internal Hilbert space $V_{\tau^{+}_1} \ot V_{\tau^{+}_1}$ 
associated with this system decomposes according to the fusion 
rule~(\ref{spreek}), which we repeat for convenience
\bea                     \label{repeat}
\tau^{+}_1 \times \tau^{+}_1 &=& 1+J_1+\bar{J}_2+\bar{J}_3.
\eea                                   
The two particle states corresponding to the different fusion channels
carry an one dimensional (irreducible) 
representation of the abelian truncated braid group
$B(2,4)=\Z_4$. We first establish the different irreducible pieces
contained in the $B(2,4)$ representation carried by the total internal 
Hilbert space $V_{\tau^{+}_1} \ot V_{\tau^{+}_1}$.
This can be done by calculating the traces of the elements 
$\{e, \tau, \tau^2, \tau^3 \}$ of $B(2,4)$ in this representation
using the standard diagrammatic techniques (see, for 
instance, the references~\cite{adams,kaufman}).
From the character vector obtained in this way, we learn that this 
representation breaks up as
\bea                    \label{brad24}
\Lambda_{B(2,4)} &=& 3\;  \Gamma^1 +   \Gamma^3,
\eea 
with $\Gamma^1$ and $\Gamma^3$ the irreducible $\Z_4$ representations 
displayed in the character table~\ref{tabcond2}, i.e.\
$\Gamma^1(\tau) := \im$ and $\Gamma^3(\tau) := -\im$. 
After some algebra,  we then arrive at the following  basis of 
mutual eigenstates under the combined action of the quantum double 
and the truncated braid group 
\begin{eqnfourarray} 
 V_{\tau^{+}_1} \ot V_{\tau^{+}_1}  \qquad 
& \qquad &  D(\bar{D}_2)   & $\qquad B(2,4) $   \nn \\
\frac{1}{\sqrt{2}} \{ 
|\uparrow \, \rangle |\downarrow \, \rangle-|\downarrow \, \rangle
|\uparrow \, \rangle \} 
&\qquad & \;\; 1 & $\qquad  \Gamma^{1} \;\;\; \,$  \label{vac11} \\
\frac{1}{\sqrt{2}} \{ 
|\uparrow \, \rangle |\downarrow \, \rangle + |\downarrow \, \rangle
|\uparrow \,  \rangle \} 
&\qquad & \;\;  J_1& $\qquad  \Gamma^{3} \;\;\;\,$  \label{j13}   \\
\frac{1}{\sqrt{2}} \{ 
|\uparrow \, \rangle |\uparrow \, \rangle+|\downarrow \, \rangle
|\downarrow \, \rangle \} 
&\qquad & \;\;  \bar{J}_2 & $\qquad   \Gamma^{1} \;\;\;\, $  \label{j21}\\
\frac{1}{\sqrt{2}} \{ 
|\uparrow \, \rangle |\uparrow \, \rangle- |\downarrow \,\rangle
|\downarrow \, \rangle  \} 
&\qquad & \;\; \bar{J}_3 & $\qquad   \Gamma^{1} , \;\;\; $   \label{j31}
\end{eqnfourarray}   
from which we conclude that 
the two particle internal Hilbert space 
$V_{\tau^{+}_1} \ot V_{\tau^{+}_1}$ decomposes
into the following  direct sum of one dimensional 
irreducible representations of the direct product  
$D(\bar{D}_2) \times B(2,4)$
\bea
(1,\Gamma^{1}) \: + \: (J_1,\Gamma^{3}) \: + \:
(\bar{J}_2,\Gamma^{1}) \: + \: (\bar{J}_3, \Gamma^{1}).
\eea
In accordance with the general discussion concerning
relation~(\ref{braidactid}), the
 two particle states contained in~(\ref{j21}) 
and~(\ref{j31}), which are given by a linear combination of two
states in which both particles are in the same internal quantum state,  
satisfy the canonical spin-statistics connection~(\ref{spist}). That is,   
$\exp (\im \Theta) = \exp(2\pi \im s_{\tau^{+}_1})= \im$.
Accidentally, the same observation appears for the two particle 
state~(\ref{vac11}). 
Finally, the two particle state displayed in~(\ref{j13})
satisfies the generalized spin-statistics connection~(\ref{gespietst}) and 
describes semion statistics with quantum statistical parameter
$\exp (\im \Theta)=-\im$.

We now extend our analysis to a system containing 
three dyons $\tau^{+}_1$. From the fusion rules~(\ref{repeat})  
and~(\ref{zendy})--(\ref{centflufu}), we infer that the decomposition of 
the total internal Hilbert space under the action of the quantum double 
becomes
\bea                  \label{dd2bardeco}
\tau^{+}_1 \times \tau^{+}_1 \times \tau^{+}_1  &=& 4 \; \tau^{+}_1.
\eea
The occurrence of four equivalent fusion channels indicates that 
nonabelian braid statistics is conceivable and it turns out that  
higher dimensional irreducible representations of the 
truncated braid group $B(3,4)$ indeed appear. The
structure of this group and its irreducible representations 
are discussed in appendix~\ref{trubra}.  A lengthy but straightforward 
diagrammatic calculation of the character vector associated with
the $B(3,4)$ representation carried by the three particle internal 
Hilbert space $V_{\tau^{+}_1} \ot V_{\tau^{+}_1} \ot V_{\tau^{+}_1}$
reveals the following irreducible pieces
\bea    \label{b34dec}
\Lambda_{B(3,4)}  &=&  4 \; \Lambda_1 + 2 \; \Lambda_5,
\eea 
with  $\Lambda_1$ and $\Lambda_5$ the irreducible representations of 
$B(3,4)$ exhibited in the character table~\ref{tab:tab4}
of appendix~\ref{trubra}.
The one dimensional representation $\Lambda_1$ describes 
abelian semion statistics,
while the two dimensional representation $\Lambda_5$ corresponds to 
nonabelian braid statistics.
From~(\ref{dd2bardeco}) and~(\ref{b34dec}), we can immediately conclude that 
this three particle internal Hilbert space breaks up into  
the following direct sum of irreducible subspaces under the action of the  
direct product $D(\bar{D}_2) \times B(3,4)$  
\bea
2 \; (\tau^{+}_1, \Lambda_1) \: + \: ( \tau^{+}_1, \Lambda_5 ),
\eea 
where $(\tau^{+}_1, \Lambda_1)$ labels a two dimensional  
and $( \tau^{+}_1, \Lambda_5 )$ a four dimensional representation.
A basis adapted to this decomposition can be cast in the following form
\begin{eqnfourarray}
 V_{\tau^{+}_1} \ot V_{\tau^{+}_1} \ot V_{\tau^{+}_1} \qquad \qquad  \qquad
& \qquad &  D(\bar{D}_2)   & $\qquad B(3,4) $   \nn \\
|\downarrow  \, \rangle |\downarrow  \, \rangle |\downarrow  \, \rangle 
& \qquad &
\; |\uparrow \, \rangle_1 & $\Lambda_1 \;\;\;\, $  \\
|\uparrow  \, \rangle|\uparrow  \, \rangle|\uparrow  \, \rangle 
& \qquad & 
\; |\downarrow \, \rangle_1 & $\Lambda_1 \;\;\;\,$  \\   
\frac{1}{\sqrt{3}}\{ |\uparrow  \, \rangle|\uparrow  \, \rangle
                     |\downarrow  \, \rangle
                    -|\uparrow  \, \rangle |\downarrow  \, \rangle
                     |\uparrow  \, \rangle
                    +|\downarrow  \, \rangle |\uparrow  \, \rangle
                     |\uparrow  \, \rangle \} 
& \qquad & 
\; |\uparrow \, \rangle_2 & $\Lambda_1 \;\;\;\,$ \\
\frac{1}{\sqrt{3}}\{ |\downarrow  \, \rangle|\downarrow  \, \rangle 
                     |\uparrow  \, \rangle
                    -|\downarrow  \, \rangle|\uparrow  \, \rangle
                     |\downarrow  \, \rangle
                    +|\uparrow  \, \rangle|\downarrow  \, \rangle
                     |\downarrow  \, \rangle\} 
&\qquad& 
\; |\downarrow \, \rangle_2 & $\Lambda_1 \;\;\;\,$ \\
\frac{1}{2} \{2|\uparrow  \, \rangle|\uparrow  \, \rangle
               |\downarrow  \, \rangle
              +|\uparrow  \, \rangle |\downarrow  \, \rangle
               |\uparrow  \, \rangle
              -|\downarrow  \, \rangle |\uparrow  \, \rangle
               |\uparrow  \, \rangle \} 
& \qquad &
\; |\uparrow \, \rangle_3 & $\Lambda_5 \;\;\;\,$     \label{la35} \\
\frac{1}{2}\{2|\downarrow  \, \rangle|\downarrow  \, \rangle 
              |\uparrow  \, \rangle
             +|\downarrow  \, \rangle|\uparrow  \, \rangle
              |\downarrow  \, \rangle
             -|\uparrow  \, \rangle|\downarrow  \, \rangle
              |\downarrow  \, \rangle\} 
& \qquad &
\; |\downarrow \, \rangle_3 & $\Lambda_5 ' \;\;\;\,$ \label{la35p}  \\
\frac{1}{\sqrt{2}} \{|\uparrow  \, \rangle |\downarrow  \, \rangle
                     |\uparrow  \, \rangle
                    +|\downarrow  \, \rangle |\uparrow  \, \rangle
                     |\uparrow  \, \rangle \} 
& \qquad &
\; |\uparrow \, \rangle_4 & $\Lambda_5 \;\;\;\, $     \label{la45}\\
\frac{1}{\sqrt{2}}\{ |\downarrow  \, \rangle|\uparrow  \, \rangle
                     |\downarrow  \, \rangle
                    +|\uparrow  \, \rangle|\downarrow  \, \rangle
                     |\downarrow  \, \rangle\}
& \qquad &
\; |\downarrow \, \rangle_4 & $\Lambda_5 ' ,\;\;\; $     \label{la45p}
\end{eqnfourarray}   
The subscript attached to the single particle states in the second column 
label the four fusion channels showing up in~(\ref{dd2bardeco}). 
In other words, these states summarize the global properties of 
the three particle states in the first column, that is, the total flux 
and charge, which are conserved under braiding. 
Each of the three particle states in the first four rows  
carry the one dimensional representation $\Lambda_1$  
of the truncated braid group $B(3,4)$.
The particles in these states obey semion statistics
with quantum statistical parameter $\exp (\im \Theta)= \im$ and 
satisfy the canonical spin-statistics connection~(\ref{spist}).
Finally, the states in the last four rows constitute a basis for the 
representation  $( \tau^{+}_1, \Lambda_5 )$. To be specific,
the states~(\ref{la35}) and~(\ref{la45}),
carrying the same total flux and charge, form a basis for 
a two dimensional irreducible representation $\Lambda_5$ of the 
truncated braid group. The same remark holds for the states~(\ref{la35p}) 
and~(\ref{la45p}).
For convenience, we have distinguished these 
two irreducible representations by a prime. 
Note that we have chosen a basis which 
diagonalizes the braid operator ${\cal R}_1$  acting on the   
first two particles with eigenvalues either $\im$ or $-\im$, 
whereas the braid operator ${\cal R}_2$ for the last 
two particles mixes the states in the different fusion channels.
Of course, this choice is quite arbitrary.
By another basis choice, we could have reversed this situation.  

Let us also comment briefly on the distinguishable particle systems
that may occur. 
The maximal order of the monodromy operator for distinguishable particles
in this $\bar{D}_2$ gauge  theory is four. 
Hence, the distinguishable particle systems 
in this theory
are governed by the truncated colored braid 
groups $P(n,8)$ and their subgroups. 
A system consisting of the three different particles 
$\sigma_1^+$, $\sigma_2^+$ and $\tau_3^{+}$, for instance, realizes a 
representation of the colored braid group $P(3,4) \subset P(3,8)$.
(The group structure of $P(3,4)$ and a classification of its irreducible 
representations are given in appendix~\ref{trubra}). 
In a similar fashion
as before, it is easily inferred that this represention 
of $P(3,4)$ breaks up into 
the following irreducible pieces 
\bea \label{lomp}
\Lambda_{P(3,4)} &=& 2 \; \Omega_8 + 2 \; \Omega_9,
\eea
with $\Omega_8$ and $\Omega_9$ the two dimensional irreducible 
representations displayed in the character 
table~\ref{char} of appendix~\ref{trubra}. 
Thus, this system obeys nonabelian `monodromy statistics': 
the three monodromy operators displayed in~(\ref{p34mongen}) 
can not be diagonalized simultaneously.
Further, from the fusion rules~(\ref{p3hop}) and~(\ref{p3hop1}) in combination 
with~(\ref{puch})--(\ref{ooknog}), 
we obtain that the internal Hilbert space of this three particle 
particle system splits up into the following irreducible components 
under the action 
of the quantum double $D(\bar{D}_2)$
\bea \label{lomp1}
\sigma_1^+ \times \sigma_2^+ \times \tau_3^{+} &=& 
(\sigma_3^+ + \sigma_3^-) \times \tau_3^{+} \; = \; 
2 \; \chi + 2 \;\bar{\chi} \, . 
\eea
It is then readily checked that a basis adapted to the simultaneous  
decomposition of the 3-particle internal Hilbert space 
$V_{\sigma^{+}_1} \ot V_{\sigma^{+}_2} \ot V_{\tau^{+}_3}$ 
into UIR's of $D(\bar{D}_2)$ and  $P(3,4)$ reads:
\begin{eqnfourarray}
\qquad V_{\sigma^{+}_1} \ot V_{\sigma^{+}_2} \ot V_{\tau^{+}_3} 
\qquad  \qquad
& \qquad &  D(\bar{D}_2)   & $\qquad P(3,4) $   \nn \\
\frac{1}{\sqrt{2}}\{|X_1\rangle |\bar{X}_2 \rangle +
|\bar{X}_1 \rangle|X_2\rangle\} |\bar{X}_3, ^1 v \rangle 
& \qquad &
\; \; \; \chi & $\Omega_9 \;\;\;\, $   \label{beginne} \\
\frac{1}{\sqrt{2}}\{|X_1\rangle |X_2 \rangle +
|\bar{X}_1 \rangle|\bar{X}_2\rangle\} |{X}_3, ^1 v \rangle 
& \qquad & 
\; \; \; \chi & $\Omega_9 ' \;\;\;\,$  \label{begga2}  \\   
\frac{1}{\sqrt{2}}\{|X_1\rangle |\bar{X}_2 \rangle -
|\bar{X}_1 \rangle|X_2\rangle\} |\bar{X}_3, ^1 v \rangle 
& \qquad & 
\; \; \; \chi \, ' & $\Omega_9 ' \;\;\;\,$ \label{begga3}   \\
\frac{1}{\sqrt{2}}\{|X_1\rangle |X_2 \rangle -
|\bar{X}_1 \rangle|\bar{X}_2\rangle\} |{X}_3, ^1 v \rangle 
&\qquad& 
\; \; \; \chi \, ' & $\Omega_9 \;\;\;\,$   \label{begga4} \\
\frac{1}{\sqrt{2}}\{|X_1\rangle |\bar{X}_2 \rangle +
|\bar{X}_1 \rangle|X_2\rangle\} |{X}_3, ^1 v \rangle 
& \qquad &
\; \;  \; \bar{\chi} & $\Omega_8 \;\;\;\,$      \\
\frac{1}{\sqrt{2}}\{|X_1\rangle |X_2 \rangle +
|\bar{X}_1 \rangle|\bar{X}_2\rangle\} |\bar{X}_3, ^1 v \rangle 
& \qquad &
\; \; \; \bar{\chi} & $\Omega_8 ' \;\;\;\,$ \label{lap35p}  \\
\frac{1}{\sqrt{2}}\{|X_1\rangle |\bar{X}_2 \rangle -
|\bar{X}_1 \rangle|X_2\rangle\} |{X}_3, ^1 v \rangle 
& \qquad & 
\; \;\; \bar{\chi} \, ' & $\Omega_8 ' \;\;\;\,$ \\
\frac{1}{\sqrt{2}}\{|X_1\rangle |X_2 \rangle -
|\bar{X}_1 \rangle|\bar{X}_2\rangle\} |\bar{X}_3, ^1 v \rangle 
&\qquad&
\; \; \; \bar{\chi} \, ' & $\Omega_8 \, . \,\;\;$ \label{einde}
\end{eqnfourarray}   
As before, the isotypical fusion channels and the equivalent
UIR's of $P(3,4)$ are distinguished by a prime. So, the 
three particle states in~(\ref{beginne}) and~(\ref{begga2}), for example, 
form a basis for one of the two fusion channels $\chi$ 
in~(\ref{lomp1}), whereas the states 
in~(\ref{begga3}) and~(\ref{begga4}) constitute a basis for the other 
fusion channel $\chi$. 
The three particle states in~(\ref{beginne}) 
and~(\ref{begga4}) then span one of the two UIR's $\Omega_9$ of $P(3,4)$ 
in~(\ref{lomp}) and the states in~(\ref{begga2}) 
and~(\ref{begga3}) the other.
Finally, from~(\ref{beginne})--(\ref{einde}), we learn that 
the two fusion channels $\chi$ in~(\ref{lomp1}) combine with 
the two UIR's  $\Omega_9$ of $P(3,4)$
in~(\ref{lomp}) and the two fusion channels $\bar{\chi}$ with the 
two UIR's $\Omega_8$.
Hence, the three particle internal Hilbert space 
$V_{\sigma^{+}_1} \ot V_{\sigma^{+}_2} \ot V_{\tau^{+}_3}$ 
breaks up into the following two 4-dimensional irreducible representations 
\bea
(\chi, \Omega_9) \: + \: (\bar{\chi}, \Omega_8),
\eea  
of the direct product $D(\bar{D}_2) \times P(3,4)$.

As a last blow, we return to the process depicted in 
figure~(\ref{ccatd2}).
After the double pair creation, we are dealing with a four particle system
consisting of a subsystem of two indistinguishable particles $\sigma_2^+$ 
and a subsystem of two indistinguishable particles $\chi$. 
Recall from the sequence~(\ref{uitzd2}) that 
the two particle state for the fluxes $\sigma_2^+$ was initially bosonic,
whereas  the two particle state for the charges $\chi$ was fermionic.
After the monodromy has taken place, the situation is reversed. 
The two particle state for the fluxes $\sigma_2^+$ has become fermionic 
and the 
two particle state for the charges $\chi$ bosonic. In other words,
the exchange of Cheshire charge is accompanied by an exchange of quantum 
statistics, see also reference~\cite{brekke} in this connection. 
The total  four particle system now realizes a two dimensional
irreducible representation of the associated 
truncated partially colored braid group. 
The two braid operators ${\cal R}_1$ and ${\cal R}_3$ 
for the indistinguisbale particle exchanges in the two subsystems 
act diagonally with eigenvalues $\pm 1$ and $\mp 1$ respectively.
Furthermore, under the repeated action of the monodromy 
operator ${\cal R}_2^2$,
the subsystems simultaneously jump back and forth between the 
fusion channels $1$ and $J_2$ with their associated 
Cheshire charge and quantum statistics.

\aanhangsel

\sectiona{Aharonov-Bohm scattering}  \label{ahboverl}

The only experiments in which 
the particles in a discrete ${ H}$ gauge theory 
leave `long range fingerprints' 
are of a quantum mechanical nature, namely quantum interference experiments, 
such as the double slit experiment~\cite{colem,preslo}
and the Aharonov-Bohm scattering experiment~\cite{ahabo}.
What we are measuring in these experiments is the way the 
particles affect their mutual internal flux/charge quantum numbers when 
they encircle each other. In other words, we are probing the content 
of the monodromy matrix ${\cal R}^2$ following from~(\ref{braidact}).   
In this appendix, we will give a concise discussion of 
two particle Aharonov-Bohm scattering and provide the details entering
the calculation of the cross sections in section~\ref{abd2b}.
For a recent review of the experimental status of the Aharonov-Bohm effect,
the reader is referred to~\cite{peshkin}.

\begin{figure}[tbh]    \epsfxsize=14cm
\centerline{\epsffile{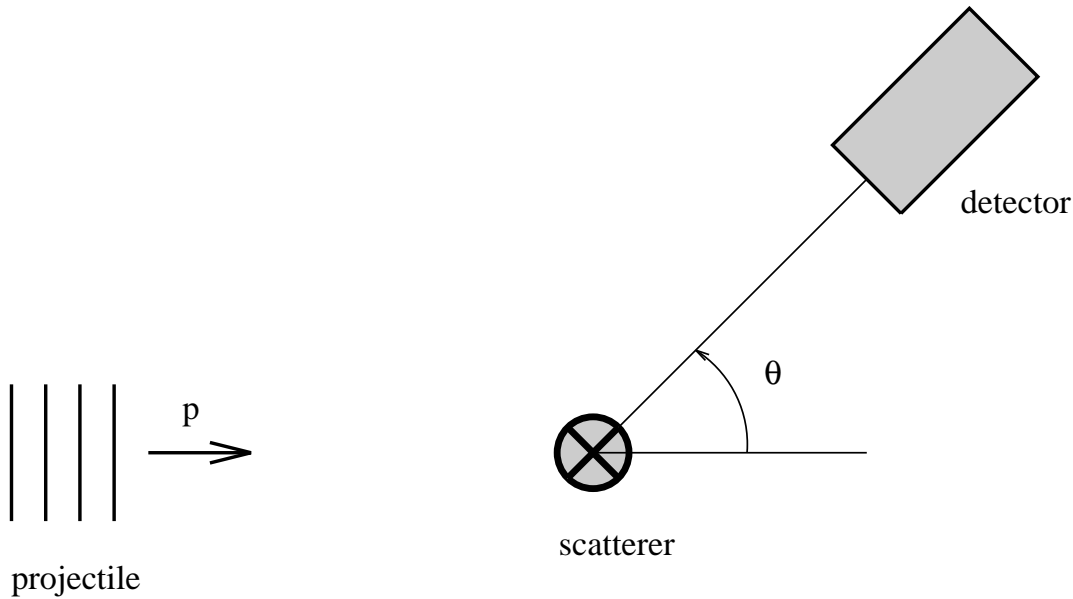}}
\caption{\sl The  geometry of the Aharonov-Bohm scattering experiment. 
The projectile comes 
in as a plane wave with momentum $p$ and scatters elastically 
from a scatterer fixed at the origin. It is assumed that the 
projectile never enters the region where
the scatterer is located. 
The cross section for the scattered projectile
is measured by a detector placed at the scattering angle $\theta$.}
\label{abverpres}
\end{figure}

The geometry of the Aharonov-Bohm scattering experiment is depicted 
in figure~\ref{abverpres}. 
It involves two particles, a projectile and a scatterer fixed at the origin.
The incoming external part of the total wave function is a plane wave 
for the projectile vanishing at the location of the scatterer.
Nontrivial scattering takes place if and only if the 
monodromy matrix ${\cal R}^2$ acting on the internal part of the 
wave function is nontrivial. 

In the abelian discrete gauge theory discussed in section~\ref{abznz},
we only encountered the abelian version.
That is,  the effect of a monodromy of the two particles 
in the internal wave function is just a phase 
\bea
{\cal R}^2 &=& e^{ 2 \pi \im \alpha}.
\eea 
The differential cross section for the quantum mechanical 
scattering experiment  involving such particles 
has been derived almost fourty years ago by Aharonov and Bohm~\cite{ahabo} 
\bea                        \label{ababcs}
\frac{{\rm d} \sigma}{{\rm d} \theta} &=&
\frac{\sin^2 {(\pi \alpha)}}{2\pi p \sin^2 (\theta/2)}    \; .
\eea
Here,  $\theta$ denotes the scattering angle and  
$p$ the momentum of the incoming plane wave of the projectile.

The particles appearing in a  nonabelian discrete 
gauge theory may in general exchange 
internal flux/charge quantum numbers when they encircle each other.
This effect is described by 
nondiagonal monodromy {\em matrices} ${\cal R}^2$ acting on multi-component 
internal wave functions. The cross section measured in 
Aharonov-Bohm scattering experiment involving these particles
is a  `nonabelian' generalization of 
the abelian one given in~(\ref{ababcs}).
An elegant closed formula for these nonabelian cross sections has been
derived by Erik Verlinde~\cite{ver1}.
The crucial insight was that the monodromy matrix ${\cal R}^2$ for 
two particles can always be diagonalized, since 
the braid group for two particles is abelian.
In the monodromy eigenbasis in which the monodromy matrix ${\cal R}^2$ 
is diagonal, the nonabelian problem then
reduces to the abelian one solved by Aharonov and Bohm.
The solution can subsequently be cast in the  basis independent form
\bea \label{Aharonov}
\frac{{\rm d} \sigma}{{\rm d} \theta}|_{\mbox{in} \rightarrow \mbox{all}} 
&=&
\frac{1}{4\pi p \sin^2(\theta/2)} \;\; 
[1-\mbox{Re} \langle \mbox{in}|{\cal R}^2|\mbox{in}\rangle],
\eea 
with $|\mbox{in} \rangle$ the normalized two particle 
incoming internal quantum state.   Note that this cross section 
boils down to~(\ref{ababcs}) for the abelian case.
We will always work in the natural two particle 
flux/charge eigenbasis being the tensor product of the single particle 
internal basis states~(\ref{quantum states}).
In fact,  in our applications the $|\mbox{in} \rangle$ state 
usually is a particular two particle flux/charge eigenstate. 
The detector  measuring the cross section~(\ref{Aharonov}) 
is a device which does not 
discriminate between the different internal `disguises' the scattered 
projectile can take. 
Specifically, in the scattering process 
discussed in section~\ref{abd2b}, the Verlinde
detector gives a signal, when the scattered pure doublet charge $\chi$ enters
the apparatus with its charge orientation either up or down.
In this sense, Verlinde's cross section~(\ref{Aharonov}) is inclusive.  

Inspired by this work, 
Lo and Preskill subsequently introduced a finer detector~\cite{preslo}.
Their device is able to distinguish between the different internal 
appearances of the projectile. 
In the scattering process studied in section~\ref{abd2b}, for example,
we can use a device, which  only gives a signal 
if the projectile enters the device with its internal charge orientation up.
The  exclusive cross section measured with such a detector can be expressed
as
\bea \label{exclus}
\frac{{\rm d} \sigma}{{\rm d} \theta}|_{\mbox{in} \rightarrow \mbox{out}} 
&=&  
\frac{1}{8 \pi p \sin^2 (\theta/2)} \; \; 
|\langle \mbox{out}|{\cal R}^{-\theta/\pi} 
({\bf 1} - {\cal R}^2) | \mbox{in} \rangle |^2,
\eea
where $|\mbox{in} \rangle$ and $|\mbox{out} \rangle$ denote
normalized two particle incoming-  and  outgoing internal quantum states.
The outgoing  state we observe depends on  
the detector we have installed, but since we only measure the projectile, 
so `half' of the out state,  
the state $| \mbox{out} \rangle$ in~(\ref{exclus})
should always be summed over a complete basis for the internal Hilbert space
of the scatterer.  
The new ingredient in the exclusive cross section~(\ref{exclus}) 
is the matrix ${\cal R}^{-\theta/\pi}$. This matrix is defined 
as the diagonal matrix in the monodromy eigenbasis, which acts as
\bea  \label{exclur}
{\cal R}^{-\theta/\pi} &:=&  e^{-\im \alpha\theta}   \qquad\qquad
\mbox{with $\alpha \in [0,1)$,}
\eea
on a monodromy eigenstate characterized by the
eigenvalue $\exp (2 \pi \im \alpha)$ under ${\cal R}^2$.
By a basis transformation, we then find the matrix elements of 
${\cal R}^{-\theta/\pi}$ in our favourite 
two particle flux/charge eigenbasis.

A peculiar property of the exclusive cross section~(\ref{exclus}) 
is that it is in general multi-valued. This is just 
a reflection of the fact that the detector can  generally 
change its nature, when it is parallel transported around
the scatterer. An apparatus that only detects projectiles with 
internal charge orientation up
in the scattering process studied in section~\ref{abd2b}, for example, 
becomes an apparatus,
which only detects projectiles with charge orientation down, after a rotation
over an angle of $ 2 \pi$ around the scatterer. 
Verlinde's detector, giving a signal independent of the internal 
`disguise' of the projectile entering the device, 
obviously does {\em not} suffer from this multi-valuedness. 
As a matter of fact, extending the aforementioned 
sum of  the $|\mbox{out} \rangle$ state  in~(\ref{exclus})
over a complete basis of the internal Hilbert space 
for the scatterer by a sum 
over a complete basis of the internal Hilbert space for the projectile
and subsequently using the partition of unity,  yields
the single valued inclusive cross section~(\ref{Aharonov}).

As a last remark, the cross sections 
for Aharonov-Bohm scattering experiments in which
the projectile and the scatterer are indistinguishable
particles contain an extra contribution 
due to conceivable exchange processes between the scatterer 
and the projectile~\cite{preslo,anyonbook}. The 
incorporation of this exchange contribution
merely amounts to diagonalizing the braid matrix ${\cal R}$ 
instead of the monodromy matrix ${\cal R}^2$.

\sectiona{$B(3,4)$ and $P(3,4)$}         \label{trubra}

In this last appendix, we give the structure of the truncated braid group 
$B(3,4)$ and the truncated colored braid group $P(3,4)$, which enter
the discussion of the nonabelian braid properties of certain three particle 
configurations in a $\bar{D}_2$ gauge theory in section~\ref{d2bqst}.

According to the general definition~(\ref{eqy})--(\ref{truncate}) given in
section~\ref{trunckbr},
the truncated braid group $B(3,4)$ for three indistinguishable particles 
is generated
by two elements $\tau_1$ and $\tau_2$ subject to the relations
\beas
\tau_1 \tau_2 \tau_1 &=& \tau_2   \tau_1 \tau_2   \\
\tau_1^4 &=& \tau_2^4 \; \, = \; \, e.
\eeas
By  explicit construction from these defining relations, which is a 
lengthy and not at all trivial job, it can be inferred
that $B(3,4)$ is a  group of order 96, which splits up into the 
following conjugacy classes 
\bea 
C_0^1 & = & \{ e \}    \\
C_0^2 & = & \{\tau_1\tau_2\tau_1\tau_2\tau_1\tau_2 \} \nn \\
C_0^3 & = & \{\tau_2^2\tau_1^2\tau_2^2\tau_1^2\} \nn \\
C_0^4 & = & \{\tau_2^2\tau_1^3\tau_2^2\tau_1^3\} \nn \\
C_1^1 & = & \{\tau_1\; , \; \tau_2\; , \; \tau_2\tau_1\tau_2^3\; , \; 
\tau_2^2\tau_1\tau_2^2\; , \; \tau_2^3\tau_1\tau_2\; , \; 
\tau_1^2\tau_2\tau_1^2\} \nn \\
C_1^2 & = & \{\tau_1^3\tau_2\tau_1^2\tau_2\; , \; 
\tau_2^3\tau_1\tau_2^2\tau_1\; , \; 
\tau_2\tau_1^3\tau_2\tau_1^2\; , \; 
\tau_2^2\tau_1^2\tau_2^2\tau_1\; , \; 
\tau_1\tau_2^3\tau_1\tau_2^2\; , \; 
\tau_1^2\tau_2^2\tau_1^2\tau_2\} \nn \\
C_1^3 & = & \{\tau_2\tau_1^3\tau_2\tau_1^3\tau_2\; , \; 
\tau_1^2\tau_2\tau_1^3\tau_2^2\tau_1\; , \; 
\tau_2^3\tau_1\tau_2^3\tau_1^2\; , \; 
\tau_1\tau_2^2\tau_1^3\tau_2^2\tau_1\; , \; 
\tau_2\tau_1\tau_2^3\tau_1^2\tau_2^2\; , \; 
\tau_2\tau_1^2\tau_2^3\tau_1^2\tau_2\} \nn \\
C_1^4 & = & \{\tau_2^2\tau_1^3\tau_2^2\; , \; 
\tau_1^2\tau_2^3\tau_1^2\; , \; 
\tau_2^3\tau_1^3\tau_2\; , \; 
\tau_1^3\; , \; \tau_2\tau_1^3\tau_2^3\; , \; 
\tau_2^3\}  \nn \\
C_2^1 & = & \{\tau_1\tau_2\; , \; \tau_2\tau_1\; , \; 
\tau_1^2\tau_2\tau_1^3\; , \; 
\tau_1^3\tau_2\tau_1^2\; , \; 
\tau_2\tau_1^2\tau_2^2\tau_1\; , \; 
\tau_2^2\tau_1\tau_2^3\; , \; 
\tau_2^3\tau_1\tau_2^2\; , \; 
\tau_1\tau_2^2\tau_1^2\tau_2\} \nn \\
C_2^2 & = & \{\tau_1^2\tau_2\tau_1^3\tau_2\tau_1\; , \; 
\tau_1\tau_2\tau_1^3\tau_2\tau_1^2\; , \; 
\tau_2\tau_1^2\tau_2^2\tau_1^3\; , \; 
\tau_1\tau_2\tau_1^2\tau_2^2\tau_1^2\; , \; \nn \\
      &   & \; \tau_2\tau_1^3\tau_2^2\tau_1^2\; , \; 
\tau_1\tau_2\tau_1^3\tau_2^2\tau_1\; , 
\; \tau_1^2\tau_2\tau_1^3\tau_2^2\; , 
\; \tau_1^2\tau_2^2\tau_1^3\tau_2\} \nn \\
C_2^3 & = & \{\tau_1^3\tau_2\tau_1^3\tau_2^2\tau_1\; , \; 
\tau_1\tau_2^2\tau_1^3\tau_2\tau_1^3\; , \; 
\tau_2\tau_1^3\tau_2^2\; , \; \tau_2^2\tau_1^3\tau_2\; , \; 
 \tau_2^3\tau_1^3\; , \; 
\tau_1\tau_2^3\tau_1^2\; , \; 
\tau_1^2\tau_2^3\tau_1\; , \; 
\tau_1^3\tau_2^3\}  \nn  \\
C_2^4 & = & \{\tau_1^3\tau_2^3\tau_1^2\; , \; 
\tau_1^2\tau_2^3\tau_1^3\; , \; 
\tau_2^3\tau_1\; , \; \tau_1\tau_2^3\; , \; 
\tau_1\tau_2\tau_1\tau_2\; , \; \tau_1^3\tau_2\; , \; 
\tau_2\tau_1^3\; , \; \tau_2\tau_1\tau_2\tau_1\} \nn \\
C_3^1 & = & \{\tau_1^2\; , \; \tau_2^2\; , \; 
\tau_1\tau_2^2\tau_1^3\; , \; 
\tau_2^2\tau_1^2\tau_2^2\; , \; 
\tau_1^2\tau_2^2\tau_1^2\; , \; 
\tau_1^3\tau_2^2\tau_1\} \nn \\
C_3^2 & = & \{\tau_2\tau_1^2\tau_2\; , \; 
\tau_1\tau_2^2\tau_1\; , \; \tau_1^2\tau_2^2\; , \; 
\tau_2^3\tau_1^2\tau_2^3\; , \; 
\tau_1^3\tau_2^2\tau_1^3\; , \; 
\tau_2^2\tau_1^2\} \nn \\
C_4^1 & = & \{\tau_1\tau_2\tau_1\; , \; \tau_1^2\tau_2\; , \; 
\tau_2^2\tau_1\; , \; \tau_2\tau_1^2\; , \; 
\tau_1\tau_2^2\; , \; \tau_1^3\tau_2\tau_1^3\; , \; 
\tau_1^3\tau_2\tau_1^3\tau_2^2\tau_1^2\; , \; \nn \\
      &   & \; \tau_2\tau_1^3\tau_2^2\tau_1\; , \; 
\tau_1^2\tau_2^2\tau_1^3\; , \; 
\tau_1\tau_2^2\tau_1^3\tau_2\; , \; 
\tau_1^3\tau_2^2\tau_1^2\; , \; 
\tau_1\tau_2\tau_1^3\tau_2^2\} \nn \\
C_4^2 & = & \{\tau_1\tau_2\tau_1\tau_2\tau_1\tau_2\tau_1\tau_2\tau_1\; , \; 
\tau_2\tau_1^2\tau_2^2\; , \; \tau_1\tau_2^2\tau_1^2\; , \; 
\tau_2^2\tau_1^2\tau_2\; , \; \tau_1^2\tau_2^2\tau_1\; , \; 
\tau_2\tau_1^3\tau_2\; , \; \nn \\
      &   & \; \tau_2^3\tau_1^3\tau_2^3\; , \; 
\tau_2^3\tau_1^2\; , \; \tau_1^3\tau_2^2\; , \; 
\tau_2^2\tau_1^3\; , \; \tau_1^2\tau_2^3\; , \; 
\tau_1\tau_2^3\tau_1\}.           \nn
\label{B34}
\eea
We organized the conjugacy classes such 
that $C_k^{i+1} = z C_k^i$, with 
$z = \tau_1\tau_2\tau_1\tau_2\tau_1\tau_2$ the 
generator of the centre of $B(3,4)$. 
The character table of the truncated braid group
$B(3,4)$ is displayed in table~\ref{tab:tab4}.

The truncated colored braid group $P(3,4)$, which  consists of
the monodromy  
operations on a configuration of three distinguishable particles,
is the subgroup of $B(3,4)$ generated by 
the elements
\bea
\gamma_{12} &=& \tau_1^2  \nn  \\
\gamma_{13} &=& \tau_1 \tau_2^2 \tau_1^{-1}= \tau_1 \tau_2^2\tau_1^3 
\label{p34mongen} \\
\gamma_{23} &=& \tau_2^2,    \nn 
\eea
which satisfy
\beas
\gamma_{12}^2 \; = \; \gamma_{13}^2 \;= \; \gamma_{23}^2 \;=\; e.
\eeas
It can be verified  that $P(3,4)$ is a group
of order 16 which splits up into the following 10 conjugacy classes
\bea          \label{p34}
\ba{lcl lcl}
C_0 &=& \{e \}  &   
\qquad  
C_1 &=& \{\gamma_{13} \gamma_{12} \gamma_{23}\} \\
C_2 &=& \{ \gamma_{23}\gamma_{12}\gamma_{23} \gamma_{12} \}           &   
\qquad
C_3 &=& \{ \gamma_{23}\gamma_{12}\gamma_{13} \} \\
C_4 &=& \{\gamma_{12} \; , \; \gamma_{23} \gamma_{12}\gamma_{23} \}   &   
\qquad
C_5 &=& \{\gamma_{23} \; , \; \gamma_{12} \gamma_{23} \gamma_{12} \}  \\
C_6 &=& \{\gamma_{13} \; , \; \gamma_{12} \gamma_{13} \gamma_{12} \}  &      
\qquad
C_7 &=& \{\gamma_{13}\gamma_{12} \; , \; \gamma_{12} \gamma_{13}  \}   \\
C_8 &=& \{\gamma_{23}\gamma_{13} \; , \; \gamma_{13}\gamma_{23}   \}  &  
\qquad
C_9 &=& \{\gamma_{12} \gamma_{23} \; , \; \gamma_{23} \gamma_{12} \} .
\ea
\eea   
It turns out that the truncated colored braid group
$P(3,4)$ is, in fact,  isomorphic to the coxeter group denoted 
as $16/8$ in~\cite{thomas}. Further, the centre
of $P(3,4)$ contained in the first four conjugacy classes 
in~(\ref{p34}) is of order four and coincides with that 
of $B(3,4)$. Finally, the character table of $P(3,4)$ is given 
in table~\ref{char}.

\newpage

\begin{table}[t]
\begin{tabular}{crrrrrrrrrrrrrrrr}
 \hline  \\[-4mm]
 & ${\ssc C_0^1}$ & ${\ssc C_0^2}$ & ${\ssc C_0^3}$ & 
            ${\ssc C_0^4}$ & ${\ssc C_1^1}$ & ${\ssc C_1^2}$ & 
            ${\ssc C_1^3}$ & ${\ssc C_1^4}$ & ${\ssc C_2^1}$ & 
            ${\ssc C^2_2}$ & ${\ssc C^3_2}$ & ${\ssc C^4_2}$ & 
            ${\ssc C_3^1}$ & ${\ssc C_3^2}$ & ${\ssc C_4^1}$ & 
            ${\ssc C_4^2}$  \\ \hline   \\[-4mm]
${\ssc \Lambda_0}$ & ${\ssc 1}$ & ${\ssc 1}$ & ${\ssc 1}$ & ${\ssc 1}$ & 
                     ${\ssc 1}$ & ${\ssc 1}$ & ${\ssc 1}$ & ${\ssc 1}$ &
                     ${\ssc 1}$ & ${\ssc 1}$ & ${\ssc 1}$ & ${\ssc 1}$ & 
                     ${\ssc 1}$ & ${\ssc 1}$ & ${\ssc 1}$ & ${\ssc 1}$   
                     \\ 
${\ssc \Lambda_1}$ & ${\ssc 1}$ & -${\ssc 1}$ & ${\ssc 1}$ & -${\ssc 1}$ & 
                     ${\ssc \im}$ & -${\ssc \im}$ & ${\ssc \im}$ & -${\ssc \im}$ & 
                    -${\ssc 1}$ &  ${\ssc 1}$ & -${\ssc 1}$ & ${\ssc 1}$ & 
                    -${\ssc 1}$ & ${\ssc 1}$ & -${\ssc \im}$ & ${\ssc \im}$ 
                     \\
${\ssc \Lambda_2}$ & ${\ssc 1}$ & ${\ssc 1}$ & ${\ssc 1}$ & ${\ssc 1}$ & 
                    -${\ssc 1}$ & -${\ssc 1}$ & -${\ssc 1}$ & -${\ssc 1}$ & 
                     ${\ssc 1}$ & ${\ssc 1}$ & ${\ssc 1}$ & ${\ssc 1}$ & 
                     ${\ssc 1}$ & ${\ssc 1}$ & -${\ssc 1}$ & -${\ssc 1}$  
                     \\
${\ssc \Lambda_3}$ & ${\ssc 1}$ & -${\ssc 1}$ & ${\ssc 1}$ & -${\ssc 1}$ & 
                    -${\ssc \im}$ & ${\ssc \im}$ & -${\ssc \im}$ & ${\ssc \im}$ & 
                    -${\ssc 1}$ & ${\ssc 1}$ & -${\ssc 1}$ & ${\ssc 1}$ & 
                    -${\ssc 1}$ & ${\ssc 1}$ & ${\ssc \im}$ & -${\ssc \im}$ 
                     \\ 
${\ssc \Lambda_4}$ & ${\ssc 2}$ & ${\ssc 2}$ & ${\ssc 2}$ & ${\ssc 2}$ & 
                     ${\ssc 0}$ & ${\ssc 0}$ & ${\ssc 0}$ & ${\ssc 0}$ & 
                    -${\ssc 1}$ & -${\ssc 1}$ & -${\ssc 1}$ & -${\ssc 1}$ & 
                     ${\ssc 2}$ & ${\ssc 2}$ & ${\ssc 0}$ & ${\ssc 0}$ \\ 
${\ssc \Lambda_5}$ & ${\ssc 2}$ & -${\ssc 2}$ & ${\ssc 2}$ & -${\ssc 2}$ & 
                     ${\ssc 0}$ & ${\ssc 0}$ & ${\ssc 0}$ & ${\ssc 0}$ & 
                     ${\ssc 1}$ & -${\ssc 1}$ & ${\ssc 1}$ & -${\ssc 1}$ & 
                    -${\ssc 2}$ & ${\ssc 2}$ & ${\ssc 0}$ & ${\ssc 0}$ \\ 
${\ssc \Lambda_6}$ & ${\ssc 2}$ & ${\ssc 2\im}$ & -${\ssc 2}$ & -${\ssc 2\im}$ & 
                   ${\ssc \oq}$ & -${\ssc \oq^*}$ & -${\ssc \oq}$ & 
                   ${\ssc \oq^*}$ & ${\ssc \im}$ & -${\ssc 1}$ & -${\ssc \im}$ & 
                   ${\ssc 1}$ & ${\ssc 0}$ & ${\ssc 0}$ & ${\ssc 0}$ & 
                   ${\ssc 0}$  \\ 
${\ssc \Lambda_7}$ & ${\ssc 2}$ & ${\ssc 2\im}$ & -${\ssc 2}$ & -${\ssc 2\im}$ & 
                    -${\ssc \oq}$ & ${\ssc \oq^*}$ & ${\ssc \oq}$ & 
                    -${\ssc \oq^*}$ & ${\ssc \im}$ & -${\ssc 1}$ & -${\ssc \im}$ & 
                     ${\ssc 1}$ & ${\ssc 0}$ & ${\ssc 0}$ & ${\ssc 0}$ & 
                     ${\ssc 0}$ \\ 
${\ssc \Lambda_8}$  & ${\ssc 2}$ & -${\ssc 2\im}$ & -${\ssc 2}$ & ${\ssc 2\im}$ & 
                      -${\ssc \oq^*}$ & ${\ssc \oq}$ & ${\ssc \oq^*}$ & 
                      -${\ssc \oq}$ & -${\ssc \im}$ & -${\ssc 1}$ & ${\ssc \im}$ & 
                      ${\ssc 1}$ & ${\ssc 0}$ & ${\ssc 0}$ & ${\ssc 0}$ & 
                      ${\ssc 0}$  \\ 
${\ssc \Lambda_9}$  & ${\ssc 2}$ & -${\ssc 2\im}$ & -${\ssc 2}$ & ${\ssc 2\im}$ & 
                      ${\ssc \oq^*}$ & -${\ssc \oq}$ & -${\ssc \oq^*}$ & 
                      ${\ssc \oq}$ & -${\ssc \im}$ & -${\ssc 1}$ & ${\ssc \im}$ & 
                      ${\ssc 1}$ & ${\ssc 0}$ & ${\ssc 0}$ & ${\ssc 0}$ & 
                      ${\ssc 0}$ \\ 
${\ssc \Lambda_{10}}$ & ${\ssc 3}$ & ${\ssc 3}$ & ${\ssc 3}$ & ${\ssc 3}$ & 
                        ${\ssc 1}$ & ${\ssc 1}$ & ${\ssc 1}$ & ${\ssc 1}$ & 
                        ${\ssc 0}$ & ${\ssc 0}$ & ${\ssc 0}$ & ${\ssc 0}$ & 
                       -${\ssc 1}$ & -${\ssc 1}$ & -${\ssc 1}$ & -${\ssc 1}$  
                        \\ 
${\ssc \Lambda_{11}}$ & ${\ssc 3}$ & -${\ssc 3}$ & ${\ssc 3}$ & -${\ssc 3}$ & 
                        ${\ssc \im}$ & -${\ssc \im}$ & ${\ssc \im}$ & -${\ssc \im}$ & 
                        ${\ssc 0}$ &  ${\ssc 0}$ & ${\ssc 0}$ & ${\ssc 0}$ & 
                        ${\ssc 1}$ & -${\ssc 1}$ & ${\ssc \im}$ & -${\ssc \im}$ 
                        \\ 
${\ssc \Lambda_{12}}$ & ${\ssc 3}$ & ${\ssc 3}$ & ${\ssc 3}$ & ${\ssc 3}$ & 
                       -${\ssc 1}$ & -${\ssc 1}$ & -${\ssc 1}$ & -${\ssc 1}$ & 
                        ${\ssc 0}$ & ${\ssc 0}$ & ${\ssc 0}$ & ${\ssc 0}$ & 
                       -${\ssc 1}$ & -${\ssc 1}$ & ${\ssc 1}$ & ${\ssc 1}$ 
                       \\ 
${\ssc \Lambda_{13}}$ & ${\ssc 3}$ & -${\ssc 3}$ & ${\ssc 3}$ & -${\ssc 3}$ & 
                        -${\ssc \im}$ & ${\ssc \im}$ & -${\ssc \im}$ & ${\ssc \im}$ & 
                         ${\ssc 0}$ & ${\ssc 0}$ & ${\ssc 0}$ & ${\ssc 0}$ & 
                         ${\ssc 1}$ & -${\ssc 1}$ & -${\ssc \im}$ & 
                         ${\ssc \im}$ \\ 
${\ssc \Lambda_{14}}$ & ${\ssc 4}$ & ${\ssc 4}$ & -${\ssc 4}$ & -${\ssc 4}$ & 
                          ${\ssc 0}$ & ${\ssc 0}$ & ${\ssc 0}$ & ${\ssc 0}$ & 
                          ${\ssc 1}$ & ${\ssc 1}$ & -${\ssc 1}$ & -${\ssc 1}$ &
                          ${\ssc 0}$ & ${\ssc 0}$ &  ${\ssc 0}$ & ${\ssc 0}$  
                          \\ 
${\ssc \Lambda_{15}}$ & ${\ssc 4}$ & -${\ssc 4}$ & -${\ssc 4}$ & ${\ssc 4}$ & 
                          ${\ssc 0}$ & ${\ssc 0}$ & ${\ssc 0}$ & ${\ssc 0}$ &
                         -${\ssc 1}$ & ${\ssc 1}$ & ${\ssc 1}$ & -${\ssc 1}$ & 
                          ${\ssc 0}$ & ${\ssc 0}$ & ${\ssc 0}$ & ${\ssc 0}$ 
                         \\[1mm] \hline
\end{tabular}  
\caption{\sl Character table of the truncated braid group 
$B(3,4)$. We used $\oq := \im+1$.}
\label{tab:tab4}
\end{table}

\begin{table}[tb]  
\begin{center}
\begin{tabular}{crrrrrrrrrrr} \hl
$P(3,4)$ & & $C_0$ & $C_1$ & $C_2$ & $C_3$ & $C_4$ & 
$C_5$ & $C_6$ & $C_7$ & $C_8$ & $C_9$    \\ \hl  \\[-4mm] 
$\Omega_0$ & & $1$ & $1$  & $1$ & $1$ & $1$ & $1$ & $1$ & $1$ & $1$ & $1$ \\ 
$\Omega_1$ & & $1$ & $1$ & $1$ & $1$ & $-1$ & $-1$ & $1$ & $-1$ & $-1$ & $1$ \\ 
$\Omega_2$ & & $1$ & $1$ & $1$ & $1$ & $-1$ & $1$ & $-1$ & $1$ & $-1$ & $-1$ \\ 
$\Omega_3$ & & $1$ & $1$ & $1$ & $1$ & $1$ & $-1$ & $-1$  & $-1$ & $1$ & $-1$ 
 \\   $\Omega_4$ & & $1$ & $-1$ & $1$ & $-1$ & $-1$ & $1$ & $1$ & $-1$ & 
$1$ & $-1$ \\ 
$\Omega_5$ & & $1$ & $-1$ & $1$ & $-1$ & $1$ & $-1$ &
$1$ & $1$ & $-1$ & $-1$ \\ 
$\Omega_6$ & & $1$ & $-1$ & $1$ & $-1$ & $1$ & $1$ & $-1$ & $-1$ &
$-1$ & $1$ \\ 
$\Omega_7$ & & $1$ & $-1$  & $1$  & $-1$  & $-1$ & $-1$ & $-1$ & $1$ & $1$ & 
             $1$ \\  
$\Omega_8$& & $2$ & $2\im$  & $-2$ & $-2\im$ & $0$ & $0$ & $0$ & $0$ & $0$
           & $0$ \\
$\Omega_{9}$& & $2$ & $-2\im$  & $-2$ & $2\im$ & $0$ & $0$ & $0$ & $0$ 
               & $0$ & $0$ \\[1mm]              \hl
\end{tabular}
\end{center}
\caption{\sl Character table of 
the truncated colored braid group $P(3,4)$.}
\label{char}
\end{table}

\chapter*{Concluding remarks and outlook}
\addcontentsline{toc}{chapter}{Concluding remarks and outlook}

In these lecture notes, we have given a thorough treatment of   
 planar gauge theories in which some continuous gauge group $G$
is broken down to a  finite subgroup $H$ by means of the Higgs mechanism.
One of the main points has been  
that the long distance physics of such a 
model is governed by a quantum group based on the
residual finite gauge group $H$, namely the quasitriangular 
Hopf algebra $D(H)$ being the result of Drinfeld's quantum double 
construction applied to the abelian algebra 
${\cal F}(H)$ of functions on the finite 
group ${ H}$.  
The different particles in the spectrum, i.e.\ magnetic vortices, 
global $H$ charges and dyonic combinations of the two, 
are in one to one correspondence with the inequivalent 
unitary irreducible representations of the quantum double $D(H)$. 
Moreover, the algebraic framework $D(H)$ provides an unified 
description of the spin, braid and fusion properties of these particles.

The implications of adding a Chern-Simons term to these spontaneously broken
models have been adressed in the references~\cite{spm1,sm,sam} 
and~\cite{thesis}. A review was beyond the scope of these notes.
Let us just briefly summarize the main results.
The distinct Chern-Simons actions $S_{\rm CS}$ 
for a compact gauge group $G$ are known to be 
in one to one correspondence with the 
elements of  the cohomology group $H^4(BG,\Z)$ of the 
classifying space $BG$ with integer
coefficients~\cite{diwi}. 
In particular, for finite groups $H$, this classification
boils down to the cohomology group $H^3(H,U(1))$ of the group $H$ 
itself with coefficients in $U(1)$.
In other words, the different Chern-Simons theories 
with finite gauge group $H$, in fact, correspond to the independent 
algebraic 3-cocycles $\omega \in H^3(H,U(1))$. 
Now suppose that we add a Chern-Simons term $S_{\rm CS} \in H^4(BG,\Z)$
to a planar gauge theory of the form~(\ref{totmod}) in which the
continuous gauge group $G$ (assumed to be simply-connected for convenience)
is spontaneously broken down to a finite 
subgroup $H$. Hence, the total action of the model becomes 
\beas
S &=& S_{\rm YMH} + S_{\rm matter} + S_{\rm CS}.
\eeas 
It can then be shown~\cite{thesis} that the long distance physics 
of this model is described by a Chern-Simons theory with finite gauge 
group $H$ and 3-cocycle $\omega \in H^3(H,U(1))$ determined by the original
Chern-Simons action $S_{\rm CS} \in H^4(BG,\Z)$ for the broken gauge group
$G$ through the natural homomorphism $H^4(BG,\Z) \rightarrow  H^3(H,U(1))$ 
induced by the inclusion $H \subset G$. The physical picture behind this  
natural homomorphism, also known as the restriction, is that the Chern-Simons 
term $S_{\rm CS}$ gives rise to additional Aharonov-Bohm interactions
for the magnetic vortices. These additional topological interactions 
are summarized by a 3-cocycle $\omega$ for the residual gauge group $H$, 
as such being `the long distance 
remnant' of the Chern-Simons term $S_{\rm CS}$ for the broken gauge group $G$.
Accordingly, the quantum double $D(H)$ related to 
the discrete $H$ gauge theory describing
the long distance physics in the absence of a Chern-Simons term is deformed
into the quasi-quantum double $D^{\omega}(H)$ in the presence
of a Chern-Simons term $S_{\rm CS}$.

For convenience, we have restricted ourselves to 2+1 dimensional Minkowski 
space time in these notes. For a discussion of discrete $H$ gauge theories
on higher genus spatial surfaces, i.e.\ surfaces with handles,
the reader is referred to~\cite{vohgs}. 
Further, most of our observations naturally extend to the 3+1 
dimensional setting in which the magnetic vortices become string-like
objects, that is, either closed or open magnetic flux tubes.

Also, we have treated the vortices, charges and dyons featuring in these 
spontaneously broken models as point particles 
in the first quantized description. Rerunning the discussion in the 
framework of canonical quantization involves the construction of 
magnetic vortex creation operators and charge creation operators 
and an analysis  of  their nontrivial commutation 
relations~\cite{bmwprep}.

An outstanding question is 
to what extent the emergence of 
the quantum double $D(H)$ is particular to the case of a 
{\em local} symmetry spontaneously broken down to a finite 
subgroup $H$. This point deserves further scrutiny,
for the discrete residual symmetries which do
arise in condensed matter systems available for experiments (such 
as nematic crystals and helium-3) are {\em global}.
In this respect, it is noteworthy that it has recently 
been pointed out~\cite{framedrag} that 
the spontaneous breakdown of a global symmetry to
a finite subgroup can lead to particles which exhibit a phenomenon 
called `internal frame dragging' when 
they are adiabatically transported around a global string. 
As a consequence, these particles scatter with Aharonov-Bohm
like cross sections off a global string.
Something similar happens in superfluid helium-3~\cite{khazan}.
See also reference~\cite{davis} in this connection.
Hence, it seems that just as in the local case,
the spectrum of a model with a residual global discrete 
symmetry group $H$ may feature $H$ charges that 
can be detected at arbitrary large distances by Aharonov-Bohm 
experiments with the global strings. Furthermore, the global strings 
labeled by the group elements of $H$ display the global analogue
of flux metamorphosis.
All this suggests that also in the global case the 
(semi-classical) long distance physics is governed by the quantum  
double $D(H)$.

To conclude, another obvious next step is to generalize the 
quantum double construction for finite groups to continuous groups.  
The quantum double related to the semidirect product group 
$U(1) \times_{\rm s.d.} \Z_2$, for instance, may be relevant 
in the discussion of Alice electrodynamics. 
Particularly interesting in this context is the case of 
2+1 dimensional gravity.
As in a discrete gauge theory, the interactions 
between the massive and/or spinning  particles 
in the spectrum of 2+1 dimensional gravity
are purely topological, e.g.\ \cite{dj,h2d,wit2dg}.
There are indications that the algebraic structure related to this 
topological field theory is the quantum double based on the  
2+1 dimensional homogeneous Lorentz group $SO(2,1)$.
These matters are currently under active investigation.

\section*{Acknowledgments} 

These notes are based on a chapter of the PhD thesis~\cite{thesis} of one
of the authors (M. de W.P.). F.A.B. would like to thank the
organizers of the CRM-CAP Banff Summer School `Particles and Fields 1994'
for a most stimulating research meeting in a terrific setting. 
We gratefully acknowledge helpful discussions with Danny Birmingham, 
Hoi-Kwong Lo, Nathalie Muller, Arjan van der Sijs, Peter van Driel 
and Alain Verberkmoes. We would also like to mention that the concept
of truncated braid groups was developed in collaboration with 
Peter van Driel~\cite{pema}.

\end{document}